\pdfoutput=1

\documentclass{entcs}
\usepackage{prentcsmacro}

\renewcommand{\firstheadline}{}
\renewcommand{\firstfootline}{}

\def\lastname{Carette, Chen, Choudhury, and Sabry}

\usepackage{bbold}
\usepackage{bussproofs}
\usepackage{keystroke}
\usepackage{comment}
\usepackage{tikz}
\usepackage[inline]{enumitem}
\usepackage{stmaryrd}

\usepackage[conor]{agda}
\usepackage[mathscr]{euscript}
\usepackage[euler]{textgreek}
\usepackage{mathabx}
\usepackage{isomath}

\usepackage{microtype}
\usepackage{etoolbox}

\makeatletter
\newcommand*\NoIndentAfterEnv[1]{%
  \AfterEndEnvironment{#1}{\par\@afterindentfalse\@afterheading}}
\makeatother

\NoIndentAfterEnv{code}
\NoIndentAfterEnv{figure}

\DeclareUnicodeCharacter{120793}{$\mathbb{1}$}
\DeclareUnicodeCharacter{120794}{$\mathbb{2}$}
\DeclareUnicodeCharacter{9726}{$\sqbullet$}
\DeclareUnicodeCharacter{120792}{$\mathbb{0}$}
\DeclareUnicodeCharacter{119932}{$\mathbfit{U}$}
\DeclareUnicodeCharacter{119984}{$\mathcal{U}$}
\DeclareUnicodeCharacter{8988}{$\ulcorner$}
\DeclareUnicodeCharacter{8989}{$\urcorner$}

\newcommand{\byiso}[1]{{\leftrightarrow₁}{\langle} ~#1~ \rangle}
\newcommand{\byisotwo}[1]{{\leftrightarrow₂}{\langle} ~#1~ \rangle}

\newcommand{\iso}{\leftrightarrow₁}
\newcommand{\isotwo}{\leftrightarrow₂}
\newcommand{\isothree}{\leftrightarrow₃}

\newcommand{\ot}{\mathbb{1}}
\newcommand{\bt}{\mathbb{2}}

\let\oldPi\Pi
\renewcommand{\Pi}{\mathrm\oldPi}
\newcommand{\PiTwo}{\ensuremath{\mathrm\Pi_{\mathbb{2}}}}

\begin{document}

\begin{frontmatter}
  \title{From Reversible Programs to \\ Univalent Universes and Back}
  \author{Jacques Carette}
  \address{McMaster University}
  \author{Chao-Hong Chen}
  \author{Vikraman Choudhury}
  \author{Amr Sabry}
  \address{Indiana University}

  \begin{abstract}
    We establish a close connection between a reversible programming
    language based on type isomorphisms and a formally presented
    univalent universe. The correspondence relates combinators
    witnessing type isomorphisms in the programming language to paths in
    the univalent universe; and combinator optimizations in the
    programming language to 2-paths in the univalent universe. The
    result suggests a simple computational interpretation of paths and
    of univalence in terms of familiar programming constructs whenever
    the universe in question is computable.
  \end{abstract}
\end{frontmatter}

\AgdaHide{
\begin{code}%
\>[0]\AgdaSymbol{\{{-}\#}\AgdaSpace{}%
\AgdaKeyword{OPTIONS}\AgdaSpace{}%
\AgdaOption{{-}{-}without{-}K}\AgdaSpace{}%
\AgdaOption{{-}{-}type{-}in{-}type}\AgdaSpace{}%
\AgdaOption{{-}{-}allow{-}unsolved{-}metas}\AgdaSpace{}%
\AgdaSymbol{\#{-}\}}\<%
\\
\>[0]\AgdaKeyword{module}\AgdaSpace{}%
\AgdaModule{upi}\AgdaSpace{}%
\AgdaKeyword{where}\<%
\\
\\
\>[0]\AgdaFunction{𝒰}\AgdaSpace{}%
\AgdaSymbol{=}\AgdaSpace{}%
\AgdaPrimitiveType{Set}\<%
\\
\\
\>[0]\AgdaKeyword{data}\AgdaSpace{}%
\AgdaDatatype{⊥}\AgdaSpace{}%
\AgdaSymbol{:}\AgdaSpace{}%
\AgdaFunction{𝒰}\AgdaSpace{}%
\AgdaKeyword{where}\<%
\\
\\
\>[0]\AgdaKeyword{record}\AgdaSpace{}%
\AgdaRecord{⊤}\AgdaSpace{}%
\AgdaSymbol{:}\AgdaSpace{}%
\AgdaFunction{𝒰}\AgdaSpace{}%
\AgdaKeyword{where}\<%
\\
\>[0][@{}l@{\AgdaIndent{0}}]%
\>[2]\AgdaKeyword{constructor}\AgdaSpace{}%
\AgdaInductiveConstructor{tt}\<%
\\
\\
\>[0]\AgdaKeyword{record}\AgdaSpace{}%
\AgdaRecord{Σ}\AgdaSpace{}%
\AgdaSymbol{(}\AgdaBound{A}\AgdaSpace{}%
\AgdaSymbol{:}\AgdaSpace{}%
\AgdaFunction{𝒰}\AgdaSymbol{)}\AgdaSpace{}%
\AgdaSymbol{(}\AgdaBound{B}\AgdaSpace{}%
\AgdaSymbol{:}\AgdaSpace{}%
\AgdaBound{A}\AgdaSpace{}%
\AgdaSymbol{→}\AgdaSpace{}%
\AgdaFunction{𝒰}\AgdaSymbol{)}\AgdaSpace{}%
\AgdaSymbol{:}\AgdaSpace{}%
\AgdaFunction{𝒰}\AgdaSpace{}%
\AgdaKeyword{where}\<%
\\
\>[0][@{}l@{\AgdaIndent{0}}]%
\>[2]\AgdaKeyword{constructor}\AgdaSpace{}%
\AgdaInductiveConstructor{\_,\_}\<%
\\
\>[0][@{}l@{\AgdaIndent{0}}]%
\>[2]\AgdaKeyword{field}\<%
\\
\>[2][@{}l@{\AgdaIndent{0}}]%
\>[4]\AgdaField{pr₁}\AgdaSpace{}%
\AgdaSymbol{:}\AgdaSpace{}%
\AgdaBound{A}\<%
\\
\>[2][@{}l@{\AgdaIndent{0}}]%
\>[4]\AgdaField{pr₂}\AgdaSpace{}%
\AgdaSymbol{:}\AgdaSpace{}%
\AgdaBound{B}\AgdaSpace{}%
\AgdaField{pr₁}\<%
\\
\\
\>[0]\AgdaKeyword{open}\AgdaSpace{}%
\AgdaModule{Σ}\AgdaSpace{}%
\AgdaKeyword{public}\<%
\\
\>[0]\AgdaKeyword{infixr}\AgdaSpace{}%
\AgdaNumber{4}\AgdaSpace{}%
\AgdaInductiveConstructor{\_,\_}\<%
\\
\>[0]\AgdaKeyword{syntax}\AgdaSpace{}%
\AgdaRecord{Σ} A \AgdaSymbol{(λ} a \AgdaSymbol{→} B\AgdaSymbol{)}\AgdaSpace{}%
\AgdaSymbol{=} Σ[ a ∶ A ] B

\AgdaKeyword{infix}\AgdaSpace{}%
\AgdaNumber{2}\AgdaSpace{}%
\AgdaFunction{\_×\_}\<%
\\
\>[0]\AgdaFunction{\_×\_}\AgdaSpace{}%
\AgdaSymbol{:}\AgdaSpace{}%
\AgdaSymbol{(}\AgdaBound{A}\AgdaSpace{}%
\AgdaBound{B}\AgdaSpace{}%
\AgdaSymbol{:}\AgdaSpace{}%
\AgdaFunction{𝒰}\AgdaSymbol{)}\AgdaSpace{}%
\AgdaSymbol{→}\AgdaSpace{}%
\AgdaFunction{𝒰}\<%
\\
\>[0]\AgdaBound{A}\AgdaSpace{}%
\AgdaFunction{×}\AgdaSpace{}%
\AgdaBound{B}\AgdaSpace{}%
\AgdaSymbol{=}\AgdaSpace{}%
\AgdaRecord{Σ}\AgdaSpace{}%
\AgdaBound{A}\AgdaSpace{}%
\AgdaSymbol{(λ}\AgdaSpace{}%
\AgdaBound{\_}\AgdaSpace{}%
\AgdaSymbol{→}\AgdaSpace{}%
\AgdaBound{B}\AgdaSymbol{)}\<%
\\
\\
\>[0]\AgdaKeyword{data}\AgdaSpace{}%
\AgdaDatatype{\_+\_}\AgdaSpace{}%
\AgdaSymbol{(}\AgdaBound{A}\AgdaSpace{}%
\AgdaBound{B}\AgdaSpace{}%
\AgdaSymbol{:}\AgdaSpace{}%
\AgdaFunction{𝒰}\AgdaSymbol{)}\AgdaSpace{}%
\AgdaSymbol{:}\AgdaSpace{}%
\AgdaFunction{𝒰}\AgdaSpace{}%
\AgdaKeyword{where}\<%
\\
\>[0][@{}l@{\AgdaIndent{0}}]%
\>[2]\AgdaInductiveConstructor{inl}\AgdaSpace{}%
\AgdaSymbol{:}\AgdaSpace{}%
\AgdaSymbol{(}\AgdaBound{x}\AgdaSpace{}%
\AgdaSymbol{:}\AgdaSpace{}%
\AgdaBound{A}\AgdaSymbol{)}\AgdaSpace{}%
\AgdaSymbol{→}\AgdaSpace{}%
\AgdaBound{A}\AgdaSpace{}%
\AgdaDatatype{+}\AgdaSpace{}%
\AgdaBound{B}\<%
\\
\>[0][@{}l@{\AgdaIndent{0}}]%
\>[2]\AgdaInductiveConstructor{inr}\AgdaSpace{}%
\AgdaSymbol{:}\AgdaSpace{}%
\AgdaSymbol{(}\AgdaBound{y}\AgdaSpace{}%
\AgdaSymbol{:}\AgdaSpace{}%
\AgdaBound{B}\AgdaSymbol{)}\AgdaSpace{}%
\AgdaSymbol{→}\AgdaSpace{}%
\AgdaBound{A}\AgdaSpace{}%
\AgdaDatatype{+}\AgdaSpace{}%
\AgdaBound{B}\<%
\\
\\
\>[0]\AgdaFunction{Π}\AgdaSpace{}%
\AgdaSymbol{:}\AgdaSpace{}%
\AgdaSymbol{(}\AgdaBound{A}\AgdaSpace{}%
\AgdaSymbol{:}\AgdaSpace{}%
\AgdaFunction{𝒰}\AgdaSymbol{)}\AgdaSpace{}%
\AgdaSymbol{(}\AgdaBound{B}\AgdaSpace{}%
\AgdaSymbol{:}\AgdaSpace{}%
\AgdaBound{A}\AgdaSpace{}%
\AgdaSymbol{→}\AgdaSpace{}%
\AgdaFunction{𝒰}\AgdaSymbol{)}\AgdaSpace{}%
\AgdaSymbol{→}\AgdaSpace{}%
\AgdaFunction{𝒰}\<%
\\
\>[0]\AgdaFunction{Π}\AgdaSpace{}%
\AgdaBound{A}\AgdaSpace{}%
\AgdaBound{B}\AgdaSpace{}%
\AgdaSymbol{=}\AgdaSpace{}%
\AgdaSymbol{(}\AgdaBound{a}\AgdaSpace{}%
\AgdaSymbol{:}\AgdaSpace{}%
\AgdaBound{A}\AgdaSymbol{)}\AgdaSpace{}%
\AgdaSymbol{→}\AgdaSpace{}%
\AgdaBound{B}\AgdaSpace{}%
\AgdaBound{a}\<%
\\
\\
\>[0]\AgdaKeyword{syntax}\AgdaSpace{}%
\AgdaFunction{Π} A \AgdaSymbol{(λ} a \AgdaSymbol{→} B\AgdaSymbol{)}\AgdaSpace{}%
\AgdaSymbol{=} Π[ a ∶ A ] B

\AgdaFunction{id}\AgdaSpace{}%
\AgdaSymbol{:}\AgdaSpace{}%
\AgdaSymbol{\{}\AgdaBound{A}\AgdaSpace{}%
\AgdaSymbol{:}\AgdaSpace{}%
\AgdaFunction{𝒰}\AgdaSymbol{\}}\AgdaSpace{}%
\AgdaSymbol{→}\AgdaSpace{}%
\AgdaBound{A}\AgdaSpace{}%
\AgdaSymbol{→}\AgdaSpace{}%
\AgdaBound{A}\<%
\\
\>[0]\AgdaFunction{id}\AgdaSpace{}%
\AgdaBound{a}\AgdaSpace{}%
\AgdaSymbol{=}\AgdaSpace{}%
\AgdaBound{a}\<%
\\
\\
\>[0]\AgdaKeyword{infix}\AgdaSpace{}%
\AgdaNumber{4}\AgdaSpace{}%
\AgdaFunction{\_∘\_}\<%
\\
\>[0]\AgdaFunction{\_∘\_}%
\>[119I]\AgdaSymbol{:}\AgdaSpace{}%
\AgdaSymbol{\{}\AgdaBound{A}\AgdaSpace{}%
\AgdaSymbol{:}\AgdaSpace{}%
\AgdaFunction{𝒰}\AgdaSymbol{\}}\AgdaSpace{}%
\AgdaSymbol{\{}\AgdaBound{B}\AgdaSpace{}%
\AgdaSymbol{:}\AgdaSpace{}%
\AgdaBound{A}\AgdaSpace{}%
\AgdaSymbol{→}\AgdaSpace{}%
\AgdaFunction{𝒰}\AgdaSymbol{\}}\AgdaSpace{}%
\AgdaSymbol{\{}\AgdaBound{C}\AgdaSpace{}%
\AgdaSymbol{:}\AgdaSpace{}%
\AgdaSymbol{\{}\AgdaBound{a}\AgdaSpace{}%
\AgdaSymbol{:}\AgdaSpace{}%
\AgdaBound{A}\AgdaSymbol{\}}\AgdaSpace{}%
\AgdaSymbol{→}\AgdaSpace{}%
\AgdaBound{B}\AgdaSpace{}%
\AgdaBound{a}\AgdaSpace{}%
\AgdaSymbol{→}\AgdaSpace{}%
\AgdaFunction{𝒰}\AgdaSymbol{\}}\<%
\\
\>[0][@{}l@{\AgdaIndent{0}}]\<[119I]%
\>[4]\AgdaSymbol{→}\AgdaSpace{}%
\AgdaSymbol{(}\AgdaBound{g}\AgdaSpace{}%
\AgdaSymbol{:}\AgdaSpace{}%
\AgdaSymbol{\{}\AgdaBound{a}\AgdaSpace{}%
\AgdaSymbol{:}\AgdaSpace{}%
\AgdaBound{A}\AgdaSymbol{\}}\AgdaSpace{}%
\AgdaSymbol{→}\AgdaSpace{}%
\AgdaSymbol{(}\AgdaBound{b}\AgdaSpace{}%
\AgdaSymbol{:}\AgdaSpace{}%
\AgdaBound{B}\AgdaSpace{}%
\AgdaBound{a}\AgdaSymbol{)}\AgdaSpace{}%
\AgdaSymbol{→}\AgdaSpace{}%
\AgdaBound{C}\AgdaSpace{}%
\AgdaBound{b}\AgdaSymbol{)}\AgdaSpace{}%
\AgdaSymbol{(}\AgdaBound{f}\AgdaSpace{}%
\AgdaSymbol{:}\AgdaSpace{}%
\AgdaSymbol{(}\AgdaBound{a}\AgdaSpace{}%
\AgdaSymbol{:}\AgdaSpace{}%
\AgdaBound{A}\AgdaSymbol{)}\AgdaSpace{}%
\AgdaSymbol{→}\AgdaSpace{}%
\AgdaBound{B}\AgdaSpace{}%
\AgdaBound{a}\AgdaSymbol{)}\<%
\\
\>[0][@{}l@{\AgdaIndent{0}}]%
\>[4]\AgdaSymbol{→}\AgdaSpace{}%
\AgdaSymbol{(}\AgdaBound{a}\AgdaSpace{}%
\AgdaSymbol{:}\AgdaSpace{}%
\AgdaBound{A}\AgdaSymbol{)}\AgdaSpace{}%
\AgdaSymbol{→}\AgdaSpace{}%
\AgdaBound{C}\AgdaSpace{}%
\AgdaSymbol{(}\AgdaBound{f}\AgdaSpace{}%
\AgdaBound{a}\AgdaSymbol{)}\<%
\\
\>[0]\AgdaBound{g}\AgdaSpace{}%
\AgdaFunction{∘}\AgdaSpace{}%
\AgdaBound{f}\AgdaSpace{}%
\AgdaSymbol{=}\AgdaSpace{}%
\AgdaSymbol{λ}\AgdaSpace{}%
\AgdaBound{a}\AgdaSpace{}%
\AgdaSymbol{→}\AgdaSpace{}%
\AgdaBound{g}\AgdaSpace{}%
\AgdaSymbol{(}\AgdaBound{f}\AgdaSpace{}%
\AgdaBound{a}\AgdaSymbol{)}\<%
\\
\\
\>[0]\AgdaKeyword{infix}\AgdaSpace{}%
\AgdaNumber{3}\AgdaSpace{}%
\AgdaDatatype{\_==\_}\<%
\\
\>[0]\AgdaKeyword{data}\AgdaSpace{}%
\AgdaDatatype{\_==\_}\AgdaSpace{}%
\AgdaSymbol{\{}\AgdaBound{A}\AgdaSpace{}%
\AgdaSymbol{:}\AgdaSpace{}%
\AgdaFunction{𝒰}\AgdaSymbol{\}}\AgdaSpace{}%
\AgdaSymbol{:}\AgdaSpace{}%
\AgdaBound{A}\AgdaSpace{}%
\AgdaSymbol{→}\AgdaSpace{}%
\AgdaBound{A}\AgdaSpace{}%
\AgdaSymbol{→}\AgdaSpace{}%
\AgdaFunction{𝒰}\AgdaSpace{}%
\AgdaKeyword{where}\<%
\\
\>[0][@{}l@{\AgdaIndent{0}}]%
\>[2]\AgdaInductiveConstructor{refl}\AgdaSpace{}%
\AgdaSymbol{:}\AgdaSpace{}%
\AgdaSymbol{(}\AgdaBound{a}\AgdaSpace{}%
\AgdaSymbol{:}\AgdaSpace{}%
\AgdaBound{A}\AgdaSymbol{)}\AgdaSpace{}%
\AgdaSymbol{→}\AgdaSpace{}%
\AgdaBound{a}\AgdaSpace{}%
\AgdaDatatype{==}\AgdaSpace{}%
\AgdaBound{a}\<%
\\
\\
\>[0]\AgdaKeyword{infix}\AgdaSpace{}%
\AgdaNumber{100}\AgdaSpace{}%
\AgdaFunction{!\_}\<%
\\
\>[0]\AgdaFunction{!\_}\AgdaSpace{}%
\AgdaSymbol{:}\AgdaSpace{}%
\AgdaSymbol{\{}\AgdaBound{A}\AgdaSpace{}%
\AgdaSymbol{:}\AgdaSpace{}%
\AgdaFunction{𝒰}\AgdaSymbol{\}}\AgdaSpace{}%
\AgdaSymbol{\{}\AgdaBound{a}\AgdaSpace{}%
\AgdaBound{b}\AgdaSpace{}%
\AgdaSymbol{:}\AgdaSpace{}%
\AgdaBound{A}\AgdaSymbol{\}}\AgdaSpace{}%
\AgdaSymbol{→}\AgdaSpace{}%
\AgdaSymbol{(}\AgdaBound{a}\AgdaSpace{}%
\AgdaDatatype{==}\AgdaSpace{}%
\AgdaBound{b}\AgdaSymbol{)}\AgdaSpace{}%
\AgdaSymbol{→}\AgdaSpace{}%
\AgdaSymbol{(}\AgdaBound{b}\AgdaSpace{}%
\AgdaDatatype{==}\AgdaSpace{}%
\AgdaBound{a}\AgdaSymbol{)}\<%
\\
\>[0]\AgdaFunction{!\_}\AgdaSpace{}%
\AgdaSymbol{(}\AgdaInductiveConstructor{refl}\AgdaSpace{}%
\AgdaSymbol{\_)}\AgdaSpace{}%
\AgdaSymbol{=}\AgdaSpace{}%
\AgdaInductiveConstructor{refl}\AgdaSpace{}%
\AgdaSymbol{\_}\<%
\\
\\
\>[0]\AgdaKeyword{infixr}\AgdaSpace{}%
\AgdaNumber{80}\AgdaSpace{}%
\AgdaFunction{\_◾\_}\<%
\\
\>[0]\AgdaFunction{\_◾\_}\AgdaSpace{}%
\AgdaSymbol{:}\AgdaSpace{}%
\AgdaSymbol{\{}\AgdaBound{A}\AgdaSpace{}%
\AgdaSymbol{:}\AgdaSpace{}%
\AgdaFunction{𝒰}\AgdaSymbol{\}}\AgdaSpace{}%
\AgdaSymbol{\{}\AgdaBound{a}\AgdaSpace{}%
\AgdaBound{b}\AgdaSpace{}%
\AgdaBound{c}\AgdaSpace{}%
\AgdaSymbol{:}\AgdaSpace{}%
\AgdaBound{A}\AgdaSymbol{\}}\AgdaSpace{}%
\AgdaSymbol{→}\AgdaSpace{}%
\AgdaSymbol{(}\AgdaBound{a}\AgdaSpace{}%
\AgdaDatatype{==}\AgdaSpace{}%
\AgdaBound{b}\AgdaSymbol{)}\AgdaSpace{}%
\AgdaSymbol{→}\AgdaSpace{}%
\AgdaSymbol{(}\AgdaBound{b}\AgdaSpace{}%
\AgdaDatatype{==}\AgdaSpace{}%
\AgdaBound{c}\AgdaSymbol{)}\AgdaSpace{}%
\AgdaSymbol{→}\AgdaSpace{}%
\AgdaSymbol{(}\AgdaBound{a}\AgdaSpace{}%
\AgdaDatatype{==}\AgdaSpace{}%
\AgdaBound{c}\AgdaSymbol{)}\<%
\\
\>[0]\AgdaFunction{\_◾\_}\AgdaSpace{}%
\AgdaSymbol{(}\AgdaInductiveConstructor{refl}\AgdaSpace{}%
\AgdaSymbol{\_)}\AgdaSpace{}%
\AgdaSymbol{(}\AgdaInductiveConstructor{refl}\AgdaSpace{}%
\AgdaSymbol{\_)}\AgdaSpace{}%
\AgdaSymbol{=}\AgdaSpace{}%
\AgdaInductiveConstructor{refl}\AgdaSpace{}%
\AgdaSymbol{\_}\<%
\\
\\
\>[0]\AgdaKeyword{infix}\AgdaSpace{}%
\AgdaNumber{3}\AgdaSpace{}%
\AgdaFunction{\_∼\_}\<%
\\
\>[0]\AgdaFunction{\_∼\_}\AgdaSpace{}%
\AgdaSymbol{:}\AgdaSpace{}%
\AgdaSymbol{\{}\AgdaBound{A}\AgdaSpace{}%
\AgdaSymbol{:}\AgdaSpace{}%
\AgdaFunction{𝒰}\AgdaSymbol{\}}\AgdaSpace{}%
\AgdaSymbol{\{}\AgdaBound{B}\AgdaSpace{}%
\AgdaSymbol{:}\AgdaSpace{}%
\AgdaBound{A}\AgdaSpace{}%
\AgdaSymbol{→}\AgdaSpace{}%
\AgdaFunction{𝒰}\AgdaSymbol{\}}\AgdaSpace{}%
\AgdaSymbol{(}\AgdaBound{f}\AgdaSpace{}%
\AgdaBound{g}\AgdaSpace{}%
\AgdaSymbol{:}\AgdaSpace{}%
\AgdaSymbol{(}\AgdaBound{a}\AgdaSpace{}%
\AgdaSymbol{:}\AgdaSpace{}%
\AgdaBound{A}\AgdaSymbol{)}\AgdaSpace{}%
\AgdaSymbol{→}\AgdaSpace{}%
\AgdaBound{B}\AgdaSpace{}%
\AgdaBound{a}\AgdaSymbol{)}\AgdaSpace{}%
\AgdaSymbol{→}\AgdaSpace{}%
\AgdaFunction{𝒰}\<%
\\
\>[0]\AgdaFunction{\_∼\_}\AgdaSpace{}%
\AgdaSymbol{\{}\AgdaBound{A}\AgdaSymbol{\}}\AgdaSpace{}%
\AgdaBound{f}\AgdaSpace{}%
\AgdaBound{g}\AgdaSpace{}%
\AgdaSymbol{=}\AgdaSpace{}%
\AgdaSymbol{(}\AgdaBound{a}\AgdaSpace{}%
\AgdaSymbol{:}\AgdaSpace{}%
\AgdaBound{A}\AgdaSymbol{)}\AgdaSpace{}%
\AgdaSymbol{→}\AgdaSpace{}%
\AgdaBound{f}\AgdaSpace{}%
\AgdaBound{a}\AgdaSpace{}%
\AgdaDatatype{==}\AgdaSpace{}%
\AgdaBound{g}\AgdaSpace{}%
\AgdaBound{a}\<%
\\
\\
\>[0]\AgdaFunction{coe}\AgdaSpace{}%
\AgdaSymbol{:}\AgdaSpace{}%
\AgdaSymbol{\{}\AgdaBound{A}\AgdaSpace{}%
\AgdaBound{B}\AgdaSpace{}%
\AgdaSymbol{:}\AgdaSpace{}%
\AgdaFunction{𝒰}\AgdaSymbol{\}}\AgdaSpace{}%
\AgdaSymbol{(}\AgdaBound{p}\AgdaSpace{}%
\AgdaSymbol{:}\AgdaSpace{}%
\AgdaBound{A}\AgdaSpace{}%
\AgdaDatatype{==}\AgdaSpace{}%
\AgdaBound{B}\AgdaSymbol{)}\AgdaSpace{}%
\AgdaSymbol{→}\AgdaSpace{}%
\AgdaBound{A}\AgdaSpace{}%
\AgdaSymbol{→}\AgdaSpace{}%
\AgdaBound{B}\<%
\\
\>[0]\AgdaFunction{coe}\AgdaSpace{}%
\AgdaSymbol{(}\AgdaInductiveConstructor{refl}\AgdaSpace{}%
\AgdaBound{A}\AgdaSymbol{)}\AgdaSpace{}%
\AgdaSymbol{=}\AgdaSpace{}%
\AgdaFunction{id}\<%
\\
\\
\>[0]\AgdaFunction{ap}\AgdaSpace{}%
\AgdaSymbol{:}\AgdaSpace{}%
\AgdaSymbol{\{}\AgdaBound{A}\AgdaSpace{}%
\AgdaBound{B}\AgdaSpace{}%
\AgdaSymbol{:}\AgdaSpace{}%
\AgdaFunction{𝒰}\AgdaSymbol{\}}\AgdaSpace{}%
\AgdaSymbol{\{}\AgdaBound{x}\AgdaSpace{}%
\AgdaBound{y}\AgdaSpace{}%
\AgdaSymbol{:}\AgdaSpace{}%
\AgdaBound{A}\AgdaSymbol{\}}\AgdaSpace{}%
\AgdaSymbol{→}\AgdaSpace{}%
\AgdaSymbol{(}\AgdaBound{f}\AgdaSpace{}%
\AgdaSymbol{:}\AgdaSpace{}%
\AgdaBound{A}\AgdaSpace{}%
\AgdaSymbol{→}\AgdaSpace{}%
\AgdaBound{B}\AgdaSymbol{)}\AgdaSpace{}%
\AgdaSymbol{(}\AgdaBound{p}\AgdaSpace{}%
\AgdaSymbol{:}\AgdaSpace{}%
\AgdaBound{x}\AgdaSpace{}%
\AgdaDatatype{==}\AgdaSpace{}%
\AgdaBound{y}\AgdaSymbol{)}\AgdaSpace{}%
\AgdaSymbol{→}\AgdaSpace{}%
\AgdaBound{f}\AgdaSpace{}%
\AgdaBound{x}\AgdaSpace{}%
\AgdaDatatype{==}\AgdaSpace{}%
\AgdaBound{f}\AgdaSpace{}%
\AgdaBound{y}\<%
\\
\>[0]\AgdaFunction{ap}\AgdaSpace{}%
\AgdaBound{f}\AgdaSpace{}%
\AgdaSymbol{(}\AgdaInductiveConstructor{refl}\AgdaSpace{}%
\AgdaBound{x}\AgdaSymbol{)}\AgdaSpace{}%
\AgdaSymbol{=}\AgdaSpace{}%
\AgdaInductiveConstructor{refl}\AgdaSpace{}%
\AgdaSymbol{(}\AgdaBound{f}\AgdaSpace{}%
\AgdaBound{x}\AgdaSymbol{)}\<%
\\
\\
\>[0]\AgdaFunction{transport}\AgdaSpace{}%
\AgdaSymbol{:}\AgdaSpace{}%
\AgdaSymbol{\{}\AgdaBound{A}\AgdaSpace{}%
\AgdaSymbol{:}\AgdaSpace{}%
\AgdaFunction{𝒰}\AgdaSymbol{\}}\AgdaSpace{}%
\AgdaSymbol{(}\AgdaBound{P}\AgdaSpace{}%
\AgdaSymbol{:}\AgdaSpace{}%
\AgdaBound{A}\AgdaSpace{}%
\AgdaSymbol{→}\AgdaSpace{}%
\AgdaFunction{𝒰}\AgdaSymbol{)}\AgdaSpace{}%
\AgdaSymbol{\{}\AgdaBound{x}\AgdaSpace{}%
\AgdaBound{y}\AgdaSpace{}%
\AgdaSymbol{:}\AgdaSpace{}%
\AgdaBound{A}\AgdaSymbol{\}}\AgdaSpace{}%
\AgdaSymbol{→}\AgdaSpace{}%
\AgdaBound{x}\AgdaSpace{}%
\AgdaDatatype{==}\AgdaSpace{}%
\AgdaBound{y}\AgdaSpace{}%
\AgdaSymbol{→}\AgdaSpace{}%
\AgdaBound{P}\AgdaSpace{}%
\AgdaBound{x}\AgdaSpace{}%
\AgdaSymbol{→}\AgdaSpace{}%
\AgdaBound{P}\AgdaSpace{}%
\AgdaBound{y}\<%
\\
\>[0]\AgdaFunction{transport}\AgdaSpace{}%
\AgdaBound{P}\AgdaSpace{}%
\AgdaSymbol{=}\AgdaSpace{}%
\AgdaFunction{coe}\AgdaSpace{}%
\AgdaFunction{∘}\AgdaSpace{}%
\AgdaFunction{ap}\AgdaSpace{}%
\AgdaBound{P}\<%
\\
\\
\>[0]\AgdaFunction{PathOver}\AgdaSpace{}%
\AgdaSymbol{:}\AgdaSpace{}%
\AgdaSymbol{\{}\AgdaBound{A}\AgdaSpace{}%
\AgdaSymbol{:}\AgdaSpace{}%
\AgdaFunction{𝒰}\AgdaSymbol{\}}\AgdaSpace{}%
\AgdaSymbol{(}\AgdaBound{P}\AgdaSpace{}%
\AgdaSymbol{:}\AgdaSpace{}%
\AgdaBound{A}\AgdaSpace{}%
\AgdaSymbol{→}\AgdaSpace{}%
\AgdaFunction{𝒰}\AgdaSymbol{)}\AgdaSpace{}%
\AgdaSymbol{\{}\AgdaBound{x}\AgdaSpace{}%
\AgdaBound{y}\AgdaSpace{}%
\AgdaSymbol{:}\AgdaSpace{}%
\AgdaBound{A}\AgdaSymbol{\}}\AgdaSpace{}%
\AgdaSymbol{(}\AgdaBound{p}\AgdaSpace{}%
\AgdaSymbol{:}\AgdaSpace{}%
\AgdaBound{x}\AgdaSpace{}%
\AgdaDatatype{==}\AgdaSpace{}%
\AgdaBound{y}\AgdaSymbol{)}\AgdaSpace{}%
\AgdaSymbol{(}\AgdaBound{u}\AgdaSpace{}%
\AgdaSymbol{:}\AgdaSpace{}%
\AgdaBound{P}\AgdaSpace{}%
\AgdaBound{x}\AgdaSymbol{)}\AgdaSpace{}%
\AgdaSymbol{(}\AgdaBound{v}\AgdaSpace{}%
\AgdaSymbol{:}\AgdaSpace{}%
\AgdaBound{P}\AgdaSpace{}%
\AgdaBound{y}\AgdaSymbol{)}\AgdaSpace{}%
\AgdaSymbol{→}\AgdaSpace{}%
\AgdaFunction{𝒰}\<%
\\
\>[0]\AgdaFunction{PathOver}\AgdaSpace{}%
\AgdaBound{P}\AgdaSpace{}%
\AgdaBound{p}\AgdaSpace{}%
\AgdaBound{u}\AgdaSpace{}%
\AgdaBound{v}\AgdaSpace{}%
\AgdaSymbol{=}\AgdaSpace{}%
\AgdaFunction{transport}\AgdaSpace{}%
\AgdaBound{P}\AgdaSpace{}%
\AgdaBound{p}\AgdaSpace{}%
\AgdaBound{u}\AgdaSpace{}%
\AgdaDatatype{==}\AgdaSpace{}%
\AgdaBound{v}\<%
\\
\\
\>[0]\AgdaKeyword{syntax}\AgdaSpace{}%
\AgdaFunction{PathOver} P p u v \AgdaSymbol{=} u == v [ P ↓ p ]

\AgdaFunction{apd}\AgdaSpace{}%
\AgdaSymbol{:}\AgdaSpace{}%
\AgdaSymbol{\{}\AgdaBound{A}\AgdaSpace{}%
\AgdaSymbol{:}\AgdaSpace{}%
\AgdaFunction{𝒰}\AgdaSymbol{\}}\AgdaSpace{}%
\AgdaSymbol{\{}\AgdaBound{P}\AgdaSpace{}%
\AgdaSymbol{:}\AgdaSpace{}%
\AgdaBound{A}\AgdaSpace{}%
\AgdaSymbol{→}\AgdaSpace{}%
\AgdaFunction{𝒰}\AgdaSymbol{\}}\AgdaSpace{}%
\AgdaSymbol{\{}\AgdaBound{x}\AgdaSpace{}%
\AgdaBound{y}\AgdaSpace{}%
\AgdaSymbol{:}\AgdaSpace{}%
\AgdaBound{A}\AgdaSymbol{\}}\AgdaSpace{}%
\AgdaSymbol{(}\AgdaBound{f}\AgdaSpace{}%
\AgdaSymbol{:}\AgdaSpace{}%
\AgdaSymbol{(}\AgdaBound{a}\AgdaSpace{}%
\AgdaSymbol{:}\AgdaSpace{}%
\AgdaBound{A}\AgdaSymbol{)}\AgdaSpace{}%
\AgdaSymbol{→}\AgdaSpace{}%
\AgdaBound{P}\AgdaSpace{}%
\AgdaBound{a}\AgdaSymbol{)}\AgdaSpace{}%
\AgdaSymbol{(}\AgdaBound{p}\AgdaSpace{}%
\AgdaSymbol{:}\AgdaSpace{}%
\AgdaBound{x}\AgdaSpace{}%
\AgdaDatatype{==}\AgdaSpace{}%
\AgdaBound{y}\AgdaSymbol{)}\AgdaSpace{}%
\AgdaSymbol{→}\AgdaSpace{}%
\AgdaBound{f}\AgdaSpace{}%
\AgdaBound{x}\AgdaSpace{}%
\AgdaFunction{==}\AgdaSpace{}%
\AgdaBound{f}\AgdaSpace{}%
\AgdaBound{y}\AgdaSpace{}%
\AgdaFunction{[}\AgdaSpace{}%
\AgdaBound{P}\AgdaSpace{}%
\AgdaFunction{↓}\AgdaSpace{}%
\AgdaBound{p}\AgdaSpace{}%
\AgdaFunction{]}\<%
\\
\>[0]\AgdaFunction{apd}\AgdaSpace{}%
\AgdaBound{f}\AgdaSpace{}%
\AgdaSymbol{(}\AgdaInductiveConstructor{refl}\AgdaSpace{}%
\AgdaBound{x}\AgdaSymbol{)}\AgdaSpace{}%
\AgdaSymbol{=}\AgdaSpace{}%
\AgdaInductiveConstructor{refl}\AgdaSpace{}%
\AgdaSymbol{(}\AgdaBound{f}\AgdaSpace{}%
\AgdaBound{x}\AgdaSymbol{)}\<%
\\
\\
\>[0]\AgdaFunction{⊥{-}elim}\AgdaSpace{}%
\AgdaSymbol{:}\AgdaSpace{}%
\AgdaSymbol{\{}\AgdaBound{C}\AgdaSpace{}%
\AgdaSymbol{:}\AgdaSpace{}%
\AgdaFunction{𝒰}\AgdaSymbol{\}}\AgdaSpace{}%
\AgdaSymbol{→}\AgdaSpace{}%
\AgdaDatatype{⊥}\AgdaSpace{}%
\AgdaSymbol{→}\AgdaSpace{}%
\AgdaBound{C}\<%
\\
\>[0]\AgdaFunction{⊥{-}elim}\AgdaSpace{}%
\AgdaSymbol{()}\<%
\\
\\
\>[0]\AgdaKeyword{module}\AgdaSpace{}%
\AgdaModule{\_}\AgdaSpace{}%
\AgdaSymbol{\{}\AgdaBound{X}\AgdaSpace{}%
\AgdaSymbol{:}\AgdaSpace{}%
\AgdaFunction{𝒰}\AgdaSymbol{\}}\AgdaSpace{}%
\AgdaKeyword{where}\<%
\\
\\
\>[0][@{}l@{\AgdaIndent{0}}]%
\>[2]\AgdaFunction{◾unitr}\AgdaSpace{}%
\AgdaSymbol{:}\AgdaSpace{}%
\AgdaSymbol{\{}\AgdaBound{x}\AgdaSpace{}%
\AgdaBound{y}\AgdaSpace{}%
\AgdaSymbol{:}\AgdaSpace{}%
\AgdaBound{X}\AgdaSymbol{\}}\AgdaSpace{}%
\AgdaSymbol{→}\AgdaSpace{}%
\AgdaSymbol{(}\AgdaBound{p}\AgdaSpace{}%
\AgdaSymbol{:}\AgdaSpace{}%
\AgdaBound{x}\AgdaSpace{}%
\AgdaDatatype{==}\AgdaSpace{}%
\AgdaBound{y}\AgdaSymbol{)}\AgdaSpace{}%
\AgdaSymbol{→}\AgdaSpace{}%
\AgdaBound{p}\AgdaSpace{}%
\AgdaFunction{◾}\AgdaSpace{}%
\AgdaInductiveConstructor{refl}\AgdaSpace{}%
\AgdaBound{y}\AgdaSpace{}%
\AgdaDatatype{==}\AgdaSpace{}%
\AgdaBound{p}\<%
\\
\>[0][@{}l@{\AgdaIndent{0}}]%
\>[2]\AgdaFunction{◾unitr}\AgdaSpace{}%
\AgdaSymbol{(}\AgdaInductiveConstructor{refl}\AgdaSpace{}%
\AgdaBound{x}\AgdaSymbol{)}\AgdaSpace{}%
\AgdaSymbol{=}\AgdaSpace{}%
\AgdaInductiveConstructor{refl}\AgdaSpace{}%
\AgdaSymbol{(}\AgdaInductiveConstructor{refl}\AgdaSpace{}%
\AgdaBound{x}\AgdaSymbol{)}\<%
\\
\\
\>[0][@{}l@{\AgdaIndent{0}}]%
\>[2]\AgdaFunction{◾unitl}\AgdaSpace{}%
\AgdaSymbol{:}\AgdaSpace{}%
\AgdaSymbol{\{}\AgdaBound{x}\AgdaSpace{}%
\AgdaBound{y}\AgdaSpace{}%
\AgdaSymbol{:}\AgdaSpace{}%
\AgdaBound{X}\AgdaSymbol{\}}\AgdaSpace{}%
\AgdaSymbol{→}\AgdaSpace{}%
\AgdaSymbol{(}\AgdaBound{p}\AgdaSpace{}%
\AgdaSymbol{:}\AgdaSpace{}%
\AgdaBound{x}\AgdaSpace{}%
\AgdaDatatype{==}\AgdaSpace{}%
\AgdaBound{y}\AgdaSymbol{)}\AgdaSpace{}%
\AgdaSymbol{→}\AgdaSpace{}%
\AgdaInductiveConstructor{refl}\AgdaSpace{}%
\AgdaBound{x}\AgdaSpace{}%
\AgdaFunction{◾}\AgdaSpace{}%
\AgdaBound{p}\AgdaSpace{}%
\AgdaDatatype{==}\AgdaSpace{}%
\AgdaBound{p}\<%
\\
\>[0][@{}l@{\AgdaIndent{0}}]%
\>[2]\AgdaFunction{◾unitl}\AgdaSpace{}%
\AgdaSymbol{(}\AgdaInductiveConstructor{refl}\AgdaSpace{}%
\AgdaBound{x}\AgdaSymbol{)}\AgdaSpace{}%
\AgdaSymbol{=}\AgdaSpace{}%
\AgdaInductiveConstructor{refl}\AgdaSpace{}%
\AgdaSymbol{(}\AgdaInductiveConstructor{refl}\AgdaSpace{}%
\AgdaBound{x}\AgdaSymbol{)}\<%
\\
\\
\>[0][@{}l@{\AgdaIndent{0}}]%
\>[2]\AgdaFunction{◾invr}\AgdaSpace{}%
\AgdaSymbol{:}\AgdaSpace{}%
\AgdaSymbol{\{}\AgdaBound{x}\AgdaSpace{}%
\AgdaBound{y}\AgdaSpace{}%
\AgdaSymbol{:}\AgdaSpace{}%
\AgdaBound{X}\AgdaSymbol{\}}\AgdaSpace{}%
\AgdaSymbol{→}\AgdaSpace{}%
\AgdaSymbol{(}\AgdaBound{p}\AgdaSpace{}%
\AgdaSymbol{:}\AgdaSpace{}%
\AgdaBound{x}\AgdaSpace{}%
\AgdaDatatype{==}\AgdaSpace{}%
\AgdaBound{y}\AgdaSymbol{)}\AgdaSpace{}%
\AgdaSymbol{→}\AgdaSpace{}%
\AgdaFunction{!}\AgdaSpace{}%
\AgdaBound{p}\AgdaSpace{}%
\AgdaFunction{◾}\AgdaSpace{}%
\AgdaBound{p}\AgdaSpace{}%
\AgdaDatatype{==}\AgdaSpace{}%
\AgdaInductiveConstructor{refl}\AgdaSpace{}%
\AgdaBound{y}\<%
\\
\>[0][@{}l@{\AgdaIndent{0}}]%
\>[2]\AgdaFunction{◾invr}\AgdaSpace{}%
\AgdaSymbol{(}\AgdaInductiveConstructor{refl}\AgdaSpace{}%
\AgdaBound{x}\AgdaSymbol{)}\AgdaSpace{}%
\AgdaSymbol{=}\AgdaSpace{}%
\AgdaInductiveConstructor{refl}\AgdaSpace{}%
\AgdaSymbol{(}\AgdaInductiveConstructor{refl}\AgdaSpace{}%
\AgdaBound{x}\AgdaSymbol{)}\<%
\\
\\
\>[0][@{}l@{\AgdaIndent{0}}]%
\>[2]\AgdaFunction{◾invl}\AgdaSpace{}%
\AgdaSymbol{:}\AgdaSpace{}%
\AgdaSymbol{\{}\AgdaBound{x}\AgdaSpace{}%
\AgdaBound{y}\AgdaSpace{}%
\AgdaSymbol{:}\AgdaSpace{}%
\AgdaBound{X}\AgdaSymbol{\}}\AgdaSpace{}%
\AgdaSymbol{→}\AgdaSpace{}%
\AgdaSymbol{(}\AgdaBound{p}\AgdaSpace{}%
\AgdaSymbol{:}\AgdaSpace{}%
\AgdaBound{x}\AgdaSpace{}%
\AgdaDatatype{==}\AgdaSpace{}%
\AgdaBound{y}\AgdaSymbol{)}\AgdaSpace{}%
\AgdaSymbol{→}\AgdaSpace{}%
\AgdaBound{p}\AgdaSpace{}%
\AgdaFunction{◾}\AgdaSpace{}%
\AgdaFunction{!}\AgdaSpace{}%
\AgdaBound{p}\AgdaSpace{}%
\AgdaDatatype{==}\AgdaSpace{}%
\AgdaInductiveConstructor{refl}\AgdaSpace{}%
\AgdaBound{x}\<%
\\
\>[0][@{}l@{\AgdaIndent{0}}]%
\>[2]\AgdaFunction{◾invl}\AgdaSpace{}%
\AgdaSymbol{(}\AgdaInductiveConstructor{refl}\AgdaSpace{}%
\AgdaBound{x}\AgdaSymbol{)}\AgdaSpace{}%
\AgdaSymbol{=}\AgdaSpace{}%
\AgdaInductiveConstructor{refl}\AgdaSpace{}%
\AgdaSymbol{(}\AgdaInductiveConstructor{refl}\AgdaSpace{}%
\AgdaBound{x}\AgdaSymbol{)}\<%
\\
\\
\>[0][@{}l@{\AgdaIndent{0}}]%
\>[2]\AgdaFunction{!!}\AgdaSpace{}%
\AgdaSymbol{:}\AgdaSpace{}%
\AgdaSymbol{\{}\AgdaBound{x}\AgdaSpace{}%
\AgdaBound{y}\AgdaSpace{}%
\AgdaSymbol{:}\AgdaSpace{}%
\AgdaBound{X}\AgdaSymbol{\}}\AgdaSpace{}%
\AgdaSymbol{→}\AgdaSpace{}%
\AgdaSymbol{(}\AgdaBound{p}\AgdaSpace{}%
\AgdaSymbol{:}\AgdaSpace{}%
\AgdaBound{x}\AgdaSpace{}%
\AgdaDatatype{==}\AgdaSpace{}%
\AgdaBound{y}\AgdaSymbol{)}\AgdaSpace{}%
\AgdaSymbol{→}\AgdaSpace{}%
\AgdaFunction{!}\AgdaSpace{}%
\AgdaSymbol{(}\AgdaFunction{!}\AgdaSpace{}%
\AgdaBound{p}\AgdaSymbol{)}\AgdaSpace{}%
\AgdaDatatype{==}\AgdaSpace{}%
\AgdaBound{p}\<%
\\
\>[0][@{}l@{\AgdaIndent{0}}]%
\>[2]\AgdaFunction{!!}\AgdaSpace{}%
\AgdaSymbol{(}\AgdaInductiveConstructor{refl}\AgdaSpace{}%
\AgdaBound{x}\AgdaSymbol{)}\AgdaSpace{}%
\AgdaSymbol{=}\AgdaSpace{}%
\AgdaInductiveConstructor{refl}\AgdaSpace{}%
\AgdaSymbol{(}\AgdaInductiveConstructor{refl}\AgdaSpace{}%
\AgdaBound{x}\AgdaSymbol{)}\<%
\\
\\
\>[0][@{}l@{\AgdaIndent{0}}]%
\>[2]\AgdaFunction{!◾}\AgdaSpace{}%
\AgdaSymbol{:}\AgdaSpace{}%
\AgdaSymbol{\{}\AgdaBound{x}\AgdaSpace{}%
\AgdaBound{y}\AgdaSpace{}%
\AgdaBound{z}\AgdaSpace{}%
\AgdaSymbol{:}\AgdaSpace{}%
\AgdaBound{X}\AgdaSymbol{\}}\AgdaSpace{}%
\AgdaSymbol{→}\AgdaSpace{}%
\AgdaSymbol{(}\AgdaBound{p}\AgdaSpace{}%
\AgdaSymbol{:}\AgdaSpace{}%
\AgdaBound{x}\AgdaSpace{}%
\AgdaDatatype{==}\AgdaSpace{}%
\AgdaBound{y}\AgdaSymbol{)}\AgdaSpace{}%
\AgdaSymbol{→}\AgdaSpace{}%
\AgdaSymbol{(}\AgdaBound{q}\AgdaSpace{}%
\AgdaSymbol{:}\AgdaSpace{}%
\AgdaBound{y}\AgdaSpace{}%
\AgdaDatatype{==}\AgdaSpace{}%
\AgdaBound{z}\AgdaSymbol{)}\AgdaSpace{}%
\AgdaSymbol{→}\AgdaSpace{}%
\AgdaFunction{!}\AgdaSpace{}%
\AgdaSymbol{(}\AgdaBound{p}\AgdaSpace{}%
\AgdaFunction{◾}\AgdaSpace{}%
\AgdaBound{q}\AgdaSymbol{)}\AgdaSpace{}%
\AgdaDatatype{==}\AgdaSpace{}%
\AgdaFunction{!}\AgdaSpace{}%
\AgdaBound{q}\AgdaSpace{}%
\AgdaFunction{◾}\AgdaSpace{}%
\AgdaFunction{!}\AgdaSpace{}%
\AgdaBound{p}\<%
\\
\>[0][@{}l@{\AgdaIndent{0}}]%
\>[2]\AgdaFunction{!◾}\AgdaSpace{}%
\AgdaSymbol{(}\AgdaInductiveConstructor{refl}\AgdaSpace{}%
\AgdaBound{y}\AgdaSymbol{)}\AgdaSpace{}%
\AgdaSymbol{(}\AgdaInductiveConstructor{refl}\AgdaSpace{}%
\AgdaSymbol{.}\AgdaBound{y}\AgdaSymbol{)}\AgdaSpace{}%
\AgdaSymbol{=}\AgdaSpace{}%
\AgdaInductiveConstructor{refl}\AgdaSpace{}%
\AgdaSymbol{(}\AgdaInductiveConstructor{refl}\AgdaSpace{}%
\AgdaBound{y}\AgdaSymbol{)}\<%
\\
\\
\>[0][@{}l@{\AgdaIndent{0}}]%
\>[2]\AgdaKeyword{infixr}\AgdaSpace{}%
\AgdaNumber{80}\AgdaSpace{}%
\AgdaFunction{\_[1,0,2]\_}\<%
\\
\>[0][@{}l@{\AgdaIndent{0}}]%
\>[2]\AgdaFunction{\_[1,0,2]\_}%
\>[622I]\AgdaSymbol{:}%
\>[623I]\AgdaSymbol{\{}\AgdaBound{x}\AgdaSpace{}%
\AgdaBound{y}\AgdaSpace{}%
\AgdaBound{z}\AgdaSpace{}%
\AgdaSymbol{:}\AgdaSpace{}%
\AgdaBound{X}\AgdaSymbol{\}}\AgdaSpace{}%
\AgdaSymbol{→}\AgdaSpace{}%
\AgdaSymbol{\{}\AgdaBound{r}\AgdaSpace{}%
\AgdaBound{s}\AgdaSpace{}%
\AgdaSymbol{:}\AgdaSpace{}%
\AgdaBound{y}\AgdaSpace{}%
\AgdaDatatype{==}\AgdaSpace{}%
\AgdaBound{z}\AgdaSymbol{\}}\<%
\\
\>[622I][@{}l@{\AgdaIndent{0}}]\<[623I]%
\>[14]\AgdaSymbol{→}\AgdaSpace{}%
\AgdaSymbol{(}\AgdaBound{p}\AgdaSpace{}%
\AgdaSymbol{:}\AgdaSpace{}%
\AgdaBound{x}\AgdaSpace{}%
\AgdaDatatype{==}\AgdaSpace{}%
\AgdaBound{y}\AgdaSymbol{)}\AgdaSpace{}%
\AgdaSymbol{→}\AgdaSpace{}%
\AgdaBound{r}\AgdaSpace{}%
\AgdaDatatype{==}\AgdaSpace{}%
\AgdaBound{s}\AgdaSpace{}%
\AgdaSymbol{→}\AgdaSpace{}%
\AgdaBound{p}\AgdaSpace{}%
\AgdaFunction{◾}\AgdaSpace{}%
\AgdaBound{r}\AgdaSpace{}%
\AgdaDatatype{==}\AgdaSpace{}%
\AgdaBound{p}\AgdaSpace{}%
\AgdaFunction{◾}\AgdaSpace{}%
\AgdaBound{s}\<%
\\
\>[0][@{}l@{\AgdaIndent{0}}]%
\>[2]\AgdaSymbol{(}\AgdaInductiveConstructor{refl}\AgdaSpace{}%
\AgdaBound{y}\AgdaSymbol{)}\AgdaSpace{}%
\AgdaFunction{[1,0,2]}\AgdaSpace{}%
\AgdaBound{γ}\AgdaSpace{}%
\AgdaSymbol{=}\AgdaSpace{}%
\AgdaFunction{◾unitl}\AgdaSpace{}%
\AgdaSymbol{\_}\AgdaSpace{}%
\AgdaFunction{◾}\AgdaSpace{}%
\AgdaBound{γ}\AgdaSpace{}%
\AgdaFunction{◾}\AgdaSpace{}%
\AgdaFunction{!}\AgdaSpace{}%
\AgdaSymbol{(}\AgdaFunction{◾unitl}\AgdaSpace{}%
\AgdaSymbol{\_)}\<%
\\
\\
\>[0][@{}l@{\AgdaIndent{0}}]%
\>[2]\AgdaFunction{◾assoc}%
\>[664I]\AgdaSymbol{:}%
\>[665I]\AgdaSymbol{\{}\AgdaBound{w}\AgdaSpace{}%
\AgdaBound{x}\AgdaSpace{}%
\AgdaBound{y}\AgdaSpace{}%
\AgdaBound{z}\AgdaSpace{}%
\AgdaSymbol{:}\AgdaSpace{}%
\AgdaBound{X}\AgdaSymbol{\}}\AgdaSpace{}%
\AgdaSymbol{→}\AgdaSpace{}%
\AgdaSymbol{(}\AgdaBound{p}\AgdaSpace{}%
\AgdaSymbol{:}\AgdaSpace{}%
\AgdaBound{w}\AgdaSpace{}%
\AgdaDatatype{==}\AgdaSpace{}%
\AgdaBound{x}\AgdaSymbol{)}\AgdaSpace{}%
\AgdaSymbol{→}\AgdaSpace{}%
\AgdaSymbol{(}\AgdaBound{q}\AgdaSpace{}%
\AgdaSymbol{:}\AgdaSpace{}%
\AgdaBound{x}\AgdaSpace{}%
\AgdaDatatype{==}\AgdaSpace{}%
\AgdaBound{y}\AgdaSymbol{)}\AgdaSpace{}%
\AgdaSymbol{→}\AgdaSpace{}%
\AgdaSymbol{(}\AgdaBound{r}\AgdaSpace{}%
\AgdaSymbol{:}\AgdaSpace{}%
\AgdaBound{y}\AgdaSpace{}%
\AgdaDatatype{==}\AgdaSpace{}%
\AgdaBound{z}\AgdaSymbol{)}\<%
\\
\>[664I][@{}l@{\AgdaIndent{0}}]\<[665I]%
\>[11]\AgdaSymbol{→}\AgdaSpace{}%
\AgdaSymbol{(}\AgdaBound{p}\AgdaSpace{}%
\AgdaFunction{◾}\AgdaSpace{}%
\AgdaBound{q}\AgdaSymbol{)}\AgdaSpace{}%
\AgdaFunction{◾}\AgdaSpace{}%
\AgdaBound{r}\AgdaSpace{}%
\AgdaDatatype{==}\AgdaSpace{}%
\AgdaBound{p}\AgdaSpace{}%
\AgdaFunction{◾}\AgdaSpace{}%
\AgdaBound{q}\AgdaSpace{}%
\AgdaFunction{◾}\AgdaSpace{}%
\AgdaBound{r}\<%
\\
\>[0][@{}l@{\AgdaIndent{0}}]%
\>[2]\AgdaFunction{◾assoc}\AgdaSpace{}%
\AgdaBound{p}\AgdaSpace{}%
\AgdaBound{q}\AgdaSpace{}%
\AgdaSymbol{(}\AgdaInductiveConstructor{refl}\AgdaSpace{}%
\AgdaBound{y}\AgdaSymbol{)}\AgdaSpace{}%
\AgdaSymbol{=}\AgdaSpace{}%
\AgdaFunction{◾unitr}\AgdaSpace{}%
\AgdaSymbol{\_}\AgdaSpace{}%
\AgdaFunction{◾}\AgdaSpace{}%
\AgdaSymbol{(}\AgdaBound{p}\AgdaSpace{}%
\AgdaFunction{[1,0,2]}\AgdaSpace{}%
\AgdaFunction{!}\AgdaSpace{}%
\AgdaSymbol{(}\AgdaFunction{◾unitr}\AgdaSpace{}%
\AgdaSymbol{\_))}\<%
\\
\\
\>[0][@{}l@{\AgdaIndent{0}}]%
\>[2]\AgdaKeyword{infixr}\AgdaSpace{}%
\AgdaNumber{80}\AgdaSpace{}%
\AgdaFunction{\_[2,0,1]\_}\<%
\\
\>[0][@{}l@{\AgdaIndent{0}}]%
\>[2]\AgdaFunction{\_[2,0,1]\_}%
\>[715I]\AgdaSymbol{:}%
\>[716I]\AgdaSymbol{\{}\AgdaBound{x}\AgdaSpace{}%
\AgdaBound{y}\AgdaSpace{}%
\AgdaBound{z}\AgdaSpace{}%
\AgdaSymbol{:}\AgdaSpace{}%
\AgdaBound{X}\AgdaSymbol{\}}\AgdaSpace{}%
\AgdaSymbol{→}\AgdaSpace{}%
\AgdaSymbol{\{}\AgdaBound{p}\AgdaSpace{}%
\AgdaBound{q}\AgdaSpace{}%
\AgdaSymbol{:}\AgdaSpace{}%
\AgdaBound{x}\AgdaSpace{}%
\AgdaDatatype{==}\AgdaSpace{}%
\AgdaBound{y}\AgdaSymbol{\}}\<%
\\
\>[715I][@{}l@{\AgdaIndent{0}}]\<[716I]%
\>[14]\AgdaSymbol{→}\AgdaSpace{}%
\AgdaBound{p}\AgdaSpace{}%
\AgdaDatatype{==}\AgdaSpace{}%
\AgdaBound{q}\AgdaSpace{}%
\AgdaSymbol{→}\AgdaSpace{}%
\AgdaSymbol{(}\AgdaBound{r}\AgdaSpace{}%
\AgdaSymbol{:}\AgdaSpace{}%
\AgdaBound{y}\AgdaSpace{}%
\AgdaDatatype{==}\AgdaSpace{}%
\AgdaBound{z}\AgdaSymbol{)}\AgdaSpace{}%
\AgdaSymbol{→}\AgdaSpace{}%
\AgdaBound{p}\AgdaSpace{}%
\AgdaFunction{◾}\AgdaSpace{}%
\AgdaBound{r}\AgdaSpace{}%
\AgdaDatatype{==}\AgdaSpace{}%
\AgdaBound{q}\AgdaSpace{}%
\AgdaFunction{◾}\AgdaSpace{}%
\AgdaBound{r}\<%
\\
\>[0][@{}l@{\AgdaIndent{0}}]%
\>[2]\AgdaBound{α}\AgdaSpace{}%
\AgdaFunction{[2,0,1]}\AgdaSpace{}%
\AgdaSymbol{(}\AgdaInductiveConstructor{refl}\AgdaSpace{}%
\AgdaBound{y}\AgdaSymbol{)}\AgdaSpace{}%
\AgdaSymbol{=}\AgdaSpace{}%
\AgdaFunction{◾unitr}\AgdaSpace{}%
\AgdaSymbol{\_}\AgdaSpace{}%
\AgdaFunction{◾}\AgdaSpace{}%
\AgdaBound{α}\AgdaSpace{}%
\AgdaFunction{◾}\AgdaSpace{}%
\AgdaFunction{!}\AgdaSpace{}%
\AgdaSymbol{(}\AgdaFunction{◾unitr}\AgdaSpace{}%
\AgdaSymbol{\_)}\<%
\\
\\
\>[0][@{}l@{\AgdaIndent{0}}]%
\>[2]\AgdaKeyword{infixr}\AgdaSpace{}%
\AgdaNumber{80}\AgdaSpace{}%
\AgdaFunction{\_[2,0,2]\_}\<%
\\
\>[0][@{}l@{\AgdaIndent{0}}]%
\>[2]\AgdaFunction{\_[2,0,2]\_}%
\>[759I]\AgdaSymbol{:}%
\>[760I]\AgdaSymbol{\{}\AgdaBound{x}\AgdaSpace{}%
\AgdaBound{y}\AgdaSpace{}%
\AgdaBound{z}\AgdaSpace{}%
\AgdaSymbol{:}\AgdaSpace{}%
\AgdaBound{X}\AgdaSymbol{\}}\AgdaSpace{}%
\AgdaSymbol{→}\AgdaSpace{}%
\AgdaSymbol{\{}\AgdaBound{p}\AgdaSpace{}%
\AgdaBound{q}\AgdaSpace{}%
\AgdaSymbol{:}\AgdaSpace{}%
\AgdaBound{x}\AgdaSpace{}%
\AgdaDatatype{==}\AgdaSpace{}%
\AgdaBound{y}\AgdaSymbol{\}}\AgdaSpace{}%
\AgdaSymbol{→}\AgdaSpace{}%
\AgdaSymbol{\{}\AgdaBound{r}\AgdaSpace{}%
\AgdaBound{s}\AgdaSpace{}%
\AgdaSymbol{:}\AgdaSpace{}%
\AgdaBound{y}\AgdaSpace{}%
\AgdaDatatype{==}\AgdaSpace{}%
\AgdaBound{z}\AgdaSymbol{\}}\<%
\\
\>[759I][@{}l@{\AgdaIndent{0}}]\<[760I]%
\>[14]\AgdaSymbol{→}\AgdaSpace{}%
\AgdaBound{p}\AgdaSpace{}%
\AgdaDatatype{==}\AgdaSpace{}%
\AgdaBound{q}\AgdaSpace{}%
\AgdaSymbol{→}\AgdaSpace{}%
\AgdaBound{r}\AgdaSpace{}%
\AgdaDatatype{==}\AgdaSpace{}%
\AgdaBound{s}\AgdaSpace{}%
\AgdaSymbol{→}\AgdaSpace{}%
\AgdaBound{p}\AgdaSpace{}%
\AgdaFunction{◾}\AgdaSpace{}%
\AgdaBound{r}\AgdaSpace{}%
\AgdaDatatype{==}\AgdaSpace{}%
\AgdaBound{q}\AgdaSpace{}%
\AgdaFunction{◾}\AgdaSpace{}%
\AgdaBound{s}\<%
\\
\>[0][@{}l@{\AgdaIndent{0}}]%
\>[2]\AgdaFunction{\_[2,0,2]\_}\AgdaSpace{}%
\AgdaSymbol{\{}\AgdaArgument{q}\AgdaSpace{}%
\AgdaSymbol{=}\AgdaSpace{}%
\AgdaBound{q}\AgdaSymbol{\}}\AgdaSpace{}%
\AgdaSymbol{\{}\AgdaBound{r}\AgdaSymbol{\}}\AgdaSpace{}%
\AgdaBound{α}\AgdaSpace{}%
\AgdaBound{β}\AgdaSpace{}%
\AgdaSymbol{=}\AgdaSpace{}%
\AgdaSymbol{(}\AgdaBound{α}\AgdaSpace{}%
\AgdaFunction{[2,0,1]}\AgdaSpace{}%
\AgdaBound{r}\AgdaSymbol{)}\AgdaSpace{}%
\AgdaFunction{◾}\AgdaSpace{}%
\AgdaSymbol{(}\AgdaBound{q}\AgdaSpace{}%
\AgdaFunction{[1,0,2]}\AgdaSpace{}%
\AgdaBound{β}\AgdaSymbol{)}\<%
\\
\>[0]\<%
\end{code}
}

\section{Introduction}

The proceedings of the $2012$ Symposium on Principles of Programming
Languages~\cite{Field:2012:2103656} included two apparently unrelated papers:
\emph{Information Effects} by James and Sabry and \emph{Canonicity for
  2-dimensional type theory} by Licata and Harper. The first paper, motivated by
the physical nature of
computation~\cite{Landauer:1961,PhysRevA.32.3266,Toffoli:1980,bennett1985fundamental,Frank:1999:REC:930275},
proposed, among other results, a reversible language $\Pi$ in which every
program is a type isomorphism. The second paper, motivated by the connections
between homotopy theory and type theory~\cite{vv06,hottbook}, proposed a
judgmental formulation of intensional dependent type theory with a
twice-iterated identity type. During the presentations and ensuing discussions
at the conference, it became apparent, at an intuitive and informal level, that
the two papers had strong similarities. Formalizing the precise connection was
far from obvious, however.

Here we report on a formal connection between appropriately formulated
reversible languages on one hand and univalent universes on the
other. In the next section, we give a rational reconstruction of the
reversible programming language $\Pi$, focusing on a small
``featherweight'' fragment~$\PiTwo$. In Sec.~\ref{sec:univalent}, we
review basic homotopy type theory (HoTT) background leading to
\emph{univalent fibrations} which allow us to give formal
presentations of ``small'' univalent universes. In
Sec.~\ref{sec:model} we define and establish the basic properties of
such a univalent subuniverse {\small\AgdaFunction{Ũ[𝟚]}} which we
prove in Sec.~\ref{sec:correspondence} as sound and complete with
respect to the reversible language $\PiTwo$.
Sec.~\ref{sec:discussion} discusses the implications of our work and
situates it into the broader context of the existing literature.

\section{Reversible Programming Languages}

Starting from the physical principle of ``conservation of
information''~\cite{Hey:1999:FCE:304763,fredkin1982conservative}, James and
Sabry~\cite{James:2012:IE:2103656.2103667} proposed a family of programming
languages $\Pi$ in which computation preserves information. Technically,
computations are \emph{type isomorphisms} which, at least in the case of finite
types, clearly preserve entropy in the information-theoretic
sense~\cite{James:2012:IE:2103656.2103667}. We illustrate the general flavor of
the family of languages with some examples and then identify a ``featherweight''
version of $\Pi$, called $\PiTwo$, to use in our formal development.

\subsection{Examples}

The examples below assume a representation of the type of booleans
${\small\bt}$ as the disjoint union ${\small\ot \oplus \ot}$ with the
left injection representing ${\small\mathsf{false}}$ and the right
injection representing ${\small\mathsf{true}}$. Given an arbitrary
reversible function {\small\AgdaFunction{f}} of type
${\small a \iso a}$, we can build the reversible function
${\small\AgdaFunction{controlled}~\AgdaFunction{f}}$ that takes a pair
of type ${\small\bt \otimes a}$ and checks the incoming boolean; if it
is false (i.e., we are in the left injection), the function behaves
like the identity; otherwise the function applies {\small\AgdaFunction{f}}
to the second argument. The incoming boolean is then reconstituted to
maintain reversibility:

{\small
\[\def\arraystretch{1.2}\begin{array}{rcll}
\AgdaFunction{controlled}  &:& \forall a.~ (a \iso a) \quad\rightarrow
                            & ~(\bt \otimes a \iso \bt \otimes a) \\
\AgdaFunction{controlled}~\AgdaFunction{f} &=&

  \bt \otimes a
    & \byiso{\AgdaFunction{unfoldBool} \otimes \AgdaFunction{id}} \\
&& (\ot \oplus \ot) \otimes a
    & \byiso{\AgdaFunction{distribute}} \\
&& (\ot \otimes a) \oplus (\ot \otimes a)
    & \byiso{\AgdaFunction{id} \oplus (\AgdaFunction{id} \otimes \AgdaFunction{f})} \\
&& (\ot \otimes a) \oplus (\ot \otimes a)
    & \byiso{\AgdaFunction{factor}} \\
&& (\ot \oplus \ot) \otimes a
    & \byiso{\AgdaFunction{foldBool} \otimes \AgdaFunction{id}} \\
&& \bt \otimes a & ~ \\
\end{array}
\]}

\noindent The left column shows the sequence of types that are visited
during the computation; the right column shows the names of the
combinators\footnote{We use names that are hopefully quite mnemonic;
  for the precise definitions of the combinators see the
  $\Pi$-papers~\cite{James:2012:IE:2103656.2103667,rc2011,rc2012,theseus,Carette2016}
  or the accompanying code at
  \url{https://git.io/v7wtW}.} that witness the
corresponding type isomorphism. The code for
{\small\AgdaFunction{controlled}~\AgdaFunction{f}} provides
constructive evidence (i.e., a program, a logic gate, or a hardware
circuit) for an automorphism on ${\small\bt \otimes a}$: it can be read
top-down or bottom-up to go back and forth.

The {\small\AgdaFunction{not}} function below is a simple lifting of
{\small\AgdaFunction{swap₊}} which swaps the left and right injections of a sum
type. Using the {\small\AgdaFunction{controlled}} building block, we can build a
controlled-not ({\small\AgdaFunction{cnot}}) gate and a controlled-controlled-not
gate, also known as the {\small\AgdaFunction{toffoli}} gate. The latter gate is a
universal function for combinational boolean circuits thus showing the
expressiveness of the language:

{\small
\[\begin{array}{rcl}
\AgdaFunction{not} &:& \bt \iso \bt \\
\AgdaFunction{not} &=&
  \AgdaFunction{unfoldBool} \odot_1 \AgdaFunction{swap₊} \odot_1 \AgdaFunction{foldBool} \\
\\
\AgdaFunction{cnot} &:& \bt \otimes \bt \iso \bt \otimes \bt \\
\AgdaFunction{cnot} &=& \AgdaFunction{controlled}~\AgdaFunction{not} \\
\\
\AgdaFunction{toffoli} &:& \bt \otimes (\bt \otimes \bt)
                           \iso \bt \otimes (\bt \otimes \bt) \\
\AgdaFunction{toffoli} &=& \AgdaFunction{controlled}~\AgdaFunction{cnot} \\
\end{array}\]}

\noindent While we wrote {\small\AgdaFunction{controlled}} in
equational-reasoning style, {\small\AgdaFunction{not}} is written in
the point-free combinator style.  These are equivalent as ${\small\byiso{-}}$
is defined in terms of the sequential composition combinator
${\small\odot_1}$.

As is customary in any semantic perspective on programming languages, we are
interested in the question of when two programs are ``equivalent.'' Consider the
following six programs of type~${\small\bt \iso \bt}$:

{\small
\[\def\arraystretch{1.2}\begin{array}{rcl}
\AgdaFunction{id₁}~\AgdaFunction{id₂}~\AgdaFunction{id₃}~
  \AgdaFunction{not₁}~\AgdaFunction{not₂}~\AgdaFunction{not₃} &:& \bt \iso \bt \\
\AgdaFunction{id₁} &=&
  \AgdaFunction{id} \odot_1 \AgdaFunction{id} \\
\AgdaFunction{id₂} &=&
  \AgdaFunction{not} \odot_1 \AgdaFunction{id} \odot_1 \AgdaFunction{not} \\
\AgdaFunction{id₃} &=&
  \AgdaFunction{uniti⋆} \odot_1 \AgdaFunction{swap⋆} \odot_1
                        (\AgdaFunction{id} \otimes \AgdaFunction{id}) \odot_1
                        \AgdaFunction{swap⋆} \odot_1
                        \AgdaFunction{unite⋆} \\
\AgdaFunction{not₁} &=&
  \AgdaFunction{id} \odot_1 \AgdaFunction{not} \\
\AgdaFunction{not₂} &=&
  \AgdaFunction{not} \odot_1 \AgdaFunction{not} \odot_1 \AgdaFunction{not} \\
\AgdaFunction{not₃} &=&
  \AgdaFunction{uniti⋆} \odot_1 \AgdaFunction{swap⋆} \odot_1
                        (\AgdaFunction{not} \otimes \AgdaFunction{id}) \odot_1
                        \AgdaFunction{swap⋆} \odot_1
                        \AgdaFunction{unite⋆}
\end{array}\]}

\begin{figure}
\begin{center}
\begin{tikzpicture}[scale=0.9,every node/.style={scale=0.9}]
  \draw (1,2) ellipse (0.5cm and 0.5cm);
  \draw[fill] (1,2) circle [radius=0.025];
  \node[below] at (1,2) {*};

  \draw (0,0) ellipse (0.5cm and 1cm);
  \draw[fill] (0,0.5) circle [radius=0.025];
  \node[below] at (0,0.5) {F};
  \draw[fill] (0,-0.5) circle [radius=0.025];
  \node[below] at (0,-0.5) {T};

  \draw     (1,2)    -- (2,2)      ; 
  \draw     (0,0.5)  -- (2,0.5)    ; 
  \draw     (0,-0.5) -- (2,-0.5)   ; 

  \draw     (2,2)    -- (3,-0.5)   ;
  \draw     (2,0.5)  -- (3,2)      ;
  \draw     (2,-0.5) -- (3,1)      ;

  \draw     (3,2)    -- (3.5,2)    ;
  \draw     (3,1)    -- (3.5,1)    ;
  \draw     (3,-0.5) -- (3.5,-0.5) ;

  \draw     (3.5,2)    -- (4.5,1)    ;
  \draw     (3.5,1)    -- (4.5,2)    ;
  \draw     (3.5,-0.5) -- (4.5,-0.5) ;

  \draw     (4.5,2)    -- (5,2)    ;
  \draw     (4.5,1)    -- (5,1)    ;
  \draw     (4.5,-0.5) -- (5,-0.5) ;

  \draw     (5,2)    -- (6,0.5)  ;
  \draw     (5,1)    -- (6,-0.5) ;
  \draw     (5,-0.5) -- (6,2)    ;

  \draw     (6,2)    -- (7,2)    ;
  \draw     (6,0.5)  -- (8,0.5)  ;
  \draw     (6,-0.5) -- (8,-0.5) ;

  \draw (7,2) ellipse (0.5cm and 0.5cm);
  \draw[fill] (7,2) circle [radius=0.025];
  \node[below] at (7,2) {*};

  \draw (8,0) ellipse (0.5cm and 1cm);
  \draw[fill] (8,0.5) circle [radius=0.025];
  \node[below] at (8,0.5) {F};
  \draw[fill] (8,-0.5) circle [radius=0.025];
  \node[below] at (8,-0.5) {T};

\end{tikzpicture}
\end{center}
\caption{\label{fig:not}Graphical representation of {\small\AgdaFunction{not₃}}}
\end{figure}
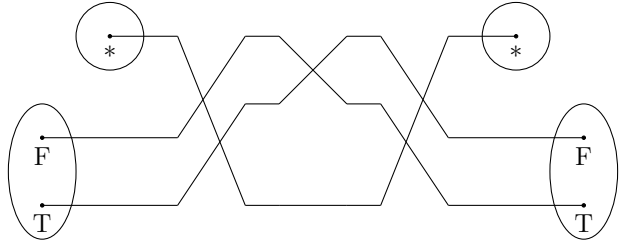

The programs are all of the same type but this is clearly not a
sufficient condition for ``equivalence.'' Thinking extensionally,
i.e., by looking at all possible input-output pairs, it is easy to
verify that the six programs split into two classes: one consisting of
the first three programs which are all equivalent to the identity
function and the other consisting of the remaining three programs
which all equivalent to boolean negation. In the context of $\Pi$, we
can provide \emph{evidence} (i.e., a reversible program of type
$\isotwo$ that manipulates lower level reversible programs of type
$\iso$ ) that can constructively identify programs in each equivalence
class. We show such a level-2 program proving that
{\small\AgdaFunction{not₃}} is equivalent to
{\small\AgdaFunction{not}}. For illustration, the program for
{\small\AgdaFunction{not₃}} is depicted in Fig.~\ref{fig:not}. We
encourage the reader to map the steps below to manipulations on the
diagram that would incrementally simplify it:

{\small
\[\def\arraystretch{1.2}\begin{array}{rcll}
\AgdaFunction{notOpt} &:& \AgdaFunction{not₃} \isotwo \AgdaFunction{not} \\
\AgdaFunction{notOpt} &=&
  \AgdaFunction{uniti⋆} \odot_1 (\AgdaFunction{swap⋆} \odot_1
                        ((\AgdaFunction{not} \otimes \AgdaFunction{id}) \odot_1
                        (\AgdaFunction{swap⋆} \odot_1 \AgdaFunction{unite⋆})))
 & \quad\byisotwo{\AgdaFunction{id} \boxdot \AgdaFunction{assocLeft}} \\
&& \AgdaFunction{uniti⋆} \odot_1 (\AgdaFunction{swap⋆} \odot_1
                        (\AgdaFunction{not} \otimes \AgdaFunction{id})) \odot_1
                        (\AgdaFunction{swap⋆} \odot_1 \AgdaFunction{unite⋆})
 & \quad\byisotwo{\AgdaFunction{id} \boxdot (\AgdaFunction{swapLeft}
                                  \boxdot \AgdaFunction{id})} \\
&& \AgdaFunction{uniti⋆} \odot_1 ((\AgdaFunction{id} \otimes \AgdaFunction{not})
                      \odot_1 \AgdaFunction{swap⋆}) \odot_1
                        (\AgdaFunction{swap⋆} \odot_1 \AgdaFunction{unite⋆})
 & \quad\byisotwo{\AgdaFunction{id} \boxdot \AgdaFunction{assocRight}} \\
&& \AgdaFunction{uniti⋆} \odot_1 ((\AgdaFunction{id} \otimes \AgdaFunction{not})
                      \odot_1 (\AgdaFunction{swap⋆} \odot_1
                        (\AgdaFunction{swap⋆} \odot_1 \AgdaFunction{unite⋆})))
 & \quad\byisotwo{\AgdaFunction{id} \boxdot (\AgdaFunction{id}
                                  \boxdot \AgdaFunction{assocLeft})} \\
&& \AgdaFunction{uniti⋆} \odot_1 ((\AgdaFunction{id} \otimes \AgdaFunction{not})
                      \odot_1 ((\AgdaFunction{swap⋆} \odot_1
                      \AgdaFunction{swap⋆}) \odot_1 \AgdaFunction{unite⋆}))
 & \quad\byisotwo{\AgdaFunction{id} \boxdot (\AgdaFunction{id}
                                  \boxdot (\AgdaFunction{leftInv} \boxdot \AgdaFunction{id}))} \\
&& \AgdaFunction{uniti⋆} \odot_1 ((\AgdaFunction{id} \otimes \AgdaFunction{not})
                      \odot_1 (\AgdaFunction{id} \odot_1 \AgdaFunction{unite⋆}))
 & \quad\byisotwo{\AgdaFunction{id} \boxdot (\AgdaFunction{id}
                                  \boxdot \AgdaFunction{idLeft})} \\
&& \AgdaFunction{uniti⋆} \odot_1 ((\AgdaFunction{id} \otimes \AgdaFunction{not})
                      \odot_1 \AgdaFunction{unite⋆})
 & \quad\byisotwo{\AgdaFunction{assocLeft}} \\
&& (\AgdaFunction{uniti⋆} \odot_1 (\AgdaFunction{id} \otimes \AgdaFunction{not}))
                      \odot_1 \AgdaFunction{unite⋆}
 & \quad\byisotwo{\AgdaFunction{unitiLeft} \boxdot \AgdaFunction{id}} \\
&& (\AgdaFunction{not} \odot_1 \AgdaFunction{uniti⋆}) \odot_1 \AgdaFunction{unite⋆}
 & \quad\byisotwo{\AgdaFunction{assocRight}} \\
&& \AgdaFunction{not} \odot_1 (\AgdaFunction{uniti⋆} \odot_1 \AgdaFunction{unite⋆})
 & \quad\byisotwo{\AgdaFunction{id} \boxdot \AgdaFunction{leftInv}} \\
&& \AgdaFunction{not} \odot_1 \AgdaFunction{id}
 & \quad\byisotwo{\AgdaFunction{idRight}} \\
&& \AgdaFunction{not}
\end{array}\]}

\noindent It is worthwhile mentioning that the above derivation could also be
drawn as one (large!) commutative diagram in an appropriate category, with each
$\byisotwo{-}$ as a $2$-arrow (and representing a natural isomorphism).  See
Shulman's draft book~\cite{shulman} for that interpretation.

\subsection{A Small Reversible Language of Booleans: \PiTwo}{\label{sec:pi2}}

Having illustrated the general flavor of the $\Pi$ family of
languages, we present in full detail an Agda-based formalization of a
small $\Pi$-based language which we will use to establish the
connection to an explicit univalent universe. The language is the
restriction of $\Pi$ to the case of just one type $\mathbb{2}$:


\begin{code}%
\>[0]\AgdaKeyword{data}\AgdaSpace{}%
\AgdaDatatype{𝟚}\AgdaSpace{}%
\AgdaSymbol{:}\AgdaSpace{}%
\AgdaFunction{𝒰}\AgdaSpace{}%
\AgdaKeyword{where}\<%
\\
\>[0][@{}l@{\AgdaIndent{0}}]%
\>[2]\AgdaInductiveConstructor{0₂}\AgdaSpace{}%
\AgdaInductiveConstructor{1₂}\AgdaSpace{}%
\AgdaSymbol{:}\AgdaSpace{}%
\AgdaDatatype{𝟚}\<%
\end{code}

The syntax of \PiTwo\ is given by the following four Agda
definitions. The first definition~{\small\AgdaFunction{Π₂}} introduces
the set of types of the language: this set contains
just~{\small\AgdaInductiveConstructor{`𝟚}} which is a name for the
type of booleans $\mathbb{2}$. The next three definitions introduce
the programs (combinators) in the language stratified by levels. The
level-1 programs of type $\iso$ map between types; the level-2
programs of type $\isotwo$ map between level-1 programs; and the
level-3 programs of type $\isothree$ map between level-2 programs:

\AgdaHide{
\begin{code}%
\>[0]\AgdaKeyword{infix}\AgdaSpace{}%
\AgdaNumber{3}\AgdaSpace{}%
\AgdaDatatype{\_⟷₁\_}\AgdaSpace{}%
\AgdaDatatype{\_⟷₂\_}\AgdaSpace{}%
\AgdaDatatype{\_⟷₃\_}\<%
\\
\>[0]\AgdaKeyword{infix}\AgdaSpace{}%
\AgdaNumber{5}\AgdaSpace{}%
\AgdaInductiveConstructor{!₁\_}\AgdaSpace{}%
\AgdaInductiveConstructor{!₂\_}\<%
\\
\>[0]\AgdaKeyword{infix}\AgdaSpace{}%
\AgdaNumber{4}\AgdaSpace{}%
\AgdaInductiveConstructor{\_⊙₁\_}\AgdaSpace{}%
\AgdaInductiveConstructor{\_⊙₂\_}\<%
\end{code}
}

\begin{code}%
\>[0]\AgdaKeyword{data}\AgdaSpace{}%
\AgdaDatatype{Π₂}\AgdaSpace{}%
\AgdaSymbol{:}\AgdaSpace{}%
\AgdaFunction{𝒰}\AgdaSpace{}%
\AgdaKeyword{where}\<%
\\
\>[0][@{}l@{\AgdaIndent{0}}]%
\>[2]\AgdaInductiveConstructor{`𝟚}\AgdaSpace{}%
\AgdaSymbol{:}\AgdaSpace{}%
\AgdaDatatype{Π₂}\<%
\\
\\
\>[0]\AgdaComment{{-}{-}{-}{-}{-}{-}{-}{-}{-}{-}{-}{-}{-}{-}{-}}\<%
\\
\>[0]\AgdaKeyword{data}\AgdaSpace{}%
\AgdaDatatype{\_⟷₁\_}\AgdaSpace{}%
\AgdaSymbol{:}\AgdaSpace{}%
\AgdaSymbol{(}\AgdaBound{A}\AgdaSpace{}%
\AgdaBound{B}\AgdaSpace{}%
\AgdaSymbol{:}\AgdaSpace{}%
\AgdaDatatype{Π₂}\AgdaSymbol{)}\AgdaSpace{}%
\AgdaSymbol{→}\AgdaSpace{}%
\AgdaFunction{𝒰}\AgdaSpace{}%
\AgdaKeyword{where}\<%
\\
\\
\>[0][@{}l@{\AgdaIndent{0}}]%
\>[2]\AgdaInductiveConstructor{`id}%
\>[9]\AgdaSymbol{:}\AgdaSpace{}%
\AgdaSymbol{∀}\AgdaSpace{}%
\AgdaSymbol{\{}\AgdaBound{A}\AgdaSymbol{\}}\AgdaSpace{}%
\AgdaSymbol{→}\AgdaSpace{}%
\AgdaBound{A}\AgdaSpace{}%
\AgdaDatatype{⟷₁}\AgdaSpace{}%
\AgdaBound{A}\<%
\\
\>[0][@{}l@{\AgdaIndent{0}}]%
\>[2]\AgdaInductiveConstructor{`not}%
\>[9]\AgdaSymbol{:}\AgdaSpace{}%
\AgdaInductiveConstructor{`𝟚}\AgdaSpace{}%
\AgdaDatatype{⟷₁}\AgdaSpace{}%
\AgdaInductiveConstructor{`𝟚}\<%
\\
\\
\>[0][@{}l@{\AgdaIndent{0}}]%
\>[2]\AgdaInductiveConstructor{!₁\_}%
\>[9]\AgdaSymbol{:}\AgdaSpace{}%
\AgdaSymbol{∀}\AgdaSpace{}%
\AgdaSymbol{\{}\AgdaBound{A}\AgdaSpace{}%
\AgdaBound{B}\AgdaSymbol{\}}\AgdaSpace{}%
\AgdaSymbol{→}\AgdaSpace{}%
\AgdaSymbol{(}\AgdaBound{A}\AgdaSpace{}%
\AgdaDatatype{⟷₁}\AgdaSpace{}%
\AgdaBound{B}\AgdaSymbol{)}\AgdaSpace{}%
\AgdaSymbol{→}\AgdaSpace{}%
\AgdaSymbol{(}\AgdaBound{B}\AgdaSpace{}%
\AgdaDatatype{⟷₁}\AgdaSpace{}%
\AgdaBound{A}\AgdaSymbol{)}\<%
\\
\>[0][@{}l@{\AgdaIndent{0}}]%
\>[2]\AgdaInductiveConstructor{\_⊙₁\_}%
\>[9]\AgdaSymbol{:}\AgdaSpace{}%
\AgdaSymbol{∀}\AgdaSpace{}%
\AgdaSymbol{\{}\AgdaBound{A}\AgdaSpace{}%
\AgdaBound{B}\AgdaSpace{}%
\AgdaBound{C}\AgdaSymbol{\}}\AgdaSpace{}%
\AgdaSymbol{→}\AgdaSpace{}%
\AgdaSymbol{(}\AgdaBound{A}\AgdaSpace{}%
\AgdaDatatype{⟷₁}\AgdaSpace{}%
\AgdaBound{B}\AgdaSymbol{)}\AgdaSpace{}%
\AgdaSymbol{→}\AgdaSpace{}%
\AgdaSymbol{(}\AgdaBound{B}\AgdaSpace{}%
\AgdaDatatype{⟷₁}\AgdaSpace{}%
\AgdaBound{C}\AgdaSymbol{)}\AgdaSpace{}%
\AgdaSymbol{→}\AgdaSpace{}%
\AgdaSymbol{(}\AgdaBound{A}\AgdaSpace{}%
\AgdaDatatype{⟷₁}\AgdaSpace{}%
\AgdaBound{C}\AgdaSymbol{)}\<%
\\
\\
\>[0]\AgdaComment{{-}{-}{-}{-}{-}{-}{-}{-}{-}{-}{-}{-}{-}{-}{-}}\<%
\\
\>[0]\AgdaKeyword{data}\AgdaSpace{}%
\AgdaDatatype{\_⟷₂\_}\AgdaSpace{}%
\AgdaSymbol{:}\AgdaSpace{}%
\AgdaSymbol{∀}\AgdaSpace{}%
\AgdaSymbol{\{}\AgdaBound{A}\AgdaSpace{}%
\AgdaBound{B}\AgdaSymbol{\}}\AgdaSpace{}%
\AgdaSymbol{(}\AgdaBound{p}\AgdaSpace{}%
\AgdaBound{q}\AgdaSpace{}%
\AgdaSymbol{:}\AgdaSpace{}%
\AgdaBound{A}\AgdaSpace{}%
\AgdaDatatype{⟷₁}\AgdaSpace{}%
\AgdaBound{B}\AgdaSymbol{)}\AgdaSpace{}%
\AgdaSymbol{→}\AgdaSpace{}%
\AgdaFunction{𝒰}\AgdaSpace{}%
\AgdaKeyword{where}\<%
\\
\\
\>[0][@{}l@{\AgdaIndent{0}}]%
\>[2]\AgdaInductiveConstructor{`id₂}%
\>[9]\AgdaSymbol{:}\AgdaSpace{}%
\AgdaSymbol{∀}\AgdaSpace{}%
\AgdaSymbol{\{}\AgdaBound{A}\AgdaSpace{}%
\AgdaBound{B}\AgdaSymbol{\}}\AgdaSpace{}%
\AgdaSymbol{\{}\AgdaBound{p}\AgdaSpace{}%
\AgdaSymbol{:}\AgdaSpace{}%
\AgdaBound{A}\AgdaSpace{}%
\AgdaDatatype{⟷₁}\AgdaSpace{}%
\AgdaBound{B}\AgdaSymbol{\}}\AgdaSpace{}%
\AgdaSymbol{→}\AgdaSpace{}%
\AgdaBound{p}\AgdaSpace{}%
\AgdaDatatype{⟷₂}\AgdaSpace{}%
\AgdaBound{p}\<%
\\
\\
\>[0][@{}l@{\AgdaIndent{0}}]%
\>[2]\AgdaInductiveConstructor{!₂\_}%
\>[9]\AgdaSymbol{:}\AgdaSpace{}%
\AgdaSymbol{∀}\AgdaSpace{}%
\AgdaSymbol{\{}\AgdaBound{A}\AgdaSpace{}%
\AgdaBound{B}\AgdaSymbol{\}}\AgdaSpace{}%
\AgdaSymbol{\{}\AgdaBound{p}\AgdaSpace{}%
\AgdaBound{q}\AgdaSpace{}%
\AgdaSymbol{:}\AgdaSpace{}%
\AgdaBound{A}\AgdaSpace{}%
\AgdaDatatype{⟷₁}\AgdaSpace{}%
\AgdaBound{B}\AgdaSymbol{\}}\AgdaSpace{}%
\AgdaSymbol{→}\AgdaSpace{}%
\AgdaSymbol{(}\AgdaBound{p}\AgdaSpace{}%
\AgdaDatatype{⟷₂}\AgdaSpace{}%
\AgdaBound{q}\AgdaSymbol{)}\AgdaSpace{}%
\AgdaSymbol{→}\AgdaSpace{}%
\AgdaSymbol{(}\AgdaBound{q}\AgdaSpace{}%
\AgdaDatatype{⟷₂}\AgdaSpace{}%
\AgdaBound{p}\AgdaSymbol{)}\<%
\\
\>[0][@{}l@{\AgdaIndent{0}}]%
\>[2]\AgdaInductiveConstructor{\_⊙₂\_}%
\>[9]\AgdaSymbol{:}\AgdaSpace{}%
\AgdaSymbol{∀}\AgdaSpace{}%
\AgdaSymbol{\{}\AgdaBound{A}\AgdaSpace{}%
\AgdaBound{B}\AgdaSymbol{\}}\AgdaSpace{}%
\AgdaSymbol{\{}\AgdaBound{p}\AgdaSpace{}%
\AgdaBound{q}\AgdaSpace{}%
\AgdaBound{r}\AgdaSpace{}%
\AgdaSymbol{:}\AgdaSpace{}%
\AgdaBound{A}\AgdaSpace{}%
\AgdaDatatype{⟷₁}\AgdaSpace{}%
\AgdaBound{B}\AgdaSymbol{\}}\AgdaSpace{}%
\AgdaSymbol{→}\AgdaSpace{}%
\AgdaSymbol{(}\AgdaBound{p}\AgdaSpace{}%
\AgdaDatatype{⟷₂}\AgdaSpace{}%
\AgdaBound{q}\AgdaSymbol{)}\AgdaSpace{}%
\AgdaSymbol{→}\AgdaSpace{}%
\AgdaSymbol{(}\AgdaBound{q}\AgdaSpace{}%
\AgdaDatatype{⟷₂}\AgdaSpace{}%
\AgdaBound{r}\AgdaSymbol{)}\AgdaSpace{}%
\AgdaSymbol{→}\AgdaSpace{}%
\AgdaSymbol{(}\AgdaBound{p}\AgdaSpace{}%
\AgdaDatatype{⟷₂}\AgdaSpace{}%
\AgdaBound{r}\AgdaSymbol{)}\<%
\\
\\
\>[0][@{}l@{\AgdaIndent{0}}]%
\>[2]\AgdaInductiveConstructor{`idl}%
\>[9]\AgdaSymbol{:}\AgdaSpace{}%
\AgdaSymbol{∀}\AgdaSpace{}%
\AgdaSymbol{\{}\AgdaBound{A}\AgdaSpace{}%
\AgdaBound{B}\AgdaSymbol{\}}\AgdaSpace{}%
\AgdaSymbol{(}\AgdaBound{p}\AgdaSpace{}%
\AgdaSymbol{:}\AgdaSpace{}%
\AgdaBound{A}\AgdaSpace{}%
\AgdaDatatype{⟷₁}\AgdaSpace{}%
\AgdaBound{B}\AgdaSymbol{)}\AgdaSpace{}%
\AgdaSymbol{→}\AgdaSpace{}%
\AgdaInductiveConstructor{`id}\AgdaSpace{}%
\AgdaInductiveConstructor{⊙₁}\AgdaSpace{}%
\AgdaBound{p}\AgdaSpace{}%
\AgdaDatatype{⟷₂}\AgdaSpace{}%
\AgdaBound{p}\<%
\\
\>[0][@{}l@{\AgdaIndent{0}}]%
\>[2]\AgdaInductiveConstructor{`idr}%
\>[9]\AgdaSymbol{:}\AgdaSpace{}%
\AgdaSymbol{∀}\AgdaSpace{}%
\AgdaSymbol{\{}\AgdaBound{A}\AgdaSpace{}%
\AgdaBound{B}\AgdaSymbol{\}}\AgdaSpace{}%
\AgdaSymbol{(}\AgdaBound{p}\AgdaSpace{}%
\AgdaSymbol{:}\AgdaSpace{}%
\AgdaBound{A}\AgdaSpace{}%
\AgdaDatatype{⟷₁}\AgdaSpace{}%
\AgdaBound{B}\AgdaSymbol{)}\AgdaSpace{}%
\AgdaSymbol{→}\AgdaSpace{}%
\AgdaBound{p}\AgdaSpace{}%
\AgdaInductiveConstructor{⊙₁}\AgdaSpace{}%
\AgdaInductiveConstructor{`id}\AgdaSpace{}%
\AgdaDatatype{⟷₂}\AgdaSpace{}%
\AgdaBound{p}\<%
\\
\>[0][@{}l@{\AgdaIndent{0}}]%
\>[2]\AgdaInductiveConstructor{`assoc}%
\>[969I]\AgdaSymbol{:}\AgdaSpace{}%
\AgdaSymbol{∀}\AgdaSpace{}%
\AgdaSymbol{\{}\AgdaBound{A}\AgdaSpace{}%
\AgdaBound{B}\AgdaSpace{}%
\AgdaBound{C}\AgdaSpace{}%
\AgdaBound{D}\AgdaSymbol{\}}\AgdaSpace{}%
\AgdaSymbol{(}\AgdaBound{p}\AgdaSpace{}%
\AgdaSymbol{:}\AgdaSpace{}%
\AgdaBound{A}\AgdaSpace{}%
\AgdaDatatype{⟷₁}\AgdaSpace{}%
\AgdaBound{B}\AgdaSymbol{)}\AgdaSpace{}%
\AgdaSymbol{(}\AgdaBound{q}\AgdaSpace{}%
\AgdaSymbol{:}\AgdaSpace{}%
\AgdaBound{B}\AgdaSpace{}%
\AgdaDatatype{⟷₁}\AgdaSpace{}%
\AgdaBound{C}\AgdaSymbol{)}\AgdaSpace{}%
\AgdaSymbol{(}\AgdaBound{r}\AgdaSpace{}%
\AgdaSymbol{:}\AgdaSpace{}%
\AgdaBound{C}\AgdaSpace{}%
\AgdaDatatype{⟷₁}\AgdaSpace{}%
\AgdaBound{D}\AgdaSymbol{)}\<%
\\
\>[2][@{}l@{\AgdaIndent{0}}]\<[969I]%
\>[9]\AgdaSymbol{→}\AgdaSpace{}%
\AgdaSymbol{(}\AgdaBound{p}\AgdaSpace{}%
\AgdaInductiveConstructor{⊙₁}\AgdaSpace{}%
\AgdaBound{q}\AgdaSymbol{)}\AgdaSpace{}%
\AgdaInductiveConstructor{⊙₁}\AgdaSpace{}%
\AgdaBound{r}\AgdaSpace{}%
\AgdaDatatype{⟷₂}\AgdaSpace{}%
\AgdaBound{p}\AgdaSpace{}%
\AgdaInductiveConstructor{⊙₁}\AgdaSpace{}%
\AgdaSymbol{(}\AgdaBound{q}\AgdaSpace{}%
\AgdaInductiveConstructor{⊙₁}\AgdaSpace{}%
\AgdaBound{r}\AgdaSymbol{)}\<%
\\
\\
\>[0][@{}l@{\AgdaIndent{0}}]%
\>[2]\AgdaInductiveConstructor{\_□₂\_}%
\>[9]\AgdaSymbol{:}\AgdaSpace{}%
\AgdaSymbol{∀}\AgdaSpace{}%
\AgdaSymbol{\{}\AgdaBound{A}\AgdaSpace{}%
\AgdaBound{B}\AgdaSpace{}%
\AgdaBound{C}\AgdaSymbol{\}}\AgdaSpace{}%
\AgdaSymbol{\{}\AgdaBound{p}\AgdaSpace{}%
\AgdaBound{q}\AgdaSpace{}%
\AgdaSymbol{:}\AgdaSpace{}%
\AgdaBound{A}\AgdaSpace{}%
\AgdaDatatype{⟷₁}\AgdaSpace{}%
\AgdaBound{B}\AgdaSymbol{\}}\AgdaSpace{}%
\AgdaSymbol{\{}\AgdaBound{r}\AgdaSpace{}%
\AgdaBound{s}\AgdaSpace{}%
\AgdaSymbol{:}\AgdaSpace{}%
\AgdaBound{B}\AgdaSpace{}%
\AgdaDatatype{⟷₁}\AgdaSpace{}%
\AgdaBound{C}\AgdaSymbol{\}}\<%
\\
\>[2][@{}l@{\AgdaIndent{0}}]%
\>[9]\AgdaSymbol{→}\AgdaSpace{}%
\AgdaSymbol{(}\AgdaBound{p}\AgdaSpace{}%
\AgdaDatatype{⟷₂}\AgdaSpace{}%
\AgdaBound{q}\AgdaSymbol{)}\AgdaSpace{}%
\AgdaSymbol{→}\AgdaSpace{}%
\AgdaSymbol{(}\AgdaBound{r}\AgdaSpace{}%
\AgdaDatatype{⟷₂}\AgdaSpace{}%
\AgdaBound{s}\AgdaSymbol{)}\AgdaSpace{}%
\AgdaSymbol{→}\AgdaSpace{}%
\AgdaSymbol{(}\AgdaBound{p}\AgdaSpace{}%
\AgdaInductiveConstructor{⊙₁}\AgdaSpace{}%
\AgdaBound{r}\AgdaSymbol{)}\AgdaSpace{}%
\AgdaDatatype{⟷₂}\AgdaSpace{}%
\AgdaSymbol{(}\AgdaBound{q}\AgdaSpace{}%
\AgdaInductiveConstructor{⊙₁}\AgdaSpace{}%
\AgdaBound{s}\AgdaSymbol{)}\<%
\\
\>[0][@{}l@{\AgdaIndent{0}}]%
\>[2]\AgdaInductiveConstructor{`!}%
\>[9]\AgdaSymbol{:}\AgdaSpace{}%
\AgdaSymbol{∀}\AgdaSpace{}%
\AgdaSymbol{\{}\AgdaBound{A}\AgdaSpace{}%
\AgdaBound{B}\AgdaSymbol{\}}\AgdaSpace{}%
\AgdaSymbol{\{}\AgdaBound{p}\AgdaSpace{}%
\AgdaBound{q}\AgdaSpace{}%
\AgdaSymbol{:}\AgdaSpace{}%
\AgdaBound{A}\AgdaSpace{}%
\AgdaDatatype{⟷₁}\AgdaSpace{}%
\AgdaBound{B}\AgdaSymbol{\}}\AgdaSpace{}%
\AgdaSymbol{→}\AgdaSpace{}%
\AgdaSymbol{(}\AgdaBound{p}\AgdaSpace{}%
\AgdaDatatype{⟷₂}\AgdaSpace{}%
\AgdaBound{q}\AgdaSymbol{)}\AgdaSpace{}%
\AgdaSymbol{→}\AgdaSpace{}%
\AgdaSymbol{(}\AgdaInductiveConstructor{!₁}\AgdaSpace{}%
\AgdaBound{p}\AgdaSpace{}%
\AgdaDatatype{⟷₂}\AgdaSpace{}%
\AgdaInductiveConstructor{!₁}\AgdaSpace{}%
\AgdaBound{q}\AgdaSymbol{)}\<%
\\
\>[0][@{}l@{\AgdaIndent{0}}]%
\>[2]\AgdaInductiveConstructor{`!l}%
\>[9]\AgdaSymbol{:}\AgdaSpace{}%
\AgdaSymbol{∀}\AgdaSpace{}%
\AgdaSymbol{\{}\AgdaBound{A}\AgdaSpace{}%
\AgdaBound{B}\AgdaSymbol{\}}\AgdaSpace{}%
\AgdaSymbol{(}\AgdaBound{p}\AgdaSpace{}%
\AgdaSymbol{:}\AgdaSpace{}%
\AgdaBound{A}\AgdaSpace{}%
\AgdaDatatype{⟷₁}\AgdaSpace{}%
\AgdaBound{B}\AgdaSymbol{)}\AgdaSpace{}%
\AgdaSymbol{→}\AgdaSpace{}%
\AgdaSymbol{(}\AgdaBound{p}\AgdaSpace{}%
\AgdaInductiveConstructor{⊙₁}\AgdaSpace{}%
\AgdaInductiveConstructor{!₁}\AgdaSpace{}%
\AgdaBound{p}\AgdaSpace{}%
\AgdaDatatype{⟷₂}\AgdaSpace{}%
\AgdaInductiveConstructor{`id}\AgdaSymbol{)}\<%
\\
\>[0][@{}l@{\AgdaIndent{0}}]%
\>[2]\AgdaInductiveConstructor{`!r}%
\>[9]\AgdaSymbol{:}\AgdaSpace{}%
\AgdaSymbol{∀}\AgdaSpace{}%
\AgdaSymbol{\{}\AgdaBound{A}\AgdaSpace{}%
\AgdaBound{B}\AgdaSymbol{\}}\AgdaSpace{}%
\AgdaSymbol{(}\AgdaBound{p}\AgdaSpace{}%
\AgdaSymbol{:}\AgdaSpace{}%
\AgdaBound{B}\AgdaSpace{}%
\AgdaDatatype{⟷₁}\AgdaSpace{}%
\AgdaBound{A}\AgdaSymbol{)}\AgdaSpace{}%
\AgdaSymbol{→}\AgdaSpace{}%
\AgdaSymbol{(}\AgdaInductiveConstructor{!₁}\AgdaSpace{}%
\AgdaBound{p}\AgdaSpace{}%
\AgdaInductiveConstructor{⊙₁}\AgdaSpace{}%
\AgdaBound{p}\AgdaSpace{}%
\AgdaDatatype{⟷₂}\AgdaSpace{}%
\AgdaInductiveConstructor{`id}\AgdaSymbol{)}\<%
\\
\\
\>[0][@{}l@{\AgdaIndent{0}}]%
\>[2]\AgdaInductiveConstructor{`!id}%
\>[9]\AgdaSymbol{:}\AgdaSpace{}%
\AgdaSymbol{∀}\AgdaSpace{}%
\AgdaSymbol{\{}\AgdaBound{A}\AgdaSymbol{\}}\AgdaSpace{}%
\AgdaSymbol{→}\AgdaSpace{}%
\AgdaInductiveConstructor{!₁}\AgdaSpace{}%
\AgdaInductiveConstructor{`id}\AgdaSpace{}%
\AgdaSymbol{\{}\AgdaBound{A}\AgdaSymbol{\}}\AgdaSpace{}%
\AgdaDatatype{⟷₂}\AgdaSpace{}%
\AgdaInductiveConstructor{`id}\AgdaSpace{}%
\AgdaSymbol{\{}\AgdaBound{A}\AgdaSymbol{\}}\<%
\\
\>[0][@{}l@{\AgdaIndent{0}}]%
\>[2]\AgdaInductiveConstructor{`!not}%
\>[9]\AgdaSymbol{:}\AgdaSpace{}%
\AgdaInductiveConstructor{!₁}\AgdaSpace{}%
\AgdaInductiveConstructor{`not}\AgdaSpace{}%
\AgdaDatatype{⟷₂}\AgdaSpace{}%
\AgdaInductiveConstructor{`not}\<%
\\
\\
\>[0][@{}l@{\AgdaIndent{0}}]%
\>[2]\AgdaInductiveConstructor{`!◾}%
\>[9]\AgdaSymbol{:}\AgdaSpace{}%
\AgdaSymbol{∀}\AgdaSpace{}%
\AgdaSymbol{\{}\AgdaBound{A}\AgdaSpace{}%
\AgdaBound{B}\AgdaSpace{}%
\AgdaBound{C}\AgdaSymbol{\}}\AgdaSpace{}%
\AgdaSymbol{\{}\AgdaBound{p}\AgdaSpace{}%
\AgdaSymbol{:}\AgdaSpace{}%
\AgdaBound{A}\AgdaSpace{}%
\AgdaDatatype{⟷₁}\AgdaSpace{}%
\AgdaBound{B}\AgdaSymbol{\}}\AgdaSpace{}%
\AgdaSymbol{\{}\AgdaBound{q}\AgdaSpace{}%
\AgdaSymbol{:}\AgdaSpace{}%
\AgdaBound{B}\AgdaSpace{}%
\AgdaDatatype{⟷₁}\AgdaSpace{}%
\AgdaBound{C}\AgdaSymbol{\}}\<%
\\
\>[2][@{}l@{\AgdaIndent{0}}]%
\>[9]\AgdaSymbol{→}\AgdaSpace{}%
\AgdaInductiveConstructor{!₁}\AgdaSpace{}%
\AgdaSymbol{(}\AgdaBound{p}\AgdaSpace{}%
\AgdaInductiveConstructor{⊙₁}\AgdaSpace{}%
\AgdaBound{q}\AgdaSymbol{)}\AgdaSpace{}%
\AgdaDatatype{⟷₂}\AgdaSpace{}%
\AgdaSymbol{(}\AgdaInductiveConstructor{!₁}\AgdaSpace{}%
\AgdaBound{q}\AgdaSymbol{)}\AgdaSpace{}%
\AgdaInductiveConstructor{⊙₁}\AgdaSpace{}%
\AgdaSymbol{(}\AgdaInductiveConstructor{!₁}\AgdaSpace{}%
\AgdaBound{p}\AgdaSymbol{)}\<%
\\
\>[0][@{}l@{\AgdaIndent{0}}]%
\>[2]\AgdaInductiveConstructor{`!!}%
\>[9]\AgdaSymbol{:}\AgdaSpace{}%
\AgdaSymbol{∀}\AgdaSpace{}%
\AgdaSymbol{\{}\AgdaBound{A}\AgdaSpace{}%
\AgdaBound{B}\AgdaSymbol{\}}\AgdaSpace{}%
\AgdaSymbol{\{}\AgdaBound{p}\AgdaSpace{}%
\AgdaSymbol{:}\AgdaSpace{}%
\AgdaBound{A}\AgdaSpace{}%
\AgdaDatatype{⟷₁}\AgdaSpace{}%
\AgdaBound{B}\AgdaSymbol{\}}\AgdaSpace{}%
\AgdaSymbol{→}\AgdaSpace{}%
\AgdaInductiveConstructor{!₁}\AgdaSpace{}%
\AgdaSymbol{(}\AgdaInductiveConstructor{!₁}\AgdaSpace{}%
\AgdaBound{p}\AgdaSymbol{)}\AgdaSpace{}%
\AgdaDatatype{⟷₂}\AgdaSpace{}%
\AgdaBound{p}\<%
\\
\\
\>[0]\AgdaComment{{-}{-}{-}{-}{-}{-}{-}{-}{-}{-}{-}{-}{-}{-}{-}}\<%
\\
\>[0]\AgdaKeyword{data}\AgdaSpace{}%
\AgdaDatatype{\_⟷₃\_}\AgdaSpace{}%
\AgdaSymbol{\{}\AgdaBound{A}\AgdaSpace{}%
\AgdaBound{B}\AgdaSymbol{\}}\AgdaSpace{}%
\AgdaSymbol{\{}\AgdaBound{p}\AgdaSpace{}%
\AgdaBound{q}\AgdaSpace{}%
\AgdaSymbol{:}\AgdaSpace{}%
\AgdaBound{A}\AgdaSpace{}%
\AgdaDatatype{⟷₁}\AgdaSpace{}%
\AgdaBound{B}\AgdaSymbol{\}}\AgdaSpace{}%
\AgdaSymbol{(}\AgdaBound{u}\AgdaSpace{}%
\AgdaBound{v}\AgdaSpace{}%
\AgdaSymbol{:}\AgdaSpace{}%
\AgdaBound{p}\AgdaSpace{}%
\AgdaDatatype{⟷₂}\AgdaSpace{}%
\AgdaBound{q}\AgdaSymbol{)}\AgdaSpace{}%
\AgdaSymbol{:}\AgdaSpace{}%
\AgdaFunction{𝒰}\AgdaSpace{}%
\AgdaKeyword{where}\<%
\\
\>[0][@{}l@{\AgdaIndent{0}}]%
\>[2]\AgdaInductiveConstructor{`trunc}\AgdaSpace{}%
\AgdaSymbol{:}\AgdaSpace{}%
\AgdaBound{u}\AgdaSpace{}%
\AgdaDatatype{⟷₃}\AgdaSpace{}%
\AgdaBound{v}\<%
\end{code}


In the previous presentations of
$\Pi$~\cite{rc2011,James:2012:IE:2103656.2103667,Carette2016}, the level-3
programs, consisting of just one trivial program
{\small\AgdaInductiveConstructor{`trunc}}, were not made explicit. The much
larger level-1 and level-2 programs of the full $\Pi$
language~\cite{Carette2016} have been specialized to our small language. For the
level-1 constructors, denoting reversible programs, type isomorphisms,
permutations between finite sets, or equivalences depending on one's favorite
interpretation, we have two canonical programs
{\small\AgdaInductiveConstructor{`id}} and
{\small\AgdaInductiveConstructor{`not}} closed under inverses
{\small\AgdaInductiveConstructor{!₁}} and sequential
composition~{\small\AgdaInductiveConstructor{⊙₁}}. For level-2 constructors,
denoting reversible program transformations, coherence conditions on type
isomorphisms, equivalences between permutations, or program optimizations
depending on one's favorite interpretation, we have the following groups: (i)
the first group contains the identity, inverses, and sequential composition;
(ii) the second group establishes the coherence laws for level-1 sequential
composition (e.g, it is associative); and (iii) finally the third group includes
general rules for inversions of level-1 constructors.

Each of the level-2 combinators of type $p \isotwo q$ is easily seen
to establish an equivalence between level-1 programs $p$ and $q$ (as
shown in previous work~\cite{Carette2016} and in
Sec.~\ref{sec:correspondence}). For example, composition of negation
is equivalent to the identity:

\begin{code}%
\>[0]\AgdaFunction{not⊙₁not⟷₂id}\AgdaSpace{}%
\AgdaSymbol{:}\AgdaSpace{}%
\AgdaInductiveConstructor{`not}\AgdaSpace{}%
\AgdaInductiveConstructor{⊙₁}\AgdaSpace{}%
\AgdaInductiveConstructor{`not}\AgdaSpace{}%
\AgdaDatatype{⟷₂}\AgdaSpace{}%
\AgdaInductiveConstructor{`id}\<%
\\
\>[0]\AgdaFunction{not⊙₁not⟷₂id}\AgdaSpace{}%
\AgdaSymbol{=}\AgdaSpace{}%
\AgdaSymbol{((}\AgdaInductiveConstructor{!₂}\AgdaSpace{}%
\AgdaInductiveConstructor{`!not}\AgdaSymbol{)}\AgdaSpace{}%
\AgdaInductiveConstructor{□₂}\AgdaSpace{}%
\AgdaInductiveConstructor{`id₂}\AgdaSymbol{)}\AgdaSpace{}%
\AgdaInductiveConstructor{⊙₂}\AgdaSpace{}%
\AgdaSymbol{(}\AgdaInductiveConstructor{`!r}\AgdaSpace{}%
\AgdaInductiveConstructor{`not}\AgdaSymbol{)}\<%
\end{code}

What is particularly interesting, however, is that the collection of
level-2 combinators above is \emph{complete} in the sense that any
equivalence between level-1 programs $p$ and $q$ can be proved using
the level-2 combinators. Formally we have two canonical level-1
programs {\small\AgdaInductiveConstructor{`id}} and
{\small\AgdaInductiveConstructor{`not}} and for any level-1 program
${\small p}$, we have evidence that either
${\small p \isotwo \AgdaInductiveConstructor{`id}}$ or
${\small p \isotwo \AgdaInductiveConstructor{`not}}$.

To prove this, we introduce a type which encodes the knowledge of
which level-1 programs are canonical. The type {\small\AgdaDatatype{Which}}
names the subset of {\small\AgdaDatatype{⟷₁}} which are canonical forms:

\begin{code}%
\>[0]\AgdaKeyword{data}\AgdaSpace{}%
\AgdaDatatype{Which}\AgdaSpace{}%
\AgdaSymbol{:}\AgdaSpace{}%
\AgdaFunction{𝒰}\AgdaSpace{}%
\AgdaKeyword{where}\<%
\\
\>[0][@{}l@{\AgdaIndent{0}}]%
\>[2]\AgdaInductiveConstructor{ID}\AgdaSpace{}%
\AgdaInductiveConstructor{NOT}\AgdaSpace{}%
\AgdaSymbol{:}\AgdaSpace{}%
\AgdaDatatype{Which}\<%
\\
\\
\>[0]\AgdaFunction{refine}\AgdaSpace{}%
\AgdaSymbol{:}\AgdaSpace{}%
\AgdaSymbol{(}\AgdaBound{w}\AgdaSpace{}%
\AgdaSymbol{:}\AgdaSpace{}%
\AgdaDatatype{Which}\AgdaSymbol{)}\AgdaSpace{}%
\AgdaSymbol{→}\AgdaSpace{}%
\AgdaInductiveConstructor{`𝟚}\AgdaSpace{}%
\AgdaDatatype{⟷₁}\AgdaSpace{}%
\AgdaInductiveConstructor{`𝟚}\<%
\\
\>[0]\AgdaFunction{refine}\AgdaSpace{}%
\AgdaInductiveConstructor{ID}\AgdaSpace{}%
\AgdaSymbol{=}\AgdaSpace{}%
\AgdaInductiveConstructor{`id}\<%
\\
\>[0]\AgdaFunction{refine}\AgdaSpace{}%
\AgdaInductiveConstructor{NOT}\AgdaSpace{}%
\AgdaSymbol{=}\AgdaSpace{}%
\AgdaInductiveConstructor{`not}\<%
\end{code}

This enables us to compute for any 2-combinator $c$ (the name of) its canonical
form, as well as a proof that $c$ is equivalent to its canonical form:

\begin{code}%
\>[0]\AgdaFunction{canonical}\AgdaSpace{}%
\AgdaSymbol{:}\AgdaSpace{}%
\AgdaSymbol{(}\AgdaBound{c}\AgdaSpace{}%
\AgdaSymbol{:}\AgdaSpace{}%
\AgdaInductiveConstructor{`𝟚}\AgdaSpace{}%
\AgdaDatatype{⟷₁}\AgdaSpace{}%
\AgdaInductiveConstructor{`𝟚}\AgdaSymbol{)}\AgdaSpace{}%
\AgdaSymbol{→}\AgdaSpace{}%
\AgdaRecord{Σ[}\AgdaSpace{}%
\AgdaBound{c'}\AgdaSpace{}%
\AgdaRecord{∶}\AgdaSpace{}%
\AgdaDatatype{Which}\AgdaSpace{}%
\AgdaRecord{]}\AgdaSpace{}%
\AgdaSymbol{(}\AgdaBound{c}\AgdaSpace{}%
\AgdaDatatype{⟷₂}\AgdaSpace{}%
\AgdaFunction{refine}\AgdaSpace{}%
\AgdaBound{c'}\AgdaSymbol{)}\<%
\\
\>[0]\AgdaFunction{canonical}\AgdaSpace{}%
\AgdaInductiveConstructor{`id}\AgdaSpace{}%
\AgdaSymbol{=}\AgdaSpace{}%
\AgdaInductiveConstructor{ID}\AgdaSpace{}%
\AgdaInductiveConstructor{,}\AgdaSpace{}%
\AgdaInductiveConstructor{`id₂}\<%
\\
\>[0]\AgdaFunction{canonical}\AgdaSpace{}%
\AgdaInductiveConstructor{`not}\AgdaSpace{}%
\AgdaSymbol{=}\AgdaSpace{}%
\AgdaInductiveConstructor{NOT}\AgdaSpace{}%
\AgdaInductiveConstructor{,}\AgdaSpace{}%
\AgdaInductiveConstructor{`id₂}\<%
\\
\>[0]\AgdaFunction{canonical}\AgdaSpace{}%
\AgdaSymbol{(}\AgdaInductiveConstructor{!₁}\AgdaSpace{}%
\AgdaBound{c}\AgdaSymbol{)}\AgdaSpace{}%
\AgdaKeyword{with}\AgdaSpace{}%
\AgdaFunction{canonical}\AgdaSpace{}%
\AgdaBound{c}\<%
\\
\>[0]\AgdaSymbol{...}\AgdaSpace{}%
\AgdaSymbol{|}\AgdaSpace{}%
\AgdaInductiveConstructor{ID}\AgdaSpace{}%
\AgdaInductiveConstructor{,}\AgdaSpace{}%
\AgdaBound{c⟷₂id}\AgdaSpace{}%
\AgdaSymbol{=}\AgdaSpace{}%
\AgdaInductiveConstructor{ID}\AgdaSpace{}%
\AgdaInductiveConstructor{,}\AgdaSpace{}%
\AgdaSymbol{(}\AgdaInductiveConstructor{`!}\AgdaSpace{}%
\AgdaBound{c⟷₂id}\AgdaSpace{}%
\AgdaInductiveConstructor{⊙₂}\AgdaSpace{}%
\AgdaInductiveConstructor{`!id}\AgdaSymbol{)}\<%
\\
\>[0]\AgdaSymbol{...}\AgdaSpace{}%
\AgdaSymbol{|}\AgdaSpace{}%
\AgdaInductiveConstructor{NOT}\AgdaSpace{}%
\AgdaInductiveConstructor{,}\AgdaSpace{}%
\AgdaBound{c⟷₂not}\AgdaSpace{}%
\AgdaSymbol{=}\AgdaSpace{}%
\AgdaInductiveConstructor{NOT}\AgdaSpace{}%
\AgdaInductiveConstructor{,}\AgdaSpace{}%
\AgdaSymbol{(}\AgdaInductiveConstructor{`!}\AgdaSpace{}%
\AgdaBound{c⟷₂not}\AgdaSpace{}%
\AgdaInductiveConstructor{⊙₂}\AgdaSpace{}%
\AgdaInductiveConstructor{`!not}\AgdaSymbol{)}\<%
\\
\>[0]\AgdaFunction{canonical}\AgdaSpace{}%
\AgdaSymbol{(}\AgdaInductiveConstructor{\_⊙₁\_}\AgdaSpace{}%
\AgdaSymbol{\{\_\}}\AgdaSpace{}%
\AgdaSymbol{\{}\AgdaInductiveConstructor{`𝟚}\AgdaSymbol{\}}\AgdaSpace{}%
\AgdaBound{c₁}\AgdaSpace{}%
\AgdaBound{c₂}\AgdaSymbol{)}\AgdaSpace{}%
\AgdaKeyword{with}\AgdaSpace{}%
\AgdaFunction{canonical}\AgdaSpace{}%
\AgdaBound{c₁}\AgdaSpace{}%
\AgdaSymbol{|}\AgdaSpace{}%
\AgdaFunction{canonical}\AgdaSpace{}%
\AgdaBound{c₂}\<%
\\
\>[0]\AgdaSymbol{...}\AgdaSpace{}%
\AgdaSymbol{|}\AgdaSpace{}%
\AgdaInductiveConstructor{ID}\AgdaSpace{}%
\AgdaInductiveConstructor{,}\AgdaSpace{}%
\AgdaBound{c₁⟷₂id}\AgdaSpace{}%
\AgdaSymbol{|}\AgdaSpace{}%
\AgdaInductiveConstructor{ID}\AgdaSpace{}%
\AgdaInductiveConstructor{,}\AgdaSpace{}%
\AgdaBound{c₂⟷₂id}\AgdaSpace{}%
\AgdaSymbol{=}\AgdaSpace{}%
\AgdaInductiveConstructor{ID}\AgdaSpace{}%
\AgdaInductiveConstructor{,}\AgdaSpace{}%
\AgdaSymbol{((}\AgdaBound{c₁⟷₂id}\AgdaSpace{}%
\AgdaInductiveConstructor{□₂}\AgdaSpace{}%
\AgdaBound{c₂⟷₂id}\AgdaSymbol{)}\AgdaSpace{}%
\AgdaInductiveConstructor{⊙₂}\AgdaSpace{}%
\AgdaInductiveConstructor{`idl}\AgdaSpace{}%
\AgdaInductiveConstructor{`id}\AgdaSymbol{)}\<%
\\
\>[0]\AgdaSymbol{...}\AgdaSpace{}%
\AgdaSymbol{|}\AgdaSpace{}%
\AgdaInductiveConstructor{ID}\AgdaSpace{}%
\AgdaInductiveConstructor{,}\AgdaSpace{}%
\AgdaBound{c₁⟷₂id}\AgdaSpace{}%
\AgdaSymbol{|}\AgdaSpace{}%
\AgdaInductiveConstructor{NOT}\AgdaSpace{}%
\AgdaInductiveConstructor{,}\AgdaSpace{}%
\AgdaBound{c₂⟷₂not}\AgdaSpace{}%
\AgdaSymbol{=}\AgdaSpace{}%
\AgdaInductiveConstructor{NOT}\AgdaSpace{}%
\AgdaInductiveConstructor{,}\AgdaSpace{}%
\AgdaSymbol{((}\AgdaBound{c₁⟷₂id}\AgdaSpace{}%
\AgdaInductiveConstructor{□₂}\AgdaSpace{}%
\AgdaBound{c₂⟷₂not}\AgdaSymbol{)}\AgdaSpace{}%
\AgdaInductiveConstructor{⊙₂}\AgdaSpace{}%
\AgdaInductiveConstructor{`idl}\AgdaSpace{}%
\AgdaInductiveConstructor{`not}\AgdaSymbol{)}\<%
\\
\>[0]\AgdaSymbol{...}\AgdaSpace{}%
\AgdaSymbol{|}\AgdaSpace{}%
\AgdaInductiveConstructor{NOT}\AgdaSpace{}%
\AgdaInductiveConstructor{,}\AgdaSpace{}%
\AgdaBound{c₁⟷₂not}\AgdaSpace{}%
\AgdaSymbol{|}\AgdaSpace{}%
\AgdaInductiveConstructor{ID}\AgdaSpace{}%
\AgdaInductiveConstructor{,}\AgdaSpace{}%
\AgdaBound{c₂⟷₂id}\AgdaSpace{}%
\AgdaSymbol{=}\AgdaSpace{}%
\AgdaInductiveConstructor{NOT}\AgdaSpace{}%
\AgdaInductiveConstructor{,}\AgdaSpace{}%
\AgdaSymbol{((}\AgdaBound{c₁⟷₂not}\AgdaSpace{}%
\AgdaInductiveConstructor{□₂}\AgdaSpace{}%
\AgdaBound{c₂⟷₂id}\AgdaSymbol{)}\AgdaSpace{}%
\AgdaInductiveConstructor{⊙₂}\AgdaSpace{}%
\AgdaInductiveConstructor{`idr}\AgdaSpace{}%
\AgdaInductiveConstructor{`not}\AgdaSymbol{)}\<%
\\
\>[0]\AgdaSymbol{...}\AgdaSpace{}%
\AgdaSymbol{|}\AgdaSpace{}%
\AgdaInductiveConstructor{NOT}\AgdaSpace{}%
\AgdaInductiveConstructor{,}\AgdaSpace{}%
\AgdaBound{c₁⟷₂not}\AgdaSpace{}%
\AgdaSymbol{|}\AgdaSpace{}%
\AgdaInductiveConstructor{NOT}\AgdaSpace{}%
\AgdaInductiveConstructor{,}\AgdaSpace{}%
\AgdaBound{c₂⟷₂not}\AgdaSpace{}%
\AgdaSymbol{=}\AgdaSpace{}%
\AgdaInductiveConstructor{ID}\AgdaSpace{}%
\AgdaInductiveConstructor{,}\AgdaSpace{}%
\AgdaSymbol{((}\AgdaBound{c₁⟷₂not}\AgdaSpace{}%
\AgdaInductiveConstructor{□₂}\AgdaSpace{}%
\AgdaBound{c₂⟷₂not}\AgdaSymbol{)}\AgdaSpace{}%
\AgdaInductiveConstructor{⊙₂}\AgdaSpace{}%
\AgdaFunction{not⊙₁not⟷₂id}\AgdaSymbol{)}\<%
\end{code}

It is worthwhile to note that the proof of
{\small\AgdaFunction{canonical}} does not use all the level-2
combinators. The larger set of 2-combinators is however useful to
establish a more direct connection with the model presented in the
next sections.


\section{HoTT Background}
\label{sec:univalent}

We work in intensional type theory with one univalent universe
{\small\AgdaPrimitiveType{𝒰}} closed under propositional truncation.  The rest
of this section is devoted to explaining what that means.  We follow
the terminology used in the HoTT book~\cite{hottbook}.  For brevity,
we will often just give type signatures and elide the term. The details
can be found in the accompanying code at
{\small\url{https://git.io/v7wtW}}.

\subsection{Equivalences}
\label{sec:eq}

Given types {\small\AgdaBound{A}} and {\small\AgdaBound{B}}, a function
{\small\AgdaBound{f}~\AgdaSymbol{:}~\AgdaBound{A}~\AgdaSymbol{→}~\AgdaBound{B}}
is a quasi-inverse, if there is another function
{\small\AgdaBound{g}~\AgdaSymbol{:}~\AgdaBound{B}~\AgdaSymbol{→}~\AgdaBound{A}}
that acts as both a left and right inverse to {\small\AgdaBound{f}}:

\begin{code}%
\>[0]\AgdaFunction{is{-}qinv}\AgdaSpace{}%
\AgdaSymbol{:}\AgdaSpace{}%
\AgdaSymbol{\{}\AgdaBound{A}\AgdaSpace{}%
\AgdaBound{B}\AgdaSpace{}%
\AgdaSymbol{:}\AgdaSpace{}%
\AgdaFunction{𝒰}\AgdaSymbol{\}}\AgdaSpace{}%
\AgdaSymbol{→}\AgdaSpace{}%
\AgdaSymbol{(}\AgdaBound{f}\AgdaSpace{}%
\AgdaSymbol{:}\AgdaSpace{}%
\AgdaBound{A}\AgdaSpace{}%
\AgdaSymbol{→}\AgdaSpace{}%
\AgdaBound{B}\AgdaSymbol{)}\AgdaSpace{}%
\AgdaSymbol{→}\AgdaSpace{}%
\AgdaFunction{𝒰}\<%
\\
\>[0]\AgdaFunction{is{-}qinv}\AgdaSpace{}%
\AgdaSymbol{\{}\AgdaBound{A}\AgdaSymbol{\}}\AgdaSpace{}%
\AgdaSymbol{\{}\AgdaBound{B}\AgdaSymbol{\}}\AgdaSpace{}%
\AgdaBound{f}\AgdaSpace{}%
\AgdaSymbol{=}\AgdaSpace{}%
\AgdaRecord{Σ[}\AgdaSpace{}%
\AgdaBound{g}\AgdaSpace{}%
\AgdaRecord{∶}\AgdaSpace{}%
\AgdaSymbol{(}\AgdaBound{B}\AgdaSpace{}%
\AgdaSymbol{→}\AgdaSpace{}%
\AgdaBound{A}\AgdaSymbol{)}\AgdaSpace{}%
\AgdaRecord{]}\AgdaSpace{}%
\AgdaSymbol{(}\AgdaBound{g}\AgdaSpace{}%
\AgdaFunction{∘}\AgdaSpace{}%
\AgdaBound{f}\AgdaSpace{}%
\AgdaFunction{∼}\AgdaSpace{}%
\AgdaFunction{id}\AgdaSpace{}%
\AgdaFunction{×}\AgdaSpace{}%
\AgdaBound{f}\AgdaSpace{}%
\AgdaFunction{∘}\AgdaSpace{}%
\AgdaBound{g}\AgdaSpace{}%
\AgdaFunction{∼}\AgdaSpace{}%
\AgdaFunction{id}\AgdaSymbol{)}\<%
\end{code}


In general, for a given ${\small\AgdaBound{f}}$, there could be several
unequal inhabitants of the type
${\small\AgdaFunction{is-qinv}~\AgdaBound{f}}$. As Ch.~4 of the HoTT
book~\cite{hottbook} details, this is problematic in the
proof-relevant setting of HoTT. To ensure that a function
${\small\AgdaBound{f}}$ can be an equivalence in at most one way, an
additional coherence condition is added to quasi-inverses to define
\emph{half adjoint equivalences}:

\begin{code}%
\>[0]\AgdaFunction{is{-}hae}\AgdaSpace{}%
\AgdaSymbol{:}\AgdaSpace{}%
\AgdaSymbol{\{}\AgdaBound{A}\AgdaSpace{}%
\AgdaBound{B}\AgdaSpace{}%
\AgdaSymbol{:}\AgdaSpace{}%
\AgdaFunction{𝒰}\AgdaSymbol{\}}\AgdaSpace{}%
\AgdaSymbol{→}\AgdaSpace{}%
\AgdaSymbol{(}\AgdaBound{f}\AgdaSpace{}%
\AgdaSymbol{:}\AgdaSpace{}%
\AgdaBound{A}\AgdaSpace{}%
\AgdaSymbol{→}\AgdaSpace{}%
\AgdaBound{B}\AgdaSymbol{)}\AgdaSpace{}%
\AgdaSymbol{→}\AgdaSpace{}%
\AgdaFunction{𝒰}\<%
\\
\>[0]\AgdaFunction{is{-}hae}\AgdaSpace{}%
\AgdaSymbol{\{}\AgdaBound{A}\AgdaSymbol{\}}\AgdaSpace{}%
\AgdaSymbol{\{}\AgdaBound{B}\AgdaSymbol{\}}\AgdaSpace{}%
\AgdaBound{f}\AgdaSpace{}%
\AgdaSymbol{=}\AgdaSpace{}%
\AgdaRecord{Σ[}\AgdaSpace{}%
\AgdaBound{g}\AgdaSpace{}%
\AgdaRecord{∶}\AgdaSpace{}%
\AgdaSymbol{(}\AgdaBound{B}\AgdaSpace{}%
\AgdaSymbol{→}\AgdaSpace{}%
\AgdaBound{A}\AgdaSymbol{)}\AgdaSpace{}%
\AgdaRecord{]}\AgdaSpace{}%
\AgdaRecord{Σ[}\AgdaSpace{}%
\AgdaBound{η}\AgdaSpace{}%
\AgdaRecord{∶}\AgdaSpace{}%
\AgdaBound{g}\AgdaSpace{}%
\AgdaFunction{∘}\AgdaSpace{}%
\AgdaBound{f}\AgdaSpace{}%
\AgdaFunction{∼}\AgdaSpace{}%
\AgdaFunction{id}\AgdaSpace{}%
\AgdaRecord{]}\AgdaSpace{}%
\AgdaRecord{Σ[}\AgdaSpace{}%
\AgdaBound{ε}\AgdaSpace{}%
\AgdaRecord{∶}\AgdaSpace{}%
\AgdaBound{f}\AgdaSpace{}%
\AgdaFunction{∘}\AgdaSpace{}%
\AgdaBound{g}\AgdaSpace{}%
\AgdaFunction{∼}\AgdaSpace{}%
\AgdaFunction{id}\AgdaSpace{}%
\AgdaRecord{]}\AgdaSpace{}%
\AgdaSymbol{(}\AgdaFunction{ap}\AgdaSpace{}%
\AgdaBound{f}\AgdaSpace{}%
\AgdaFunction{∘}\AgdaSpace{}%
\AgdaBound{η}\AgdaSpace{}%
\AgdaFunction{∼}\AgdaSpace{}%
\AgdaBound{ε}\AgdaSpace{}%
\AgdaFunction{∘}\AgdaSpace{}%
\AgdaBound{f}\AgdaSymbol{)}\<%
\\
\\
\>[0]\AgdaFunction{qinv{-}is{-}hae}\AgdaSpace{}%
\AgdaSymbol{:}\AgdaSpace{}%
\AgdaSymbol{\{}\AgdaBound{A}\AgdaSpace{}%
\AgdaBound{B}\AgdaSpace{}%
\AgdaSymbol{:}\AgdaSpace{}%
\AgdaFunction{𝒰}\AgdaSymbol{\}}\AgdaSpace{}%
\AgdaSymbol{\{}\AgdaBound{f}\AgdaSpace{}%
\AgdaSymbol{:}\AgdaSpace{}%
\AgdaBound{A}\AgdaSpace{}%
\AgdaSymbol{→}\AgdaSpace{}%
\AgdaBound{B}\AgdaSymbol{\}}\AgdaSpace{}%
\AgdaSymbol{→}\AgdaSpace{}%
\AgdaFunction{is{-}qinv}\AgdaSpace{}%
\AgdaBound{f}\AgdaSpace{}%
\AgdaSymbol{→}\AgdaSpace{}%
\AgdaFunction{is{-}hae}\AgdaSpace{}%
\AgdaBound{f}\<%
\end{code}
\AgdaHide{
\begin{code}%
\>[0]\AgdaFunction{qinv{-}is{-}hae}\AgdaSpace{}%
\AgdaSymbol{=}\AgdaSpace{}%
\AgdaSymbol{\{!!\}}\<%
\end{code}
}

Using this latter notion, we can define a well-behaved notion of
equivalences between two types:

\begin{code}%
\>[0]\AgdaFunction{is{-}equiv}\AgdaSpace{}%
\AgdaSymbol{=}\AgdaSpace{}%
\AgdaFunction{is{-}hae}\<%
\\
\\
\>[0]\AgdaFunction{\_≃\_}\AgdaSpace{}%
\AgdaSymbol{:}\AgdaSpace{}%
\AgdaSymbol{(}\AgdaBound{A}\AgdaSpace{}%
\AgdaBound{B}\AgdaSpace{}%
\AgdaSymbol{:}\AgdaSpace{}%
\AgdaFunction{𝒰}\AgdaSymbol{)}\AgdaSpace{}%
\AgdaSymbol{→}\AgdaSpace{}%
\AgdaFunction{𝒰}\<%
\\
\>[0]\AgdaBound{A}\AgdaSpace{}%
\AgdaFunction{≃}\AgdaSpace{}%
\AgdaBound{B}\AgdaSpace{}%
\AgdaSymbol{=}\AgdaSpace{}%
\AgdaRecord{Σ[}\AgdaSpace{}%
\AgdaBound{f}\AgdaSpace{}%
\AgdaRecord{∶}\AgdaSpace{}%
\AgdaSymbol{(}\AgdaBound{A}\AgdaSpace{}%
\AgdaSymbol{→}\AgdaSpace{}%
\AgdaBound{B}\AgdaSymbol{)}\AgdaSpace{}%
\AgdaRecord{]}\AgdaSpace{}%
\AgdaSymbol{(}\AgdaFunction{is{-}equiv}\AgdaSpace{}%
\AgdaBound{f}\AgdaSymbol{)}\<%
\end{code}

It is straightforward to lift paths to equivalences as shown below:


\AgdaHide{
\begin{code}%
\>[0]\AgdaFunction{idh}\AgdaSpace{}%
\AgdaSymbol{:}\AgdaSpace{}%
\AgdaSymbol{\{}\AgdaBound{A}\AgdaSpace{}%
\AgdaSymbol{:}\AgdaSpace{}%
\AgdaFunction{𝒰}\AgdaSymbol{\}}\AgdaSpace{}%
\AgdaSymbol{\{}\AgdaBound{P}\AgdaSpace{}%
\AgdaSymbol{:}\AgdaSpace{}%
\AgdaBound{A}\AgdaSpace{}%
\AgdaSymbol{→}\AgdaSpace{}%
\AgdaFunction{𝒰}\AgdaSymbol{\}}\AgdaSpace{}%
\AgdaSymbol{→}\AgdaSpace{}%
\AgdaSymbol{(}\AgdaBound{f}\AgdaSpace{}%
\AgdaSymbol{:}\AgdaSpace{}%
\AgdaFunction{Π[}\AgdaSpace{}%
\AgdaBound{a}\AgdaSpace{}%
\AgdaFunction{∶}\AgdaSpace{}%
\AgdaBound{A}\AgdaSpace{}%
\AgdaFunction{]}\AgdaSpace{}%
\AgdaBound{P}\AgdaSpace{}%
\AgdaBound{a}\AgdaSymbol{)}\AgdaSpace{}%
\AgdaSymbol{→}\AgdaSpace{}%
\AgdaBound{f}\AgdaSpace{}%
\AgdaFunction{∼}\AgdaSpace{}%
\AgdaBound{f}\<%
\\
\>[0]\AgdaFunction{idh}\AgdaSpace{}%
\AgdaBound{f}\AgdaSpace{}%
\AgdaBound{a}\AgdaSpace{}%
\AgdaSymbol{=}\AgdaSpace{}%
\AgdaInductiveConstructor{refl}\AgdaSpace{}%
\AgdaSymbol{(}\AgdaBound{f}\AgdaSpace{}%
\AgdaBound{a}\AgdaSymbol{)}\<%
\end{code}
}

\begin{code}%
\>[0]\AgdaFunction{ide}\AgdaSpace{}%
\AgdaSymbol{:}\AgdaSpace{}%
\AgdaSymbol{(}\AgdaBound{A}\AgdaSpace{}%
\AgdaSymbol{:}\AgdaSpace{}%
\AgdaFunction{𝒰}\AgdaSymbol{)}\AgdaSpace{}%
\AgdaSymbol{→}\AgdaSpace{}%
\AgdaBound{A}\AgdaSpace{}%
\AgdaFunction{≃}\AgdaSpace{}%
\AgdaBound{A}\<%
\\
\>[0]\AgdaFunction{ide}\AgdaSpace{}%
\AgdaBound{A}\AgdaSpace{}%
\AgdaSymbol{=}\AgdaSpace{}%
\AgdaFunction{id}\AgdaSpace{}%
\AgdaInductiveConstructor{,}\AgdaSpace{}%
\AgdaFunction{id}\AgdaSpace{}%
\AgdaInductiveConstructor{,}\AgdaSpace{}%
\AgdaInductiveConstructor{refl}\AgdaSpace{}%
\AgdaInductiveConstructor{,}\AgdaSpace{}%
\AgdaInductiveConstructor{refl}\AgdaSpace{}%
\AgdaInductiveConstructor{,}\AgdaSpace{}%
\AgdaSymbol{(}\AgdaInductiveConstructor{refl}\AgdaSpace{}%
\AgdaFunction{∘}\AgdaSpace{}%
\AgdaInductiveConstructor{refl}\AgdaSymbol{)}\<%
\\
\\
\>[0]\AgdaFunction{transport{-}equiv}\AgdaSpace{}%
\AgdaSymbol{:}\AgdaSpace{}%
\AgdaSymbol{\{}\AgdaBound{A}\AgdaSpace{}%
\AgdaSymbol{:}\AgdaSpace{}%
\AgdaFunction{𝒰}\AgdaSymbol{\}}\AgdaSpace{}%
\AgdaSymbol{(}\AgdaBound{P}\AgdaSpace{}%
\AgdaSymbol{:}\AgdaSpace{}%
\AgdaBound{A}\AgdaSpace{}%
\AgdaSymbol{→}\AgdaSpace{}%
\AgdaFunction{𝒰}\AgdaSymbol{)}\AgdaSpace{}%
\AgdaSymbol{→}\AgdaSpace{}%
\AgdaSymbol{\{}\AgdaBound{a}\AgdaSpace{}%
\AgdaBound{b}\AgdaSpace{}%
\AgdaSymbol{:}\AgdaSpace{}%
\AgdaBound{A}\AgdaSymbol{\}}\AgdaSpace{}%
\AgdaSymbol{→}\AgdaSpace{}%
\AgdaBound{a}\AgdaSpace{}%
\AgdaDatatype{==}\AgdaSpace{}%
\AgdaBound{b}\AgdaSpace{}%
\AgdaSymbol{→}\AgdaSpace{}%
\AgdaBound{P}\AgdaSpace{}%
\AgdaBound{a}\AgdaSpace{}%
\AgdaFunction{≃}\AgdaSpace{}%
\AgdaBound{P}\AgdaSpace{}%
\AgdaBound{b}\<%
\\
\>[0]\AgdaFunction{transport{-}equiv}\AgdaSpace{}%
\AgdaBound{P}\AgdaSpace{}%
\AgdaSymbol{(}\AgdaInductiveConstructor{refl}\AgdaSpace{}%
\AgdaBound{a}\AgdaSymbol{)}\AgdaSpace{}%
\AgdaSymbol{=}\AgdaSpace{}%
\AgdaFunction{ide}\AgdaSpace{}%
\AgdaSymbol{(}\AgdaBound{P}\AgdaSpace{}%
\AgdaBound{a}\AgdaSymbol{)}\<%
\\
\\
\>[0]\AgdaFunction{id{-}to{-}equiv}\AgdaSpace{}%
\AgdaSymbol{:}\AgdaSpace{}%
\AgdaSymbol{\{}\AgdaBound{A}\AgdaSpace{}%
\AgdaBound{B}\AgdaSpace{}%
\AgdaSymbol{:}\AgdaSpace{}%
\AgdaFunction{𝒰}\AgdaSymbol{\}}\AgdaSpace{}%
\AgdaSymbol{→}\AgdaSpace{}%
\AgdaBound{A}\AgdaSpace{}%
\AgdaDatatype{==}\AgdaSpace{}%
\AgdaBound{B}\AgdaSpace{}%
\AgdaSymbol{→}\AgdaSpace{}%
\AgdaBound{A}\AgdaSpace{}%
\AgdaFunction{≃}\AgdaSpace{}%
\AgdaBound{B}\<%
\\
\>[0]\AgdaFunction{id{-}to{-}equiv}\AgdaSpace{}%
\AgdaSymbol{=}\AgdaSpace{}%
\AgdaFunction{transport{-}equiv}\AgdaSpace{}%
\AgdaFunction{id}\<%
\end{code}

Dually, univalence allows us to construct paths from equivalences. We postulate
univalence as an axiom in our Agda library:

\begin{code}%
\>[0]\AgdaKeyword{postulate}\<%
\\
\>[0][@{}l@{\AgdaIndent{0}}]%
\>[2]\AgdaPostulate{univalence}\AgdaSpace{}%
\AgdaSymbol{:}\AgdaSpace{}%
\AgdaSymbol{(}\AgdaBound{A}\AgdaSpace{}%
\AgdaBound{B}\AgdaSpace{}%
\AgdaSymbol{:}\AgdaSpace{}%
\AgdaFunction{𝒰}\AgdaSymbol{)}\AgdaSpace{}%
\AgdaSymbol{→}\AgdaSpace{}%
\AgdaFunction{is{-}equiv}\AgdaSpace{}%
\AgdaSymbol{(}\AgdaFunction{id{-}to{-}equiv}\AgdaSpace{}%
\AgdaSymbol{\{}\AgdaBound{A}\AgdaSymbol{\}}\AgdaSpace{}%
\AgdaSymbol{\{}\AgdaBound{B}\AgdaSymbol{\})}\<%
\end{code}

We also give a short form {\small\AgdaFunction{ua}} for getting a path from an
equivalence, and prove some computation rules for it:

\begin{code}%
\>[0]\AgdaKeyword{module}\AgdaSpace{}%
\AgdaModule{\_}\AgdaSpace{}%
\AgdaSymbol{\{}\AgdaBound{A}\AgdaSpace{}%
\AgdaBound{B}\AgdaSpace{}%
\AgdaSymbol{:}\AgdaSpace{}%
\AgdaFunction{𝒰}\AgdaSymbol{\}}\AgdaSpace{}%
\AgdaKeyword{where}\<%
\\
\>[0][@{}l@{\AgdaIndent{0}}]%
\>[2]\AgdaFunction{ua}\AgdaSpace{}%
\AgdaSymbol{:}\AgdaSpace{}%
\AgdaBound{A}\AgdaSpace{}%
\AgdaFunction{≃}\AgdaSpace{}%
\AgdaBound{B}\AgdaSpace{}%
\AgdaSymbol{→}\AgdaSpace{}%
\AgdaBound{A}\AgdaSpace{}%
\AgdaDatatype{==}\AgdaSpace{}%
\AgdaBound{B}\<%
\\
\>[0][@{}l@{\AgdaIndent{0}}]%
\>[2]\AgdaFunction{ua}\AgdaSpace{}%
\AgdaSymbol{=}\AgdaSpace{}%
\AgdaField{pr₁}\AgdaSpace{}%
\AgdaSymbol{(}\AgdaPostulate{univalence}\AgdaSpace{}%
\AgdaBound{A}\AgdaSpace{}%
\AgdaBound{B}\AgdaSymbol{)}\<%
\\
\\
\>[0][@{}l@{\AgdaIndent{0}}]%
\>[2]\AgdaFunction{ua{-}β}\AgdaSpace{}%
\AgdaSymbol{:}\AgdaSpace{}%
\AgdaFunction{id{-}to{-}equiv}\AgdaSpace{}%
\AgdaFunction{∘}\AgdaSpace{}%
\AgdaFunction{ua}\AgdaSpace{}%
\AgdaFunction{∼}\AgdaSpace{}%
\AgdaFunction{id}\<%
\\
\>[0][@{}l@{\AgdaIndent{0}}]%
\>[2]\AgdaFunction{ua{-}β}\AgdaSpace{}%
\AgdaSymbol{=}\AgdaSpace{}%
\AgdaField{pr₁}\AgdaSpace{}%
\AgdaSymbol{(}\AgdaField{pr₂}\AgdaSpace{}%
\AgdaSymbol{(}\AgdaField{pr₂}\AgdaSpace{}%
\AgdaSymbol{(}\AgdaPostulate{univalence}\AgdaSpace{}%
\AgdaBound{A}\AgdaSpace{}%
\AgdaBound{B}\AgdaSymbol{)))}\<%
\\
\\
\>[0][@{}l@{\AgdaIndent{0}}]%
\>[2]\AgdaFunction{ua{-}β₁}\AgdaSpace{}%
\AgdaSymbol{:}\AgdaSpace{}%
\AgdaFunction{transport}\AgdaSpace{}%
\AgdaFunction{id}\AgdaSpace{}%
\AgdaFunction{∘}\AgdaSpace{}%
\AgdaFunction{ua}\AgdaSpace{}%
\AgdaFunction{∼}\AgdaSpace{}%
\AgdaField{pr₁}\<%
\\
\>[0][@{}l@{\AgdaIndent{0}}]%
\>[2]\AgdaFunction{ua{-}β₁}\AgdaSpace{}%
\AgdaBound{eqv}\AgdaSpace{}%
\AgdaSymbol{=}\AgdaSpace{}%
\AgdaFunction{transport}\AgdaSpace{}%
\AgdaSymbol{\_}\AgdaSpace{}%
\AgdaSymbol{(}\AgdaFunction{ua{-}β}\AgdaSpace{}%
\AgdaBound{eqv}\AgdaSymbol{)}\AgdaSpace{}%
\AgdaSymbol{(}\AgdaFunction{ap}\AgdaSpace{}%
\AgdaField{pr₁}\AgdaSymbol{)}\<%
\\
\\
\>[0][@{}l@{\AgdaIndent{0}}]%
\>[2]\AgdaFunction{ua{-}η}\AgdaSpace{}%
\AgdaSymbol{:}\AgdaSpace{}%
\AgdaFunction{ua}\AgdaSpace{}%
\AgdaFunction{∘}\AgdaSpace{}%
\AgdaFunction{id{-}to{-}equiv}\AgdaSpace{}%
\AgdaFunction{∼}\AgdaSpace{}%
\AgdaFunction{id}\<%
\\
\>[0][@{}l@{\AgdaIndent{0}}]%
\>[2]\AgdaFunction{ua{-}η}\AgdaSpace{}%
\AgdaSymbol{=}\AgdaSpace{}%
\AgdaField{pr₁}\AgdaSpace{}%
\AgdaSymbol{(}\AgdaField{pr₂}\AgdaSpace{}%
\AgdaSymbol{(}\AgdaPostulate{univalence}\AgdaSpace{}%
\AgdaBound{A}\AgdaSpace{}%
\AgdaBound{B}\AgdaSymbol{))}\<%
\end{code}

\subsection{Propositional Truncation}

A type {\small\AgdaBound{A}} is \emph{contractible} (h-level 0 or
(-2)-truncated), if it has a center of contraction, and all other
terms of {\small\AgdaBound{A}} are connected to it by a path:


\begin{code}%
\>[0]\AgdaFunction{is{-}contr}\AgdaSpace{}%
\AgdaSymbol{:}\AgdaSpace{}%
\AgdaSymbol{(}\AgdaBound{A}\AgdaSpace{}%
\AgdaSymbol{:}\AgdaSpace{}%
\AgdaFunction{𝒰}\AgdaSymbol{)}\AgdaSpace{}%
\AgdaSymbol{→}\AgdaSpace{}%
\AgdaFunction{𝒰}\<%
\\
\>[0]\AgdaFunction{is{-}contr}\AgdaSpace{}%
\AgdaBound{A}\AgdaSpace{}%
\AgdaSymbol{=}\AgdaSpace{}%
\AgdaRecord{Σ[}\AgdaSpace{}%
\AgdaBound{a}\AgdaSpace{}%
\AgdaRecord{∶}\AgdaSpace{}%
\AgdaBound{A}\AgdaSpace{}%
\AgdaRecord{]}\AgdaSpace{}%
\AgdaFunction{Π[}\AgdaSpace{}%
\AgdaBound{b}\AgdaSpace{}%
\AgdaFunction{∶}\AgdaSpace{}%
\AgdaBound{A}\AgdaSpace{}%
\AgdaFunction{]}\AgdaSpace{}%
\AgdaSymbol{(}\AgdaBound{a}\AgdaSpace{}%
\AgdaDatatype{==}\AgdaSpace{}%
\AgdaBound{b}\AgdaSymbol{)}\<%
\end{code}

As alluded to in the previous section, equivalences are contractible
(4.2.13 in~\cite{hottbook}):

\begin{code}%
\>[0]\AgdaFunction{is{-}equiv{-}is{-}contr}\AgdaSpace{}%
\AgdaSymbol{:}\AgdaSpace{}%
\AgdaSymbol{\{}\AgdaBound{A}\AgdaSpace{}%
\AgdaBound{B}\AgdaSpace{}%
\AgdaSymbol{:}\AgdaSpace{}%
\AgdaFunction{𝒰}\AgdaSymbol{\}}\AgdaSpace{}%
\AgdaSymbol{\{}\AgdaBound{f}\AgdaSpace{}%
\AgdaSymbol{:}\AgdaSpace{}%
\AgdaBound{A}\AgdaSpace{}%
\AgdaSymbol{→}\AgdaSpace{}%
\AgdaBound{B}\AgdaSymbol{\}}\AgdaSpace{}%
\AgdaSymbol{→}\AgdaSpace{}%
\AgdaFunction{is{-}equiv}\AgdaSpace{}%
\AgdaBound{f}\AgdaSpace{}%
\AgdaSymbol{→}\AgdaSpace{}%
\AgdaFunction{is{-}contr}\AgdaSpace{}%
\AgdaSymbol{(}\AgdaFunction{is{-}equiv}\AgdaSpace{}%
\AgdaBound{f}\AgdaSymbol{)}\<%
\end{code}
\AgdaHide{
\begin{code}%
\>[0]\AgdaFunction{is{-}equiv{-}is{-}contr}\AgdaSpace{}%
\AgdaSymbol{=}\AgdaSpace{}%
\AgdaSymbol{\{!!\}}\<%
\end{code}
}

A type {\small\AgdaBound{A}} is a \emph{proposition} (h-level 1 or
(-1)-truncated) if all pairs of terms of {\small\AgdaBound{A}} are
connected by a path. Such a type can have at most one inhabitant; it is
``contractible if inhabited.'' Finally, a type {\small\AgdaBound{A}} is
a \emph{set} if for any two terms {\small\AgdaBound{a}} and
{\small\AgdaBound{b}} of {\small\AgdaBound{A}}, its type of paths
{\small\AgdaBound{a}~\AgdaFunction{==}~\AgdaBound{b}} is a proposition:

\begin{code}%
\>[0]\AgdaFunction{is{-}prop}\AgdaSpace{}%
\AgdaSymbol{:}\AgdaSpace{}%
\AgdaSymbol{(}\AgdaBound{A}\AgdaSpace{}%
\AgdaSymbol{:}\AgdaSpace{}%
\AgdaFunction{𝒰}\AgdaSymbol{)}\AgdaSpace{}%
\AgdaSymbol{→}\AgdaSpace{}%
\AgdaFunction{𝒰}\<%
\\
\>[0]\AgdaFunction{is{-}prop}\AgdaSpace{}%
\AgdaBound{A}\AgdaSpace{}%
\AgdaSymbol{=}\AgdaSpace{}%
\AgdaFunction{Π[}\AgdaSpace{}%
\AgdaBound{a}\AgdaSpace{}%
\AgdaFunction{∶}\AgdaSpace{}%
\AgdaBound{A}\AgdaSpace{}%
\AgdaFunction{]}\AgdaSpace{}%
\AgdaFunction{Π[}\AgdaSpace{}%
\AgdaBound{b}\AgdaSpace{}%
\AgdaFunction{∶}\AgdaSpace{}%
\AgdaBound{A}\AgdaSpace{}%
\AgdaFunction{]}\AgdaSpace{}%
\AgdaSymbol{(}\AgdaBound{a}\AgdaSpace{}%
\AgdaDatatype{==}\AgdaSpace{}%
\AgdaBound{b}\AgdaSymbol{)}\<%
\\
\\
\>[0]\AgdaFunction{is{-}set}\AgdaSpace{}%
\AgdaSymbol{:}\AgdaSpace{}%
\AgdaSymbol{(}\AgdaBound{A}\AgdaSpace{}%
\AgdaSymbol{:}\AgdaSpace{}%
\AgdaFunction{𝒰}\AgdaSymbol{)}\AgdaSpace{}%
\AgdaSymbol{→}\AgdaSpace{}%
\AgdaFunction{𝒰}\<%
\\
\>[0]\AgdaFunction{is{-}set}\AgdaSpace{}%
\AgdaBound{A}\AgdaSpace{}%
\AgdaSymbol{=}\AgdaSpace{}%
\AgdaFunction{Π[}\AgdaSpace{}%
\AgdaBound{a}\AgdaSpace{}%
\AgdaFunction{∶}\AgdaSpace{}%
\AgdaBound{A}\AgdaSpace{}%
\AgdaFunction{]}\AgdaSpace{}%
\AgdaFunction{Π[}\AgdaSpace{}%
\AgdaBound{b}\AgdaSpace{}%
\AgdaFunction{∶}\AgdaSpace{}%
\AgdaBound{A}\AgdaSpace{}%
\AgdaFunction{]}\AgdaSpace{}%
\AgdaFunction{is{-}prop}\AgdaSpace{}%
\AgdaSymbol{(}\AgdaBound{a}\AgdaSpace{}%
\AgdaDatatype{==}\AgdaSpace{}%
\AgdaBound{b}\AgdaSymbol{)}\<%
\end{code}

Any type can be truncated to a proposition by freely adding
paths. This is the propositional truncation (or (-1)-truncation) which
can be expressed as a higher inductive type (HIT). The type
constructor {\small\AgdaInductiveConstructor{∥\_∥}} takes a type
{\small\AgdaBound{A}} as a parameter; the point constructor
{\small\AgdaInductiveConstructor{∣\_∣}} coerces terms of type
{\small\AgdaBound{A}} to terms in the truncation; and the path
constructor {\small\AgdaInductiveConstructor{ident}} identifies any
two points in the truncation, making it a proposition. We must do this
as a {\small\AgdaKeyword{postulate}} as Agda does not yet support
HITs:

\begin{code}%
\>[0]\AgdaKeyword{postulate}\<%
\\
\>[0][@{}l@{\AgdaIndent{0}}]%
\>[4]\AgdaPostulate{∥\_∥}%
\>[11]\AgdaSymbol{:}\AgdaSpace{}%
\AgdaSymbol{(}\AgdaBound{A}\AgdaSpace{}%
\AgdaSymbol{:}\AgdaSpace{}%
\AgdaFunction{𝒰}\AgdaSymbol{)}\AgdaSpace{}%
\AgdaSymbol{→}\AgdaSpace{}%
\AgdaFunction{𝒰}\<%
\\
\>[0][@{}l@{\AgdaIndent{0}}]%
\>[4]\AgdaPostulate{∣\_∣}%
\>[11]\AgdaSymbol{:}\AgdaSpace{}%
\AgdaSymbol{\{}\AgdaBound{A}\AgdaSpace{}%
\AgdaSymbol{:}\AgdaSpace{}%
\AgdaFunction{𝒰}\AgdaSymbol{\}}\AgdaSpace{}%
\AgdaSymbol{→}\AgdaSpace{}%
\AgdaSymbol{(}\AgdaBound{a}\AgdaSpace{}%
\AgdaSymbol{:}\AgdaSpace{}%
\AgdaBound{A}\AgdaSymbol{)}\AgdaSpace{}%
\AgdaSymbol{→}\AgdaSpace{}%
\AgdaPostulate{∥}\AgdaSpace{}%
\AgdaBound{A}\AgdaSpace{}%
\AgdaPostulate{∥}\<%
\\
\>[0][@{}l@{\AgdaIndent{0}}]%
\>[4]\AgdaPostulate{ident}%
\>[11]\AgdaSymbol{:}\AgdaSpace{}%
\AgdaSymbol{\{}\AgdaBound{A}\AgdaSpace{}%
\AgdaSymbol{:}\AgdaSpace{}%
\AgdaFunction{𝒰}\AgdaSymbol{\}}\AgdaSpace{}%
\AgdaSymbol{\{}\AgdaBound{a}\AgdaSpace{}%
\AgdaBound{b}\AgdaSpace{}%
\AgdaSymbol{:}\AgdaSpace{}%
\AgdaPostulate{∥}\AgdaSpace{}%
\AgdaBound{A}\AgdaSpace{}%
\AgdaPostulate{∥}\AgdaSymbol{\}}\AgdaSpace{}%
\AgdaSymbol{→}\AgdaSpace{}%
\AgdaBound{a}\AgdaSpace{}%
\AgdaDatatype{==}\AgdaSpace{}%
\AgdaBound{b}\<%
\\
\\
\>[0]\AgdaFunction{∥{-}∥{-}is{-}prop}\AgdaSpace{}%
\AgdaSymbol{:}\AgdaSpace{}%
\AgdaSymbol{\{}\AgdaBound{A}\AgdaSpace{}%
\AgdaSymbol{:}\AgdaSpace{}%
\AgdaFunction{𝒰}\AgdaSymbol{\}}\AgdaSpace{}%
\AgdaSymbol{→}\AgdaSpace{}%
\AgdaFunction{is{-}prop}\AgdaSpace{}%
\AgdaPostulate{∥}\AgdaSpace{}%
\AgdaBound{A}\AgdaSpace{}%
\AgdaPostulate{∥}\<%
\\
\>[0]\AgdaFunction{∥{-}∥{-}is{-}prop}\AgdaSpace{}%
\AgdaSymbol{\_}\AgdaSpace{}%
\AgdaSymbol{\_}\AgdaSpace{}%
\AgdaSymbol{=}\AgdaSpace{}%
\AgdaPostulate{ident}\<%
\end{code}

This makes
{\small\AgdaInductiveConstructor{∥}\AgdaBound{A}\AgdaInductiveConstructor{∥}}
the ``free'' proposition on any type {\small\AgdaBound{A}}. The
recursion principle (below) ensures that we can only eliminate a
propositional truncation to a type that is a proposition:

\begin{code}%
\>[0]\AgdaKeyword{module}\AgdaSpace{}%
\AgdaModule{\_}\AgdaSpace{}%
\AgdaSymbol{\{}\AgdaBound{A}\AgdaSpace{}%
\AgdaSymbol{:}\AgdaSpace{}%
\AgdaFunction{𝒰}\AgdaSymbol{\}}\AgdaSpace{}%
\AgdaSymbol{(}\AgdaBound{P}\AgdaSpace{}%
\AgdaSymbol{:}\AgdaSpace{}%
\AgdaFunction{𝒰}\AgdaSymbol{)}\AgdaSpace{}%
\AgdaSymbol{(}\AgdaBound{f}\AgdaSpace{}%
\AgdaSymbol{:}\AgdaSpace{}%
\AgdaBound{A}\AgdaSpace{}%
\AgdaSymbol{→}\AgdaSpace{}%
\AgdaBound{P}\AgdaSymbol{)}\AgdaSpace{}%
\AgdaSymbol{(}\AgdaBound{φ}\AgdaSpace{}%
\AgdaSymbol{:}\AgdaSpace{}%
\AgdaFunction{is{-}prop}\AgdaSpace{}%
\AgdaBound{P}\AgdaSymbol{)}\AgdaSpace{}%
\AgdaKeyword{where}\<%
\\
\>[0][@{}l@{\AgdaIndent{0}}]%
\>[2]\AgdaKeyword{postulate}\<%
\\
\>[2][@{}l@{\AgdaIndent{0}}]%
\>[4]\AgdaPostulate{rec{-}∥{-}∥}\AgdaSpace{}%
\AgdaSymbol{:}\AgdaSpace{}%
\AgdaPostulate{∥}\AgdaSpace{}%
\AgdaBound{A}\AgdaSpace{}%
\AgdaPostulate{∥}\AgdaSpace{}%
\AgdaSymbol{→}\AgdaSpace{}%
\AgdaBound{P}\<%
\\
\>[2][@{}l@{\AgdaIndent{0}}]%
\>[4]\AgdaPostulate{rec{-}∥{-}∥{-}β}\AgdaSpace{}%
\AgdaSymbol{:}\AgdaSpace{}%
\AgdaFunction{Π[}\AgdaSpace{}%
\AgdaBound{a}\AgdaSpace{}%
\AgdaFunction{∶}\AgdaSpace{}%
\AgdaBound{A}\AgdaSpace{}%
\AgdaFunction{]}\AgdaSpace{}%
\AgdaSymbol{(}\AgdaPostulate{rec{-}∥{-}∥}\AgdaSpace{}%
\AgdaPostulate{∣}\AgdaSpace{}%
\AgdaBound{a}\AgdaSpace{}%
\AgdaPostulate{∣}\AgdaSpace{}%
\AgdaDatatype{==}\AgdaSpace{}%
\AgdaBound{f}\AgdaSpace{}%
\AgdaBound{a}\AgdaSymbol{)}\<%
\end{code}

\begin{figure}
\begin{tabular}{c@{\qquad\qquad}c}
\begin{tikzpicture}[scale=0.7,every node/.style={scale=0.7}]]
  \draw (0,-5) ellipse (2cm and 0.8cm);
  \node[below] at (0,-6) {Base Space $A$};
  \draw[fill] (-1,-5) circle [radius=0.025];
  \node[below] at (-1,-5) {$x$};
  \draw[fill] (1,-5) circle [radius=0.025];
  \node[below] at (1,-5) {$y$};
  \draw (-1,-2) ellipse (0.5cm and 2cm);
  \node[left] at (-1.5,-2) {Fiber $P(x)$};
  \draw (1,-2) ellipse (0.5cm and 2cm);
  \node[right] at (1.5,-2) {Fiber $P(y)$};
\end{tikzpicture}
&
\begin{tikzpicture}[scale=0.7,every node/.style={scale=0.7}]]
  \draw (0,-5) ellipse (2cm and 0.8cm);
  \node[below] at (0,-6) {Base Space $A$};
  \draw[fill] (-1,-5) circle [radius=0.025];
  \node[below] at (-1,-5) {$x$};
  \draw[fill] (1,-5) circle [radius=0.025];
  \node[below] at (1,-5) {$y$};
  \draw (-1.3,-2) ellipse (0.5cm and 2cm);
  \node[left] at (-1.8,-2) {Fiber $P(x)$};
  \draw (1.3,-2) ellipse (0.5cm and 2cm);
  \node[right] at (1.8,-2) {Fiber $P(y)$};
  \draw[below,cyan,thick] (-1,-5) -- (1,-5);
  \node[below,cyan,thick] at (0,-5) {$==$};
  \draw[->,red,thick] (-0.8,-1.7) to [out=45, in=135] (0.8,-1.7);
  \draw[->,red,thick] (0.8,-2.3) to [out=-135, in=-45] (-0.8,-2.3);
  \node[red,thick] at (0,-2) {$\simeq$};
\end{tikzpicture}
\end{tabular}
\caption{\label{fig:fib}(left) Type family $P : A \rightarrow \mathcal{U}$ as a
  fibration with total space $\Sigma_{(x:A)} P(x)$;\newline
 (right) a path $x == y$
  in the base space induces an equivalence between the spaces (fibers)
  $P(x)$ and $P(y)$}
\end{figure}
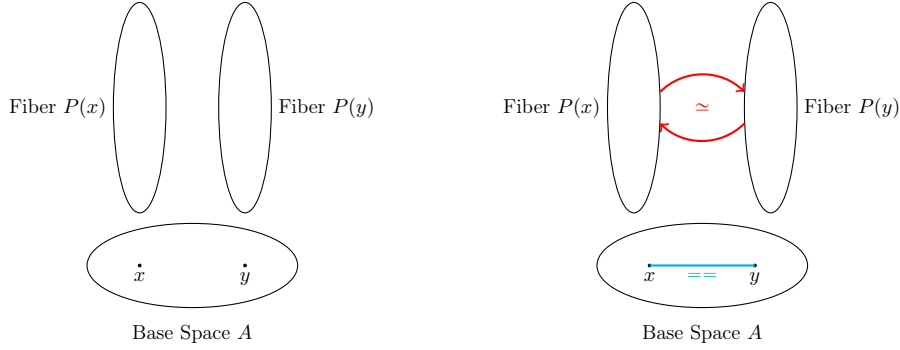

\subsection{Type Families are Fibrations}

As illustrated in Fig.~\ref{fig:fib}, a type family
{\small\AgdaBound{P}} over a type~{\small\AgdaBound{A}} is a fibration
with base space~{\small\AgdaBound{A}}, with every
{\small\AgdaBound{x}} in {\small\AgdaBound{A}} inducing a fiber
{\small\AgdaBound{P}~\AgdaBound{x}}, and with total space
{\small\AgdaPrimitiveType{Σ[}~\AgdaBound{x}~\AgdaSymbol{∶}~\AgdaBound{A}~\AgdaSymbol{]}~\AgdaSymbol{(}\AgdaBound{P}~\AgdaBound{x}\AgdaSymbol{)}}.\footnote{In
this and following figures, we color paths in blue and functions
in red.}

The path lifting property mapping a path in the base space to a path
in the total space can be defined as follows:

\begin{code}%
\>[0]\AgdaFunction{lift}\AgdaSpace{}%
\AgdaSymbol{:}\AgdaSpace{}%
\AgdaSymbol{\{}\AgdaBound{A}\AgdaSpace{}%
\AgdaSymbol{:}\AgdaSpace{}%
\AgdaFunction{𝒰}\AgdaSymbol{\}}\AgdaSpace{}%
\AgdaSymbol{\{}\AgdaBound{P}\AgdaSpace{}%
\AgdaSymbol{:}\AgdaSpace{}%
\AgdaBound{A}\AgdaSpace{}%
\AgdaSymbol{→}\AgdaSpace{}%
\AgdaFunction{𝒰}\AgdaSymbol{\}}\AgdaSpace{}%
\AgdaSymbol{\{}\AgdaBound{x}\AgdaSpace{}%
\AgdaBound{y}\AgdaSpace{}%
\AgdaSymbol{:}\AgdaSpace{}%
\AgdaBound{A}\AgdaSymbol{\}}\AgdaSpace{}%
\AgdaSymbol{→}\AgdaSpace{}%
\AgdaSymbol{(}\AgdaBound{u}\AgdaSpace{}%
\AgdaSymbol{:}\AgdaSpace{}%
\AgdaBound{P}\AgdaSpace{}%
\AgdaBound{x}\AgdaSymbol{)}\AgdaSpace{}%
\AgdaSymbol{(}\AgdaBound{p}\AgdaSpace{}%
\AgdaSymbol{:}\AgdaSpace{}%
\AgdaBound{x}\AgdaSpace{}%
\AgdaDatatype{==}\AgdaSpace{}%
\AgdaBound{y}\AgdaSymbol{)}\AgdaSpace{}%
\AgdaSymbol{→}\AgdaSpace{}%
\AgdaSymbol{(}\AgdaBound{x}\AgdaSpace{}%
\AgdaInductiveConstructor{,}\AgdaSpace{}%
\AgdaBound{u}\AgdaSymbol{)}\AgdaSpace{}%
\AgdaDatatype{==}\AgdaSpace{}%
\AgdaSymbol{(}\AgdaBound{y}\AgdaSpace{}%
\AgdaInductiveConstructor{,}\AgdaSpace{}%
\AgdaFunction{transport}\AgdaSpace{}%
\AgdaBound{P}\AgdaSpace{}%
\AgdaBound{p}\AgdaSpace{}%
\AgdaBound{u}\AgdaSymbol{)}\<%
\\
\>[0]\AgdaFunction{lift}\AgdaSpace{}%
\AgdaBound{u}\AgdaSpace{}%
\AgdaSymbol{(}\AgdaInductiveConstructor{refl}\AgdaSpace{}%
\AgdaBound{x}\AgdaSymbol{)}\AgdaSpace{}%
\AgdaSymbol{=}\AgdaSpace{}%
\AgdaInductiveConstructor{refl}\AgdaSpace{}%
\AgdaSymbol{(}\AgdaBound{x}\AgdaSpace{}%
\AgdaInductiveConstructor{,}\AgdaSpace{}%
\AgdaBound{u}\AgdaSymbol{)}\<%
\end{code}

As illustrated in the figure below, the point
{\small\AgdaFunction{transport}~\AgdaBound{P}~\AgdaBound{p}~\AgdaBound{u}}
is in the space {\small\AgdaBound{P}~\AgdaBound{y}}. A path from that
point to another point {\small\AgdaBound{v}} in
{\small\AgdaBound{P}~\AgdaBound{y}} can be viewed as a virtual
``path'' between {\small\AgdaBound{u}} and {\small\AgdaBound{v}} that
``lies over'' {\small\AgdaBound{p}}. Following Licata and
Brunerie~\cite{licata2015cubical}, we often use the syntax
{\small\AgdaBound{u} \AgdaFunction{==} \AgdaBound{v} \AgdaFunction{[}
  \AgdaBound{P} \AgdaFunction{↓} \AgdaBound{p} \AgdaFunction{]}} for
the path
{\small\AgdaFunction{transport}~\AgdaBound{P}~\AgdaBound{p}~\AgdaBound{u}
  \AgdaFunction{==} \AgdaBound{v}} to reinforce this perspective. In
other words, the curved ``path'' between {\small\AgdaBound{u}} and
{\small\AgdaBound{v}} below consists of first transporting
{\small\AgdaBound{u}} to the space {\small\AgdaBound{P}~\AgdaBound{y}}
along {\small\AgdaBound{p}} and then following the straight path in
{\small\AgdaBound{P}~\AgdaBound{y}} to {\small\AgdaBound{v}}:

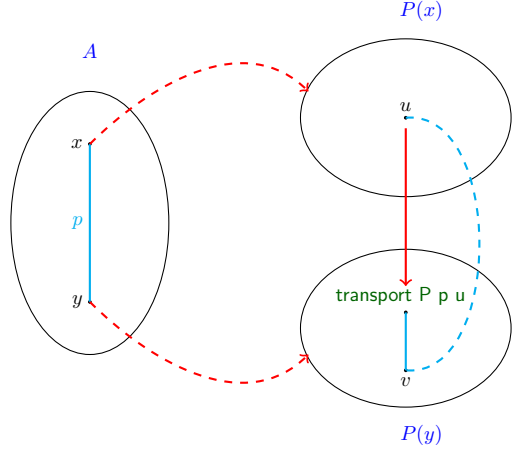
\begin{center}
\begin{tikzpicture}[scale=0.7,every node/.style={scale=0.7}]]
  \draw (-3,0) ellipse (1.5cm and 2.5cm);
  \draw (3,2) ellipse (2cm and 1.5cm);
  \draw (3,-2) ellipse (2cm and 1.5cm);
  \node[blue,thick,above] at (-3,3) {$A$};
  \node[blue,thick,above] at (3.3,3.7) {$P(x)$};
  \node[blue,thick,below] at (3.3,-3.7) {$P(y)$};
  \draw[fill] (-3,1.5) circle [radius=0.025];
  \draw[fill] (-3,-1.5) circle [radius=0.025];
  \draw[left,cyan,thick] (-3,1.5) -- (-3,-1.5);
  \node[left] (X) at (-3,1.5) {$x$};
  \node[left] at (-3,-1.5) {$y$};
  \draw[fill] (3,-1.7) circle [radius=0.025];
  \draw[fill] (3,2) circle [radius=0.025];
  \node[above] at (3,2) {$u$};
  \draw[fill] (3,-2.8) circle [radius=0.025];
  \node[below] at (3,-2.8) {$v$};
  \node[left,above] at (3,-1.7) {$\!\!\!\AgdaFunction{transport}~\AgdaFunction{P}~\AgdaFunction{p}~\AgdaFunction{u}$};
  \node[left,cyan] at (-3,0) {$p$};
  \draw[->,red,thick] (3,1.8) -- (3,-1.2);
  \draw[->,red,dashed,thick] (-3,1.5) to [out=45, in=135] (1.15,2.5);
  \draw[->,red,dashed,thick] (-3,-1.5) to [out=-45, in=-135] (1.15,-2.5);
  \draw[cyan,thick] (3,-1.7) to (3,-2.8);
  \draw[cyan,dashed,thick] (3,2) to [out=5, in=-5] (3,-2.8);
\end{tikzpicture}
\end{center}

\noindent Given a fibration ${\small\AgdaBound{P}}$ and points
{\small\AgdaBound{x}}, {\small\AgdaBound{y}}, {\small\AgdaBound{u}}, and
{\small\AgdaBound{v}} as above, we have the following characterization of
dependent paths in the total space:

\begin{code}%
\>[0]\AgdaKeyword{module}\AgdaSpace{}%
\AgdaModule{\_}\AgdaSpace{}%
\AgdaSymbol{\{}\AgdaBound{A}\AgdaSpace{}%
\AgdaSymbol{:}\AgdaSpace{}%
\AgdaFunction{𝒰}\AgdaSymbol{\}}\AgdaSpace{}%
\AgdaSymbol{\{}\AgdaBound{P}\AgdaSpace{}%
\AgdaSymbol{:}\AgdaSpace{}%
\AgdaBound{A}\AgdaSpace{}%
\AgdaSymbol{→}\AgdaSpace{}%
\AgdaFunction{𝒰}\AgdaSymbol{\}}\AgdaSpace{}%
\AgdaSymbol{\{}\AgdaBound{x}\AgdaSpace{}%
\AgdaBound{y}\AgdaSpace{}%
\AgdaSymbol{:}\AgdaSpace{}%
\AgdaBound{A}\AgdaSymbol{\}}\AgdaSpace{}%
\AgdaSymbol{\{}\AgdaBound{u}\AgdaSpace{}%
\AgdaSymbol{:}\AgdaSpace{}%
\AgdaBound{P}\AgdaSpace{}%
\AgdaBound{x}\AgdaSymbol{\}}\AgdaSpace{}%
\AgdaSymbol{\{}\AgdaBound{v}\AgdaSpace{}%
\AgdaSymbol{:}\AgdaSpace{}%
\AgdaBound{P}\AgdaSpace{}%
\AgdaBound{y}\AgdaSymbol{\}}\AgdaSpace{}%
\AgdaKeyword{where}\<%
\\
\\
\>[0][@{}l@{\AgdaIndent{0}}]%
\>[2]\AgdaFunction{dpair=}\AgdaSpace{}%
\AgdaSymbol{:}\AgdaSpace{}%
\AgdaRecord{Σ[}\AgdaSpace{}%
\AgdaBound{p}\AgdaSpace{}%
\AgdaRecord{∶}\AgdaSpace{}%
\AgdaBound{x}\AgdaSpace{}%
\AgdaDatatype{==}\AgdaSpace{}%
\AgdaBound{y}\AgdaSpace{}%
\AgdaRecord{]}\AgdaSpace{}%
\AgdaSymbol{(}\AgdaBound{u}\AgdaSpace{}%
\AgdaFunction{==}\AgdaSpace{}%
\AgdaBound{v}\AgdaSpace{}%
\AgdaFunction{[}\AgdaSpace{}%
\AgdaBound{P}\AgdaSpace{}%
\AgdaFunction{↓}\AgdaSpace{}%
\AgdaBound{p}\AgdaSpace{}%
\AgdaFunction{]}\AgdaSymbol{)}\AgdaSpace{}%
\AgdaSymbol{→}\AgdaSpace{}%
\AgdaSymbol{(}\AgdaBound{x}\AgdaSpace{}%
\AgdaInductiveConstructor{,}\AgdaSpace{}%
\AgdaBound{u}\AgdaSymbol{)}\AgdaSpace{}%
\AgdaDatatype{==}\AgdaSpace{}%
\AgdaSymbol{(}\AgdaBound{y}\AgdaSpace{}%
\AgdaInductiveConstructor{,}\AgdaSpace{}%
\AgdaBound{v}\AgdaSymbol{)}\<%
\\
\>[0][@{}l@{\AgdaIndent{0}}]%
\>[2]\AgdaFunction{dpair=}\AgdaSpace{}%
\AgdaSymbol{(}\AgdaInductiveConstructor{refl}\AgdaSpace{}%
\AgdaBound{x}\AgdaSpace{}%
\AgdaInductiveConstructor{,}\AgdaSpace{}%
\AgdaInductiveConstructor{refl}\AgdaSpace{}%
\AgdaBound{u}\AgdaSymbol{)}\AgdaSpace{}%
\AgdaSymbol{=}\AgdaSpace{}%
\AgdaInductiveConstructor{refl}\AgdaSpace{}%
\AgdaSymbol{(}\AgdaBound{x}\AgdaSpace{}%
\AgdaInductiveConstructor{,}\AgdaSpace{}%
\AgdaBound{u}\AgdaSymbol{)}\<%
\\
\\
\>[0][@{}l@{\AgdaIndent{0}}]%
\>[2]\AgdaFunction{dpair={-}β}\AgdaSpace{}%
\AgdaSymbol{:}\AgdaSpace{}%
\AgdaSymbol{(}\AgdaBound{w}\AgdaSpace{}%
\AgdaSymbol{:}\AgdaSpace{}%
\AgdaRecord{Σ[}\AgdaSpace{}%
\AgdaBound{p}\AgdaSpace{}%
\AgdaRecord{∶}\AgdaSpace{}%
\AgdaBound{x}\AgdaSpace{}%
\AgdaDatatype{==}\AgdaSpace{}%
\AgdaBound{y}\AgdaSpace{}%
\AgdaRecord{]}\AgdaSpace{}%
\AgdaSymbol{(}\AgdaBound{u}\AgdaSpace{}%
\AgdaFunction{==}\AgdaSpace{}%
\AgdaBound{v}\AgdaSpace{}%
\AgdaFunction{[}\AgdaSpace{}%
\AgdaBound{P}\AgdaSpace{}%
\AgdaFunction{↓}\AgdaSpace{}%
\AgdaBound{p}\AgdaSpace{}%
\AgdaFunction{]}\AgdaSymbol{))}\AgdaSpace{}%
\AgdaSymbol{→}\AgdaSpace{}%
\AgdaSymbol{(}\AgdaFunction{ap}\AgdaSpace{}%
\AgdaField{pr₁}\AgdaSpace{}%
\AgdaFunction{∘}\AgdaSpace{}%
\AgdaFunction{dpair=}\AgdaSymbol{)}\AgdaSpace{}%
\AgdaBound{w}\AgdaSpace{}%
\AgdaDatatype{==}\AgdaSpace{}%
\AgdaField{pr₁}\AgdaSpace{}%
\AgdaBound{w}\<%
\\
\>[0][@{}l@{\AgdaIndent{0}}]%
\>[2]\AgdaFunction{dpair={-}β}\AgdaSpace{}%
\AgdaSymbol{(}\AgdaInductiveConstructor{refl}\AgdaSpace{}%
\AgdaBound{x}\AgdaSpace{}%
\AgdaInductiveConstructor{,}\AgdaSpace{}%
\AgdaInductiveConstructor{refl}\AgdaSpace{}%
\AgdaBound{u}\AgdaSymbol{)}\AgdaSpace{}%
\AgdaSymbol{=}\AgdaSpace{}%
\AgdaInductiveConstructor{refl}\AgdaSpace{}%
\AgdaSymbol{(}\AgdaInductiveConstructor{refl}\AgdaSpace{}%
\AgdaBound{x}\AgdaSymbol{)}\<%
\\
\\
\>[0][@{}l@{\AgdaIndent{0}}]%
\>[2]\AgdaFunction{dpair={-}e}\AgdaSpace{}%
\AgdaSymbol{:}\AgdaSpace{}%
\AgdaSymbol{(}\AgdaBound{x}\AgdaSpace{}%
\AgdaInductiveConstructor{,}\AgdaSpace{}%
\AgdaBound{u}\AgdaSymbol{)}\AgdaSpace{}%
\AgdaDatatype{==}\AgdaSpace{}%
\AgdaSymbol{(}\AgdaBound{y}\AgdaSpace{}%
\AgdaInductiveConstructor{,}\AgdaSpace{}%
\AgdaBound{v}\AgdaSymbol{)}\AgdaSpace{}%
\AgdaSymbol{→}\AgdaSpace{}%
\AgdaBound{x}\AgdaSpace{}%
\AgdaDatatype{==}\AgdaSpace{}%
\AgdaBound{y}\<%
\\
\>[0][@{}l@{\AgdaIndent{0}}]%
\>[2]\AgdaFunction{dpair={-}e}\AgdaSpace{}%
\AgdaSymbol{=}\AgdaSpace{}%
\AgdaFunction{ap}\AgdaSpace{}%
\AgdaField{pr₁}\<%
\end{code}

\AgdaHide{
\begin{code}%
\>[0]\AgdaFunction{prop{-}is{-}set}\AgdaSpace{}%
\AgdaSymbol{:}\AgdaSpace{}%
\AgdaSymbol{\{}\AgdaBound{A}\AgdaSpace{}%
\AgdaSymbol{:}\AgdaSpace{}%
\AgdaFunction{𝒰}\AgdaSymbol{\}}\AgdaSpace{}%
\AgdaSymbol{→}\AgdaSpace{}%
\AgdaFunction{is{-}prop}\AgdaSpace{}%
\AgdaBound{A}\AgdaSpace{}%
\AgdaSymbol{→}\AgdaSpace{}%
\AgdaFunction{is{-}set}\AgdaSpace{}%
\AgdaBound{A}\<%
\\
\>[0]\AgdaFunction{prop{-}is{-}set}\AgdaSpace{}%
\AgdaBound{φ}\AgdaSpace{}%
\AgdaBound{a}\AgdaSpace{}%
\AgdaBound{b}\AgdaSpace{}%
\AgdaBound{p}\AgdaSpace{}%
\AgdaBound{q}\AgdaSpace{}%
\AgdaSymbol{=}\AgdaSpace{}%
\AgdaSymbol{\{!!\}}\<%
\\
\\
\>[0]\AgdaFunction{is{-}equiv{-}is{-}prop}\AgdaSpace{}%
\AgdaSymbol{:}\AgdaSpace{}%
\AgdaSymbol{\{}\AgdaBound{A}\AgdaSpace{}%
\AgdaBound{B}\AgdaSpace{}%
\AgdaSymbol{:}\AgdaSpace{}%
\AgdaFunction{𝒰}\AgdaSymbol{\}}\AgdaSpace{}%
\AgdaSymbol{\{}\AgdaBound{f}\AgdaSpace{}%
\AgdaSymbol{:}\AgdaSpace{}%
\AgdaBound{A}\AgdaSpace{}%
\AgdaSymbol{→}\AgdaSpace{}%
\AgdaBound{B}\AgdaSymbol{\}}\AgdaSpace{}%
\AgdaSymbol{→}\AgdaSpace{}%
\AgdaFunction{is{-}prop}\AgdaSpace{}%
\AgdaSymbol{(}\AgdaFunction{is{-}equiv}\AgdaSpace{}%
\AgdaBound{f}\AgdaSymbol{)}\<%
\\
\>[0]\AgdaFunction{is{-}equiv{-}is{-}prop}\AgdaSpace{}%
\AgdaSymbol{=}\AgdaSpace{}%
\AgdaSymbol{\{!!\}}\<%
\\
\\
\>[0]\AgdaFunction{eqv=}\AgdaSpace{}%
\AgdaSymbol{:}\AgdaSpace{}%
\AgdaSymbol{\{}\AgdaBound{A}\AgdaSpace{}%
\AgdaBound{B}\AgdaSpace{}%
\AgdaSymbol{:}\AgdaSpace{}%
\AgdaFunction{𝒰}\AgdaSymbol{\}}\AgdaSpace{}%
\AgdaSymbol{\{}\AgdaBound{eqv}\AgdaSpace{}%
\AgdaBound{eqv'}\AgdaSpace{}%
\AgdaSymbol{:}\AgdaSpace{}%
\AgdaBound{A}\AgdaSpace{}%
\AgdaFunction{≃}\AgdaSpace{}%
\AgdaBound{B}\AgdaSymbol{\}}\AgdaSpace{}%
\AgdaSymbol{→}\AgdaSpace{}%
\AgdaSymbol{(}\AgdaField{pr₁}\AgdaSpace{}%
\AgdaBound{eqv}\AgdaSpace{}%
\AgdaDatatype{==}\AgdaSpace{}%
\AgdaField{pr₁}\AgdaSpace{}%
\AgdaBound{eqv'}\AgdaSymbol{)}\AgdaSpace{}%
\AgdaSymbol{→}\AgdaSpace{}%
\AgdaBound{eqv}\AgdaSpace{}%
\AgdaDatatype{==}\AgdaSpace{}%
\AgdaBound{eqv'}\<%
\\
\>[0]\AgdaFunction{eqv=}\AgdaSpace{}%
\AgdaBound{φ}\AgdaSpace{}%
\AgdaSymbol{=}\AgdaSpace{}%
\AgdaFunction{dpair=}\AgdaSpace{}%
\AgdaSymbol{(}\AgdaBound{φ}\AgdaSpace{}%
\AgdaInductiveConstructor{,}\AgdaSpace{}%
\AgdaFunction{is{-}equiv{-}is{-}prop}\AgdaSpace{}%
\AgdaSymbol{\_}\AgdaSpace{}%
\AgdaSymbol{\_)}\<%
\end{code}
}

The first function builds a path in the total space given a path between
{\small\AgdaBound{u}} and {\small\AgdaBound{v}} that lies over a path
{\small\AgdaBound{p}} in the base space; the second function is a computation
rule for this path; and the third function eliminates a path in the total space
to a path in the base space.

\subsection{Univalent Fibrations}

Univalent fibrations are defined by Kapulkin and
Lumsdaine~\cite{SimplicialModel} in the simplicial set (sSet) model.  In
our context, a type family (fibration)
{\small\AgdaBound{P}~\AgdaSymbol{:}~\AgdaBound{A}~\AgdaSymbol{→}~\AgdaFunction{𝒰}}
is univalent if the map
{\small\AgdaFunction{transport-equiv}~\AgdaBound{P}} defined in
Sec.~\ref{sec:eq} is an equivalence, that is, if the space of paths in
the base space is \emph{equivalent} to the space of equivalences between
the corresponding fibers. Fig.~\ref{fig:fib} (right) illustrates the
situation: we know that for any fibration {\small\AgdaBound{P}} that a
path~{\small\AgdaBound{p}} in the base space induces via
{\small\AgdaFunction{transport-equiv}~\AgdaBound{P}~\AgdaBound{p}} an
equivalence between the fibers. For a fibration to be univalent, the
reverse must also be true: every equivalence between the fibers must
induce a path in the base space. Formally, we have the following
definition:

\begin{code}%
\>[0]\AgdaFunction{is{-}univ{-}fib}\AgdaSpace{}%
\AgdaSymbol{:}\AgdaSpace{}%
\AgdaSymbol{\{}\AgdaBound{A}\AgdaSpace{}%
\AgdaSymbol{:}\AgdaSpace{}%
\AgdaFunction{𝒰}\AgdaSymbol{\}}\AgdaSpace{}%
\AgdaSymbol{(}\AgdaBound{P}\AgdaSpace{}%
\AgdaSymbol{:}\AgdaSpace{}%
\AgdaBound{A}\AgdaSpace{}%
\AgdaSymbol{→}\AgdaSpace{}%
\AgdaFunction{𝒰}\AgdaSymbol{)}\AgdaSpace{}%
\AgdaSymbol{→}\AgdaSpace{}%
\AgdaFunction{𝒰}\<%
\\
\>[0]\AgdaFunction{is{-}univ{-}fib}\AgdaSpace{}%
\AgdaSymbol{\{}\AgdaBound{A}\AgdaSymbol{\}}\AgdaSpace{}%
\AgdaBound{P}\AgdaSpace{}%
\AgdaSymbol{=}\AgdaSpace{}%
\AgdaSymbol{∀}\AgdaSpace{}%
\AgdaSymbol{(}\AgdaBound{a}\AgdaSpace{}%
\AgdaBound{b}\AgdaSpace{}%
\AgdaSymbol{:}\AgdaSpace{}%
\AgdaBound{A}\AgdaSymbol{)}\AgdaSpace{}%
\AgdaSymbol{→}\AgdaSpace{}%
\AgdaFunction{is{-}equiv}\AgdaSpace{}%
\AgdaSymbol{(}\AgdaFunction{transport{-}equiv}\AgdaSpace{}%
\AgdaBound{P}\AgdaSpace{}%
\AgdaSymbol{\{}\AgdaBound{a}\AgdaSymbol{\}}\AgdaSpace{}%
\AgdaSymbol{\{}\AgdaBound{b}\AgdaSymbol{\})}\<%
\end{code}

We note that the univalence axiom (for {\small\AgdaFunction{𝒰}}) is a
specialization of {\small\AgdaFunction{is-univ-fib}} to the identity
fibration, {\small\AgdaFunction{id}}. More generally, we can define
universes \`{a} la Tarski by having a code {\small\AgdaFunction{U}} for
the universe and an interpretation function {\small\AgdaFunction{El}}
into {\small\AgdaFunction{𝒰}}. Such a presented universe is univalent if
{\small\AgdaFunction{El}} is a univalent fibration:

\begin{code}%
\>[0]\AgdaFunction{Ũ}\AgdaSpace{}%
\AgdaSymbol{=}\AgdaSpace{}%
\AgdaRecord{Σ[}\AgdaSpace{}%
\AgdaBound{U}\AgdaSpace{}%
\AgdaRecord{∶}\AgdaSpace{}%
\AgdaFunction{𝒰}\AgdaSpace{}%
\AgdaRecord{]}\AgdaSpace{}%
\AgdaSymbol{(}\AgdaBound{U}\AgdaSpace{}%
\AgdaSymbol{→}\AgdaSpace{}%
\AgdaFunction{𝒰}\AgdaSymbol{)}\<%
\\
\\
\>[0]\AgdaFunction{is{-}univalent}\AgdaSpace{}%
\AgdaSymbol{:}\AgdaSpace{}%
\AgdaFunction{Ũ}\AgdaSpace{}%
\AgdaSymbol{→}\AgdaSpace{}%
\AgdaFunction{𝒰}\<%
\\
\>[0]\AgdaFunction{is{-}univalent}\AgdaSpace{}%
\AgdaSymbol{(}\AgdaBound{U}\AgdaSpace{}%
\AgdaInductiveConstructor{,}\AgdaSpace{}%
\AgdaBound{El}\AgdaSymbol{)}\AgdaSpace{}%
\AgdaSymbol{=}\AgdaSpace{}%
\AgdaFunction{is{-}univ{-}fib}\AgdaSpace{}%
\AgdaBound{El}\<%
\end{code}

\noindent As Christensen~\cite{christensen} explains, a type
{\small\AgdaBound{U}} is \emph{rarely} the base of a univalent
fibration. Yet, in that same paper, Christensen characterizes a class
of types that is always the base of univalent fibrations. We explain
this point and exploit it to build a custom univalent subuniverse in
the next section.


\section{The Subuniverse {\normalfont\AgdaFunction{Ũ[} \AgdaDatatype{𝟚} \AgdaFunction{]}}}
\label{sec:model}

We now have all the ingredients necessary to define the class of
univalent subuniverses we are interested in. Given any type
{\small\AgdaBound{T}}, we can build a propositional predicate that
picks out from among all the types in the universe exactly those which
are identified with~{\small\AgdaBound{T}}. This lets us build up a
``singleton'' subuniverse of~{\small\AgdaFunction{𝒰}} as follows:

\begin{code}%
\>[0]\AgdaFunction{Ũ[\_]}\AgdaSpace{}%
\AgdaSymbol{:}\AgdaSpace{}%
\AgdaSymbol{(}\AgdaBound{T}\AgdaSpace{}%
\AgdaSymbol{:}\AgdaSpace{}%
\AgdaFunction{𝒰}\AgdaSymbol{)}\AgdaSpace{}%
\AgdaSymbol{→}\AgdaSpace{}%
\AgdaFunction{Ũ}\<%
\\
\>[0]\AgdaFunction{Ũ[}\AgdaSpace{}%
\AgdaBound{T}\AgdaSpace{}%
\AgdaFunction{]}\AgdaSpace{}%
\AgdaSymbol{=}\AgdaSpace{}%
\AgdaFunction{U}\AgdaSpace{}%
\AgdaInductiveConstructor{,}\AgdaSpace{}%
\AgdaFunction{El}\<%
\\
\>[0][@{}l@{\AgdaIndent{0}}]%
\>[2]\AgdaKeyword{where}\<%
\\
\>[2][@{}l@{\AgdaIndent{0}}]%
\>[4]\AgdaFunction{U}%
\>[8]\AgdaSymbol{=}\AgdaSpace{}%
\AgdaRecord{Σ[}\AgdaSpace{}%
\AgdaBound{X}\AgdaSpace{}%
\AgdaRecord{∶}\AgdaSpace{}%
\AgdaFunction{𝒰}\AgdaSpace{}%
\AgdaRecord{]}\AgdaSpace{}%
\AgdaPostulate{∥}\AgdaSpace{}%
\AgdaBound{X}\AgdaSpace{}%
\AgdaDatatype{==}\AgdaSpace{}%
\AgdaBound{T}\AgdaSpace{}%
\AgdaPostulate{∥}\<%
\\
\>[2][@{}l@{\AgdaIndent{0}}]%
\>[4]\AgdaFunction{El}%
\>[8]\AgdaSymbol{=}\AgdaSpace{}%
\AgdaField{pr₁}\<%
\end{code}

We will prove in this section and the next that choosing
{\small\AgdaBound{T}} to be {\small\AgdaDatatype{𝟚}} produces a
universe that is sound and complete with respect to the language
$\PiTwo$. The bulk of the argument consists of establishing that
{\small\AgdaFunction{Ũ[} \AgdaDatatype{𝟚} \AgdaFunction{]}} is a
univalent universe. We focus on this argument in the first subsection.
In the next two subsections, we use this result to characterize the
points and paths in the type of codes for this universe. In
Sec.~\ref{sec:correspondence} this characterization of points and
paths will be shown to match the types and combinators of $\PiTwo$.


\subsection{The Fibration {\normalfont\AgdaFunction{El𝟚}} is Univalent}

The universe {\small\AgdaFunction{Ũ[} \AgdaDatatype{𝟚}
  \AgdaFunction{]}} consists of a base space
{\small\AgdaFunction{U[𝟚]}} of the codes for the elements, and an
interpretation function {\small\AgdaFunction{El𝟚}}, defined as follows:

\begin{code}%
\>[0]\AgdaFunction{U[𝟚]}\AgdaSpace{}%
\AgdaSymbol{:}\AgdaSpace{}%
\AgdaFunction{𝒰}\<%
\\
\>[0]\AgdaFunction{U[𝟚]}%
\>[6]\AgdaSymbol{=}\AgdaSpace{}%
\AgdaField{pr₁}\AgdaSpace{}%
\AgdaFunction{Ũ[}\AgdaSpace{}%
\AgdaDatatype{𝟚}\AgdaSpace{}%
\AgdaFunction{]}%
\>[21]\AgdaComment{{-}{-}{-}{-}{-}  = Σ[ X ∶ 𝒰 ] ∥ X == 𝟚 ∥}\<%
\\
\\
\>[0]\AgdaFunction{El𝟚}%
\>[5]\AgdaSymbol{:}\AgdaSpace{}%
\AgdaRecord{Σ[}\AgdaSpace{}%
\AgdaBound{X}\AgdaSpace{}%
\AgdaRecord{∶}\AgdaSpace{}%
\AgdaFunction{𝒰}\AgdaSpace{}%
\AgdaRecord{]}\AgdaSpace{}%
\AgdaPostulate{∥}\AgdaSpace{}%
\AgdaBound{X}\AgdaSpace{}%
\AgdaDatatype{==}\AgdaSpace{}%
\AgdaDatatype{𝟚}\AgdaSpace{}%
\AgdaPostulate{∥}\AgdaSpace{}%
\AgdaSymbol{→}\AgdaSpace{}%
\AgdaFunction{𝒰}\<%
\\
\>[0]\AgdaFunction{El𝟚}%
\>[5]\AgdaSymbol{=}\AgdaSpace{}%
\AgdaField{pr₁}\<%
\end{code}

The type family {\small\AgdaFunction{El𝟚}} defines a fibration with
base space {\small\AgdaFunction{U[𝟚]}} as shown below:

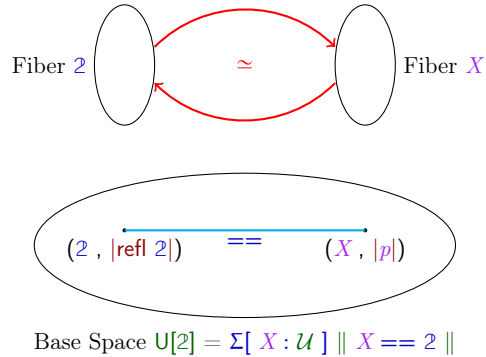
\begin{center}
\begin{tikzpicture}[scale=0.8,every node/.style={scale=0.8}]]
  \draw (0,-5) ellipse (3.5cm and 1.2cm);
  \node[below] at (0,-6.3) {Base Space \AgdaFunction{U[𝟚]} = \AgdaRecord{Σ[} \AgdaBound{X} \AgdaRecord{∶} \AgdaFunction{𝒰} \AgdaRecord{]} \AgdaPostulate{∥} \AgdaBound{X} \AgdaDatatype{==} \AgdaDatatype{𝟚} \AgdaPostulate{∥}};
  \draw[fill] (-2,-4.75) circle [radius=0.025];
  \node[below] at (-2,-4.75) {\AgdaSymbol{(}\AgdaDatatype{𝟚}~\AgdaSymbol{,}~\AgdaInductiveConstructor{∣refl}~\AgdaDatatype{𝟚}\AgdaInductiveConstructor{∣}\AgdaSymbol{)}};
  \draw[fill] (2,-4.75) circle [radius=0.025];
  \node[below] at (2,-4.75) {\AgdaSymbol{(}\AgdaBound{X}~\AgdaSymbol{,}~\AgdaInductiveConstructor{∣}\AgdaBound{p}\AgdaInductiveConstructor{∣}\AgdaSymbol{)}};
  \draw[below,cyan,thick] (-2,-4.75) -- (2,-4.75);
  \node[below,cyan,thick] at (0,-4.75) {\AgdaDatatype{==}};

  \draw (-2,-2) ellipse (0.5cm and 1cm);
  \node[left] at (-2.5,-2) {Fiber \AgdaDatatype{𝟚}};
  \draw (2,-2) ellipse (0.5cm and 1cm);
  \node[right] at (2.5,-2) {Fiber \AgdaBound{X}};
  \draw[->,red,thick] (-1.5,-1.7) to [out=45, in=135] (1.5,-1.7);
  \draw[->,red,thick] (1.5,-2.3) to [out=-135, in=-45] (-1.5,-2.3);
  \node[red,thick] at (0,-2) {$\simeq$};
\end{tikzpicture}
\end{center}

Our goal is to show that {\small\AgdaFunction{El𝟚}} is a univalent
fibration. We establish this by chaining two equivalences. The first
equivalence is a simple appeal to univalence in order to establish that
{\small (\AgdaBound{X}~\AgdaFunction{==}~\AgdaDatatype{𝟚})
  ~\AgdaFunction{≃}~ (\AgdaBound{X}~\AgdaFunction{≃}~\AgdaDatatype{𝟚})},
i.e., our base space is equivalent to the space
\mbox{\small\AgdaRecord{Σ[} ~\AgdaBound{X} ~\AgdaRecord{∶}
  ~\AgdaFunction{𝒰}~ \AgdaRecord{]} ~\AgdaPostulate{∥} ~\AgdaBound{X}
  ~\AgdaFunction{≃}~ \AgdaDatatype{𝟚} ~\AgdaPostulate{∥}}.  We name this
space {\small\AgdaFunction{BAut}~\AgdaDatatype{𝟚}}. Generally,
{\small\AgdaFunction{BAut}~\AgdaBound{T}} is the ``classifying space''
of all types that are (merely) equivalent to {\small\AgdaBound{T}}.  The
second equivalence consists of proving that the first projection on
{\small\AgdaFunction{BAut}~\AgdaDatatype{T}} is in fact a univalent
fibration, for all spaces with shape \mbox{\small\AgdaRecord{Σ[}
  ~\AgdaBound{X} ~\AgdaRecord{∶} ~\AgdaFunction{𝒰}~ \AgdaRecord{]}
  ~\AgdaPostulate{∥} ~\AgdaBound{X} ~\AgdaFunction{≃}~ \AgdaDatatype{T}
  ~\AgdaPostulate{∥}} for any type {\small\AgdaDatatype{T}}.  This is
the lemma {\small\AgdaFunction{is-univ-fib-ElB}} below whose original
formulation is due to Christensen~\cite{christensen}:

\begin{code}%
\>[0]\AgdaFunction{BAut}\AgdaSpace{}%
\AgdaSymbol{:}\AgdaSpace{}%
\AgdaSymbol{(}\AgdaBound{T}\AgdaSpace{}%
\AgdaSymbol{:}\AgdaSpace{}%
\AgdaFunction{𝒰}\AgdaSymbol{)}\AgdaSpace{}%
\AgdaSymbol{→}\AgdaSpace{}%
\AgdaFunction{𝒰}\<%
\\
\>[0]\AgdaFunction{BAut}\AgdaSpace{}%
\AgdaBound{T}\AgdaSpace{}%
\AgdaSymbol{=}\AgdaSpace{}%
\AgdaRecord{Σ[}\AgdaSpace{}%
\AgdaBound{X}\AgdaSpace{}%
\AgdaRecord{∶}\AgdaSpace{}%
\AgdaFunction{𝒰}\AgdaSpace{}%
\AgdaRecord{]}\AgdaSpace{}%
\AgdaPostulate{∥}\AgdaSpace{}%
\AgdaBound{X}\AgdaSpace{}%
\AgdaFunction{≃}\AgdaSpace{}%
\AgdaBound{T}\AgdaSpace{}%
\AgdaPostulate{∥}\<%
\\
\\
\>[0]\AgdaFunction{ElB}\AgdaSpace{}%
\AgdaSymbol{:}\AgdaSpace{}%
\AgdaSymbol{\{}\AgdaBound{T}\AgdaSpace{}%
\AgdaSymbol{:}\AgdaSpace{}%
\AgdaFunction{𝒰}\AgdaSymbol{\}}\AgdaSpace{}%
\AgdaSymbol{→}\AgdaSpace{}%
\AgdaFunction{BAut}\AgdaSpace{}%
\AgdaBound{T}\AgdaSpace{}%
\AgdaSymbol{→}\AgdaSpace{}%
\AgdaFunction{𝒰}\<%
\\
\>[0]\AgdaFunction{ElB}\AgdaSpace{}%
\AgdaSymbol{=}\AgdaSpace{}%
\AgdaField{pr₁}\<%
\\
\\
\>[0]\AgdaFunction{transport{-}equiv{-}ElB}%
\>[2097I]\AgdaSymbol{:}\AgdaSpace{}%
\AgdaSymbol{\{}\AgdaBound{T}\AgdaSpace{}%
\AgdaSymbol{:}\AgdaSpace{}%
\AgdaFunction{𝒰}\AgdaSymbol{\}}\AgdaSpace{}%
\AgdaSymbol{\{}\AgdaBound{v}\AgdaSpace{}%
\AgdaBound{w}\AgdaSpace{}%
\AgdaSymbol{:}\AgdaSpace{}%
\AgdaFunction{BAut}\AgdaSpace{}%
\AgdaBound{T}\AgdaSymbol{\}}\AgdaSpace{}%
\AgdaSymbol{(}\AgdaBound{p}\AgdaSpace{}%
\AgdaSymbol{:}\AgdaSpace{}%
\AgdaBound{v}\AgdaSpace{}%
\AgdaDatatype{==}\AgdaSpace{}%
\AgdaBound{w}\AgdaSymbol{)}\<%
\\
\>[0][@{}l@{\AgdaIndent{0}}]\<[2097I]%
\>[20]\AgdaSymbol{→}\AgdaSpace{}%
\AgdaField{pr₁}\AgdaSpace{}%
\AgdaSymbol{(}\AgdaFunction{transport{-}equiv}\AgdaSpace{}%
\AgdaFunction{ElB}\AgdaSpace{}%
\AgdaBound{p}\AgdaSymbol{)}\AgdaSpace{}%
\AgdaDatatype{==}\AgdaSpace{}%
\AgdaFunction{transport}\AgdaSpace{}%
\AgdaFunction{id}\AgdaSpace{}%
\AgdaSymbol{(}\AgdaFunction{dpair={-}e}\AgdaSpace{}%
\AgdaBound{p}\AgdaSymbol{)}\<%
\\
\>[0]\AgdaFunction{transport{-}equiv{-}ElB}\AgdaSpace{}%
\AgdaSymbol{(}\AgdaInductiveConstructor{refl}\AgdaSpace{}%
\AgdaBound{v}\AgdaSymbol{)}\AgdaSpace{}%
\AgdaSymbol{=}\AgdaSpace{}%
\AgdaInductiveConstructor{refl}\AgdaSpace{}%
\AgdaFunction{id}\<%
\\
\\
\>[0]\AgdaFunction{is{-}univ{-}fib{-}ElB}\AgdaSpace{}%
\AgdaSymbol{:}\AgdaSpace{}%
\AgdaSymbol{\{}\AgdaBound{T}\AgdaSpace{}%
\AgdaSymbol{:}\AgdaSpace{}%
\AgdaFunction{𝒰}\AgdaSymbol{\}}\AgdaSpace{}%
\AgdaSymbol{→}\AgdaSpace{}%
\AgdaFunction{is{-}univ{-}fib}\AgdaSpace{}%
\AgdaFunction{ElB}\<%
\\
\>[0]\AgdaFunction{is{-}univ{-}fib{-}ElB}\AgdaSpace{}%
\AgdaSymbol{(}\AgdaBound{T}\AgdaSpace{}%
\AgdaInductiveConstructor{,}\AgdaSpace{}%
\AgdaBound{q}\AgdaSymbol{)}\AgdaSpace{}%
\AgdaSymbol{(}\AgdaBound{T'}\AgdaSpace{}%
\AgdaInductiveConstructor{,}\AgdaSpace{}%
\AgdaBound{q'}\AgdaSymbol{)}\AgdaSpace{}%
\AgdaSymbol{=}\AgdaSpace{}%
\AgdaFunction{qinv{-}is{-}hae}\AgdaSpace{}%
\AgdaSymbol{(}\AgdaFunction{g}\AgdaSpace{}%
\AgdaInductiveConstructor{,}\AgdaSpace{}%
\AgdaFunction{η}\AgdaSpace{}%
\AgdaInductiveConstructor{,}\AgdaSpace{}%
\AgdaFunction{ε}\AgdaSymbol{)}\<%
\\
\>[0][@{}l@{\AgdaIndent{0}}]%
\>[2]\AgdaKeyword{where}%
\>[9]\AgdaFunction{g}\AgdaSpace{}%
\AgdaSymbol{:}\AgdaSpace{}%
\AgdaBound{T}\AgdaSpace{}%
\AgdaFunction{≃}\AgdaSpace{}%
\AgdaBound{T'}\AgdaSpace{}%
\AgdaSymbol{→}\AgdaSpace{}%
\AgdaBound{T}\AgdaSpace{}%
\AgdaInductiveConstructor{,}\AgdaSpace{}%
\AgdaBound{q}\AgdaSpace{}%
\AgdaDatatype{==}\AgdaSpace{}%
\AgdaBound{T'}\AgdaSpace{}%
\AgdaInductiveConstructor{,}\AgdaSpace{}%
\AgdaBound{q'}\<%
\\
\>[2][@{}l@{\AgdaIndent{0}}]%
\>[9]\AgdaFunction{g}\AgdaSpace{}%
\AgdaBound{eqv}\AgdaSpace{}%
\AgdaSymbol{=}\AgdaSpace{}%
\AgdaFunction{dpair=}\AgdaSpace{}%
\AgdaSymbol{(}\AgdaFunction{ua}\AgdaSpace{}%
\AgdaBound{eqv}\AgdaSpace{}%
\AgdaInductiveConstructor{,}\AgdaSpace{}%
\AgdaPostulate{ident}\AgdaSymbol{)}\<%
\\
\\
\>[2][@{}l@{\AgdaIndent{0}}]%
\>[9]\AgdaFunction{η}\AgdaSpace{}%
\AgdaSymbol{:}\AgdaSpace{}%
\AgdaFunction{g}\AgdaSpace{}%
\AgdaFunction{∘}\AgdaSpace{}%
\AgdaFunction{transport{-}equiv}\AgdaSpace{}%
\AgdaFunction{ElB}\AgdaSpace{}%
\AgdaFunction{∼}\AgdaSpace{}%
\AgdaFunction{id}\<%
\\
\>[2][@{}l@{\AgdaIndent{0}}]%
\>[9]\AgdaFunction{η}\AgdaSpace{}%
\AgdaSymbol{(}\AgdaInductiveConstructor{refl}\AgdaSpace{}%
\AgdaSymbol{.\_}\AgdaSymbol{)}\AgdaSpace{}%
\AgdaSymbol{=}\AgdaSpace{}%
\AgdaFunction{ap}%
\>[2175I]\AgdaFunction{dpair=}%
\>[35]\AgdaSymbol{(}\AgdaFunction{dpair=}\AgdaSpace{}%
\AgdaSymbol{(}\AgdaSpace{}%
\AgdaFunction{ua{-}η}\AgdaSpace{}%
\AgdaSymbol{(}\AgdaInductiveConstructor{refl}\AgdaSpace{}%
\AgdaSymbol{\_)}\<%
\\
\>[2175I][@{}l@{\AgdaIndent{0}}]%
\>[35]\AgdaInductiveConstructor{,}\AgdaSpace{}%
\AgdaFunction{prop{-}is{-}set}\AgdaSpace{}%
\AgdaSymbol{(λ}\AgdaSpace{}%
\AgdaBound{\_}\AgdaSpace{}%
\AgdaBound{\_}\AgdaSpace{}%
\AgdaSymbol{→}\AgdaSpace{}%
\AgdaPostulate{ident}\AgdaSymbol{)}\AgdaSpace{}%
\AgdaSymbol{\_}\AgdaSpace{}%
\AgdaSymbol{\_}\AgdaSpace{}%
\AgdaSymbol{\_}\AgdaSpace{}%
\AgdaSymbol{\_))}\<%
\\
\\
\>[2][@{}l@{\AgdaIndent{0}}]%
\>[9]\AgdaFunction{ε}\AgdaSpace{}%
\AgdaSymbol{:}\AgdaSpace{}%
\AgdaFunction{transport{-}equiv}\AgdaSpace{}%
\AgdaFunction{ElB}\AgdaSpace{}%
\AgdaFunction{∘}\AgdaSpace{}%
\AgdaFunction{g}\AgdaSpace{}%
\AgdaFunction{∼}\AgdaSpace{}%
\AgdaFunction{id}\<%
\\
\>[2][@{}l@{\AgdaIndent{0}}]%
\>[9]\AgdaFunction{ε}\AgdaSpace{}%
\AgdaBound{eqv}\AgdaSpace{}%
\AgdaSymbol{=}%
\>[2199I]\AgdaFunction{eqv=}%
\>[23]\AgdaSymbol{(}\AgdaFunction{transport{-}equiv{-}ElB}\AgdaSpace{}%
\AgdaSymbol{(}\AgdaFunction{dpair=}\AgdaSpace{}%
\AgdaSymbol{(}\AgdaFunction{ua}\AgdaSpace{}%
\AgdaBound{eqv}\AgdaSpace{}%
\AgdaInductiveConstructor{,}\AgdaSpace{}%
\AgdaPostulate{ident}\AgdaSymbol{))}\<%
\\
\>[2199I][@{}l@{\AgdaIndent{0}}]%
\>[23]\AgdaFunction{◾}\AgdaSpace{}%
\AgdaFunction{ap}\AgdaSpace{}%
\AgdaSymbol{(}\AgdaFunction{transport}\AgdaSpace{}%
\AgdaFunction{id}\AgdaSymbol{)}\AgdaSpace{}%
\AgdaSymbol{(}\AgdaFunction{dpair={-}β}\AgdaSpace{}%
\AgdaSymbol{(}\AgdaFunction{ua}\AgdaSpace{}%
\AgdaBound{eqv}\AgdaSpace{}%
\AgdaInductiveConstructor{,}\AgdaSpace{}%
\AgdaPostulate{ident}\AgdaSymbol{))}\<%
\\
\>[2199I][@{}l@{\AgdaIndent{0}}]%
\>[23]\AgdaFunction{◾}\AgdaSpace{}%
\AgdaFunction{ua{-}β₁}\AgdaSpace{}%
\AgdaBound{eqv}\AgdaSpace{}%
\AgdaSymbol{)}\<%
\end{code}

This establishes that {\small\AgdaFunction{El𝟚}} is a univalent
fibration, giving us a characterization of paths in
{\small\AgdaFunction{U[𝟚]}} in terms of equivalences on booleans which
we exploit next.

\subsection{The Base Space {\normalfont\AgdaFunction{U[𝟚]}}}

The points in the base space {\small\AgdaFunction{U[𝟚]}} are all of
the form
{\small\AgdaSymbol{(}\AgdaBound{X}~\AgdaSymbol{,}~\AgdaInductiveConstructor{∣}\AgdaBound{p}\AgdaInductiveConstructor{∣}\AgdaSymbol{)}}
where {\small\AgdaBound{p}} is of type
{\small\AgdaBound{X}~\AgdaDatatype{==}~\AgdaDatatype{𝟚}}. We evidently
have a canonical point {\small\AgdaFunction{𝟚₀}}:

\begin{code}%
\>[0]\AgdaFunction{𝟚₀}%
\>[4]\AgdaSymbol{:}\AgdaSpace{}%
\AgdaFunction{U[𝟚]}\<%
\\
\>[0]\AgdaFunction{𝟚₀}%
\>[4]\AgdaSymbol{=}\AgdaSpace{}%
\AgdaSymbol{(}\AgdaDatatype{𝟚}\AgdaSpace{}%
\AgdaInductiveConstructor{,}\AgdaSpace{}%
\AgdaPostulate{∣}\AgdaSpace{}%
\AgdaInductiveConstructor{refl}\AgdaSpace{}%
\AgdaDatatype{𝟚}\AgdaSpace{}%
\AgdaPostulate{∣}\AgdaSymbol{)}\<%
\end{code}

which directly corresponds to the boolean type in $\PiTwo$. We remind
the reader that, by construction, {\small\AgdaFunction{U[𝟚]}} is
path-connected.  What remains is to characterize the 1-paths, 2-paths,
and possibly higher paths in {\small\AgdaFunction{U[𝟚]}} and to relate
them to the 1-combinators, 2-combinators, etc. in $\PiTwo$.

To conveniently refer to the paths in {\small\AgdaFunction{U[𝟚]}}, we
define the loop space on a (pointed) type, and show that the loop
space on {\small\AgdaFunction{BAut}~\AgdaDatatype{𝟚}} is equivalent to
{\small\AgdaDatatype{𝟚}~\AgdaFunction{≃}~\AgdaDatatype{𝟚}}:

\begin{code}%
\>[0]\AgdaFunction{Ω}\AgdaSpace{}%
\AgdaSymbol{:}\AgdaSpace{}%
\AgdaRecord{Σ[}\AgdaSpace{}%
\AgdaBound{T}\AgdaSpace{}%
\AgdaRecord{∶}\AgdaSpace{}%
\AgdaFunction{𝒰}\AgdaSpace{}%
\AgdaRecord{]}\AgdaSpace{}%
\AgdaBound{T}\AgdaSpace{}%
\AgdaSymbol{→}\AgdaSpace{}%
\AgdaFunction{𝒰}\<%
\\
\>[0]\AgdaFunction{Ω}\AgdaSpace{}%
\AgdaSymbol{(}\AgdaBound{T}\AgdaSpace{}%
\AgdaInductiveConstructor{,}\AgdaSpace{}%
\AgdaBound{t₀}\AgdaSymbol{)}\AgdaSpace{}%
\AgdaSymbol{=}\AgdaSpace{}%
\AgdaBound{t₀}\AgdaSpace{}%
\AgdaDatatype{==}\AgdaSpace{}%
\AgdaBound{t₀}\<%
\\
\\
\>[0]\AgdaFunction{Aut}\AgdaSpace{}%
\AgdaSymbol{:}\AgdaSpace{}%
\AgdaSymbol{(}\AgdaBound{T}\AgdaSpace{}%
\AgdaSymbol{:}\AgdaSpace{}%
\AgdaFunction{𝒰}\AgdaSymbol{)}\AgdaSpace{}%
\AgdaSymbol{→}\AgdaSpace{}%
\AgdaFunction{𝒰}\<%
\\
\>[0]\AgdaFunction{Aut}\AgdaSpace{}%
\AgdaBound{T}\AgdaSpace{}%
\AgdaSymbol{=}\AgdaSpace{}%
\AgdaBound{T}\AgdaSpace{}%
\AgdaFunction{≃}\AgdaSpace{}%
\AgdaBound{T}\<%
\\
\\
\>[0]\AgdaFunction{b₀}\AgdaSpace{}%
\AgdaSymbol{:}\AgdaSpace{}%
\AgdaSymbol{\{}\AgdaBound{T}\AgdaSpace{}%
\AgdaSymbol{:}\AgdaSpace{}%
\AgdaFunction{𝒰}\AgdaSymbol{\}}\AgdaSpace{}%
\AgdaSymbol{→}\AgdaSpace{}%
\AgdaFunction{BAut}\AgdaSpace{}%
\AgdaBound{T}\<%
\\
\>[0]\AgdaFunction{b₀}\AgdaSpace{}%
\AgdaSymbol{\{}\AgdaBound{T}\AgdaSymbol{\}}\AgdaSpace{}%
\AgdaSymbol{=}\AgdaSpace{}%
\AgdaBound{T}\AgdaSpace{}%
\AgdaInductiveConstructor{,}\AgdaSpace{}%
\AgdaPostulate{∣}\AgdaSpace{}%
\AgdaFunction{ide}\AgdaSpace{}%
\AgdaBound{T}\AgdaSpace{}%
\AgdaPostulate{∣}\<%
\\
\\
\>[0]\AgdaFunction{ΩBAut≃Aut[\_]}\AgdaSpace{}%
\AgdaSymbol{:}\AgdaSpace{}%
\AgdaSymbol{(}\AgdaBound{T}\AgdaSpace{}%
\AgdaSymbol{:}\AgdaSpace{}%
\AgdaFunction{𝒰}\AgdaSymbol{)}\AgdaSpace{}%
\AgdaSymbol{→}\AgdaSpace{}%
\AgdaFunction{Ω}\AgdaSpace{}%
\AgdaSymbol{(}\AgdaFunction{BAut}\AgdaSpace{}%
\AgdaBound{T}\AgdaSpace{}%
\AgdaInductiveConstructor{,}\AgdaSpace{}%
\AgdaFunction{b₀}\AgdaSymbol{)}\AgdaSpace{}%
\AgdaFunction{≃}\AgdaSpace{}%
\AgdaFunction{Aut}\AgdaSpace{}%
\AgdaBound{T}\<%
\\
\>[0]\AgdaFunction{ΩBAut≃Aut[}\AgdaSpace{}%
\AgdaBound{T}\AgdaSpace{}%
\AgdaFunction{]}\AgdaSpace{}%
\AgdaSymbol{=}\AgdaSpace{}%
\AgdaFunction{transport{-}equiv}\AgdaSpace{}%
\AgdaFunction{ElB}\AgdaSpace{}%
\AgdaInductiveConstructor{,}\AgdaSpace{}%
\AgdaFunction{is{-}univ{-}fib{-}ElB}\AgdaSpace{}%
\AgdaFunction{b₀}\AgdaSpace{}%
\AgdaFunction{b₀}\<%
\end{code}

The above results states that, in general, the loop space of the
classifying space of a type {\small\AgdaBound{T}} is equivalent to the
type of automorphisms of {\small\AgdaBound{T}}.  In particular, it
follows that
{\small\AgdaFunction{Ω}~\AgdaSymbol{(}\AgdaFunction{BAut}~\AgdaDatatype{𝟚}
  \AgdaInductiveConstructor{,} \AgdaFunction{𝟚₀}\AgdaSymbol{)}
  \AgdaFunction{≃} \AgdaFunction{Aut} \AgdaDatatype{𝟚}} which reduces
the problem of characterizing paths on {\small\AgdaFunction{U[𝟚]}} to
the much simpler problem of characterizing automorphisms on the type
of booleans.  We now turn our attention to solving that problem.





\AgdaHide{
\begin{code}%
\>[0]\AgdaFunction{BAut≃Ũ[\_]}\AgdaSpace{}%
\AgdaSymbol{:}\AgdaSpace{}%
\AgdaSymbol{(}\AgdaBound{T}\AgdaSpace{}%
\AgdaSymbol{:}\AgdaSpace{}%
\AgdaFunction{𝒰}\AgdaSymbol{)}\AgdaSpace{}%
\AgdaSymbol{→}\AgdaSpace{}%
\AgdaFunction{BAut}\AgdaSpace{}%
\AgdaBound{T}\AgdaSpace{}%
\AgdaFunction{≃}\AgdaSpace{}%
\AgdaField{pr₁}\AgdaSpace{}%
\AgdaFunction{Ũ[}\AgdaSpace{}%
\AgdaBound{T}\AgdaSpace{}%
\AgdaFunction{]}\<%
\\
\>[0]\AgdaFunction{BAut≃Ũ[}\AgdaSpace{}%
\AgdaBound{T}\AgdaSpace{}%
\AgdaFunction{]}\AgdaSpace{}%
\AgdaSymbol{=}\AgdaSpace{}%
\AgdaSymbol{\{!!\}}\<%
\end{code}
}


\subsection{Automorphisms on {\normalfont\AgdaDatatype{𝟚}}}

The type {\small\AgdaFunction{𝟚}} has two point constructors, and no
path constructors, which means it has no non-trivial paths on its
points except {\small\AgdaFunction{refl}}. In fact, we can prove in
intensional type theory using large elimination, that the two
constructors are disjoint. This is reflected in the absurd pattern
when using dependent pattern matching in Agda. More generally,
{\small\AgdaFunction{𝟚 ≃ 𝟙 ⊎ 𝟙}} and the disjoint union of two sets is
a set:

\begin{code}%
\>[0]\AgdaFunction{0₂≠1₂}\AgdaSpace{}%
\AgdaSymbol{:}\AgdaSpace{}%
\AgdaInductiveConstructor{0₂}\AgdaSpace{}%
\AgdaDatatype{==}\AgdaSpace{}%
\AgdaInductiveConstructor{1₂}\AgdaSpace{}%
\AgdaSymbol{→}\AgdaSpace{}%
\AgdaDatatype{⊥}\<%
\\
\>[0]\AgdaFunction{0₂≠1₂}\AgdaSpace{}%
\AgdaBound{p}\AgdaSpace{}%
\AgdaSymbol{=}\AgdaSpace{}%
\AgdaFunction{transport}\AgdaSpace{}%
\AgdaFunction{code}\AgdaSpace{}%
\AgdaBound{p}\AgdaSpace{}%
\AgdaInductiveConstructor{tt}\<%
\\
\>[0][@{}l@{\AgdaIndent{0}}]%
\>[2]\AgdaKeyword{where}%
\>[9]\AgdaFunction{code}\AgdaSpace{}%
\AgdaSymbol{:}\AgdaSpace{}%
\AgdaDatatype{𝟚}\AgdaSpace{}%
\AgdaSymbol{→}\AgdaSpace{}%
\AgdaFunction{𝒰}\<%
\\
\>[2][@{}l@{\AgdaIndent{0}}]%
\>[9]\AgdaFunction{code}\AgdaSpace{}%
\AgdaInductiveConstructor{0₂}\AgdaSpace{}%
\AgdaSymbol{=}\AgdaSpace{}%
\AgdaRecord{⊤}\<%
\\
\>[2][@{}l@{\AgdaIndent{0}}]%
\>[9]\AgdaFunction{code}\AgdaSpace{}%
\AgdaInductiveConstructor{1₂}\AgdaSpace{}%
\AgdaSymbol{=}\AgdaSpace{}%
\AgdaDatatype{⊥}\<%
\end{code}

Using {\small\AgdaFunction{0₂≠1₂}} and function extensionality
(derivable from univalence) we can prove that there are exactly two
different equivalences between {\small\AgdaFunction{𝟚}} and
{\small\AgdaFunction{𝟚}}.  Furthermore, for any equivalence
{\small\AgdaFunction{f}}, using the fact that
{\small\AgdaFunction{is-equiv f}} is a proposition, we can show that
there are exactly two inhabitants of {\small\AgdaFunction{𝟚 ≃ 𝟚}}:

\begin{code}%
\>[0]\AgdaFunction{id≃}\AgdaSpace{}%
\AgdaFunction{not≃}\AgdaSpace{}%
\AgdaSymbol{:}\AgdaSpace{}%
\AgdaDatatype{𝟚}\AgdaSpace{}%
\AgdaFunction{≃}\AgdaSpace{}%
\AgdaDatatype{𝟚}\<%
\\
\>[0]\AgdaFunction{id≃}%
\>[6]\AgdaSymbol{=}\AgdaSpace{}%
\AgdaFunction{id}%
\>[12]\AgdaInductiveConstructor{,}\AgdaSpace{}%
\AgdaFunction{qinv{-}is{-}hae}\AgdaSpace{}%
\AgdaSymbol{(}\AgdaFunction{id}\AgdaSpace{}%
\AgdaInductiveConstructor{,}\AgdaSpace{}%
\AgdaInductiveConstructor{refl}\AgdaSpace{}%
\AgdaInductiveConstructor{,}\AgdaSpace{}%
\AgdaInductiveConstructor{refl}\AgdaSymbol{)}\<%
\\
\>[0]\AgdaFunction{not≃}%
\>[6]\AgdaSymbol{=}\AgdaSpace{}%
\AgdaFunction{not}\AgdaSpace{}%
\AgdaInductiveConstructor{,}\AgdaSpace{}%
\AgdaFunction{qinv{-}is{-}hae}%
\>[2340I]\AgdaSymbol{(}\AgdaFunction{not}%
\>[2341I]\AgdaInductiveConstructor{,}\AgdaSpace{}%
\AgdaSymbol{(λ}\AgdaSpace{}%
\AgdaSymbol{\{}\AgdaInductiveConstructor{0₂}\AgdaSpace{}%
\AgdaSymbol{→}\AgdaSpace{}%
\AgdaInductiveConstructor{refl}\AgdaSpace{}%
\AgdaInductiveConstructor{0₂}\AgdaSpace{}%
\AgdaSymbol{;}\AgdaSpace{}%
\AgdaInductiveConstructor{1₂}\AgdaSpace{}%
\AgdaSymbol{→}\AgdaSpace{}%
\AgdaInductiveConstructor{refl}\AgdaSpace{}%
\AgdaInductiveConstructor{1₂}\AgdaSymbol{\})}\<%
\\
\>[2340I][@{}l@{\AgdaIndent{0}}]\<[2341I]%
\>[31]\AgdaInductiveConstructor{,}\AgdaSpace{}%
\AgdaSymbol{(λ}\AgdaSpace{}%
\AgdaSymbol{\{}\AgdaInductiveConstructor{0₂}\AgdaSpace{}%
\AgdaSymbol{→}\AgdaSpace{}%
\AgdaInductiveConstructor{refl}\AgdaSpace{}%
\AgdaInductiveConstructor{0₂}\AgdaSpace{}%
\AgdaSymbol{;}\AgdaSpace{}%
\AgdaInductiveConstructor{1₂}\AgdaSpace{}%
\AgdaSymbol{→}\AgdaSpace{}%
\AgdaInductiveConstructor{refl}\AgdaSpace{}%
\AgdaInductiveConstructor{1₂}\AgdaSymbol{\}))}\<%
\\
\>[0][@{}l@{\AgdaIndent{0}}]%
\>[2]\AgdaKeyword{where}%
\>[9]\AgdaFunction{not}\AgdaSpace{}%
\AgdaSymbol{:}\AgdaSpace{}%
\AgdaDatatype{𝟚}\AgdaSpace{}%
\AgdaSymbol{→}\AgdaSpace{}%
\AgdaDatatype{𝟚}\<%
\\
\>[2][@{}l@{\AgdaIndent{0}}]%
\>[9]\AgdaFunction{not}\AgdaSpace{}%
\AgdaInductiveConstructor{0₂}\AgdaSpace{}%
\AgdaSymbol{=}\AgdaSpace{}%
\AgdaInductiveConstructor{1₂}\<%
\\
\>[2][@{}l@{\AgdaIndent{0}}]%
\>[9]\AgdaFunction{not}\AgdaSpace{}%
\AgdaInductiveConstructor{1₂}\AgdaSpace{}%
\AgdaSymbol{=}\AgdaSpace{}%
\AgdaInductiveConstructor{0₂}\<%
\end{code}

Here something very special happens: although in general the type
formed by taking~$n$ disjoint unions of {\small\AgdaFunction{𝟙}} has a
space of automorphisms of size $n!$, in our case we have that
{\small\AgdaFunction{𝟚}} and {\small\AgdaFunction{𝟚 ≃ 𝟚}} are of the
same size. This combinatorial accident can actually be lifted to show
that there is an equivalence between {\small\AgdaFunction{𝟚 ≃ 𝟚}} and
{\small\AgdaFunction{𝟚}}.  By composing the chain of equivalences
{\small\AgdaFunction{Ω (Ũ , 𝟚₀) ≃ Ω (BAut(𝟚) , b₀) ≃ (𝟚 ≃ 𝟚) ≃ 𝟚}} we
obtain:

\AgdaHide{\begin{code}%
\>[0]\AgdaKeyword{postulate}\<%
\end{code}}

\begin{code}%
\>[0][@{}l@{\AgdaIndent{1}}]%
\>[2]\AgdaPostulate{𝟚≃Ω𝟚₀}\AgdaSpace{}%
\AgdaSymbol{:}\AgdaSpace{}%
\AgdaDatatype{𝟚}\AgdaSpace{}%
\AgdaFunction{≃}\AgdaSpace{}%
\AgdaSymbol{(}\AgdaFunction{𝟚₀}\AgdaSpace{}%
\AgdaDatatype{==}\AgdaSpace{}%
\AgdaFunction{𝟚₀}\AgdaSymbol{)}\<%
\end{code}

Thus there are only two distinct 1-loops in
{\small\AgdaFunction{U[𝟚]}}. Calling them {\small\AgdaFunction{id𝟚}}
and {\small\AgdaFunction{not𝟚}}, we obtain a decomposition:

\AgdaHide{\begin{code}%
\>[0]\AgdaFunction{id𝟚}\AgdaSpace{}%
\AgdaSymbol{:}\AgdaSpace{}%
\AgdaSymbol{\{}\AgdaBound{A}\AgdaSpace{}%
\AgdaSymbol{:}\AgdaSpace{}%
\AgdaFunction{U[𝟚]}\AgdaSymbol{\}}\AgdaSpace{}%
\AgdaSymbol{→}\AgdaSpace{}%
\AgdaBound{A}\AgdaSpace{}%
\AgdaDatatype{==}\AgdaSpace{}%
\AgdaBound{A}\<%
\\
\>[0]\AgdaFunction{id𝟚}\AgdaSpace{}%
\AgdaSymbol{\{}\AgdaBound{A}\AgdaSymbol{\}}\AgdaSpace{}%
\AgdaSymbol{=}\AgdaSpace{}%
\AgdaInductiveConstructor{refl}\AgdaSpace{}%
\AgdaBound{A}\<%
\\
\\
\>[0]\AgdaFunction{not𝟚}\AgdaSpace{}%
\AgdaSymbol{:}\AgdaSpace{}%
\AgdaFunction{𝟚₀}\AgdaSpace{}%
\AgdaDatatype{==}\AgdaSpace{}%
\AgdaFunction{𝟚₀}\<%
\\
\>[0]\AgdaFunction{not𝟚}\AgdaSpace{}%
\AgdaSymbol{=}\AgdaSpace{}%
\AgdaFunction{dpair=}\AgdaSpace{}%
\AgdaSymbol{(}\AgdaFunction{ua}\AgdaSpace{}%
\AgdaFunction{not≃}\AgdaSpace{}%
\AgdaInductiveConstructor{,}\AgdaSpace{}%
\AgdaPostulate{ident}\AgdaSymbol{)}\<%
\\
\\
\>[0]\AgdaKeyword{postulate}\<%
\end{code}}

\begin{code}%
\>[0][@{}l@{\AgdaIndent{1}}]%
\>[2]\AgdaPostulate{all{-}1{-}loops}\AgdaSpace{}%
\AgdaSymbol{:}\AgdaSpace{}%
\AgdaSymbol{(}\AgdaBound{p}\AgdaSpace{}%
\AgdaSymbol{:}\AgdaSpace{}%
\AgdaFunction{𝟚₀}\AgdaSpace{}%
\AgdaDatatype{==}\AgdaSpace{}%
\AgdaFunction{𝟚₀}\AgdaSymbol{)}\AgdaSpace{}%
\AgdaSymbol{→}\AgdaSpace{}%
\AgdaSymbol{(}\AgdaBound{p}\AgdaSpace{}%
\AgdaDatatype{==}\AgdaSpace{}%
\AgdaFunction{id𝟚}\AgdaSymbol{)}\AgdaSpace{}%
\AgdaDatatype{+}\AgdaSpace{}%
\AgdaSymbol{(}\AgdaBound{p}\AgdaSpace{}%
\AgdaDatatype{==}\AgdaSpace{}%
\AgdaFunction{not𝟚}\AgdaSymbol{)}\<%
\end{code}

that every loop in {\small\AgdaFunction{U[𝟚]}} is identifiable with
either the identity or boolean negation.

For 2-loops in {\small\AgdaFunction{U[𝟚]}}, the following analysis shows
that they are identifiable with the trivial path. First, by applying the
induction principle for disjoint unions, and path induction, we can
prove {\small\AgdaFunction{𝟚}} is a set:

\begin{code}%
\>[0]\AgdaFunction{𝟚{-}is{-}set}\AgdaSpace{}%
\AgdaSymbol{:}\AgdaSpace{}%
\AgdaFunction{is{-}set}\AgdaSpace{}%
\AgdaDatatype{𝟚}\<%
\\
\>[0]\AgdaFunction{𝟚{-}is{-}set}\AgdaSpace{}%
\AgdaInductiveConstructor{0₂}\AgdaSpace{}%
\AgdaInductiveConstructor{0₂}\AgdaSpace{}%
\AgdaSymbol{(}\AgdaInductiveConstructor{refl}\AgdaSpace{}%
\AgdaSymbol{.}\AgdaInductiveConstructor{0₂}\AgdaSymbol{)}\AgdaSpace{}%
\AgdaSymbol{(}\AgdaInductiveConstructor{refl}\AgdaSpace{}%
\AgdaSymbol{.}\AgdaInductiveConstructor{0₂}\AgdaSymbol{)}\AgdaSpace{}%
\AgdaSymbol{=}\AgdaSpace{}%
\AgdaInductiveConstructor{refl}\AgdaSpace{}%
\AgdaSymbol{(}\AgdaInductiveConstructor{refl}\AgdaSpace{}%
\AgdaInductiveConstructor{0₂}\AgdaSymbol{)}\<%
\\
\>[0]\AgdaFunction{𝟚{-}is{-}set}\AgdaSpace{}%
\AgdaInductiveConstructor{0₂}\AgdaSpace{}%
\AgdaInductiveConstructor{1₂}\AgdaSpace{}%
\AgdaSymbol{()}\<%
\\
\>[0]\AgdaFunction{𝟚{-}is{-}set}\AgdaSpace{}%
\AgdaInductiveConstructor{1₂}\AgdaSpace{}%
\AgdaInductiveConstructor{0₂}\AgdaSpace{}%
\AgdaSymbol{()}\<%
\\
\>[0]\AgdaFunction{𝟚{-}is{-}set}\AgdaSpace{}%
\AgdaInductiveConstructor{1₂}\AgdaSpace{}%
\AgdaInductiveConstructor{1₂}\AgdaSpace{}%
\AgdaSymbol{(}\AgdaInductiveConstructor{refl}\AgdaSpace{}%
\AgdaSymbol{.}\AgdaInductiveConstructor{1₂}\AgdaSymbol{)}\AgdaSpace{}%
\AgdaSymbol{(}\AgdaInductiveConstructor{refl}\AgdaSpace{}%
\AgdaSymbol{.}\AgdaInductiveConstructor{1₂}\AgdaSymbol{)}\AgdaSpace{}%
\AgdaSymbol{=}\AgdaSpace{}%
\AgdaInductiveConstructor{refl}\AgdaSpace{}%
\AgdaSymbol{(}\AgdaInductiveConstructor{refl}\AgdaSpace{}%
\AgdaInductiveConstructor{1₂}\AgdaSymbol{)}\<%
\end{code}

From this, we obtain that {\small\AgdaFunction{𝟚₀ == 𝟚₀}} is also a
set by using {\small\AgdaFunction{ua}} and
{\small\AgdaFunction{transport}}. This in turns shows the
contractibility of 2-loops:

\begin{code}%
\>[0]\AgdaFunction{Ω𝟚₀{-}is{-}set}\AgdaSpace{}%
\AgdaSymbol{:}\AgdaSpace{}%
\AgdaFunction{is{-}set}\AgdaSpace{}%
\AgdaSymbol{(}\AgdaFunction{𝟚₀}\AgdaSpace{}%
\AgdaDatatype{==}\AgdaSpace{}%
\AgdaFunction{𝟚₀}\AgdaSymbol{)}\<%
\\
\>[0]\AgdaFunction{Ω𝟚₀{-}is{-}set}\AgdaSpace{}%
\AgdaSymbol{=}\AgdaSpace{}%
\AgdaFunction{transport}\AgdaSpace{}%
\AgdaFunction{is{-}set}\AgdaSpace{}%
\AgdaSymbol{(}\AgdaFunction{ua}\AgdaSpace{}%
\AgdaPostulate{𝟚≃Ω𝟚₀}\AgdaSymbol{)}\AgdaSpace{}%
\AgdaFunction{𝟚{-}is{-}set}\<%
\\
\\
\>[0]\AgdaFunction{all{-}2{-}loops}\AgdaSpace{}%
\AgdaSymbol{:}\AgdaSpace{}%
\AgdaSymbol{\{}\AgdaBound{p}\AgdaSpace{}%
\AgdaSymbol{:}\AgdaSpace{}%
\AgdaFunction{𝟚₀}\AgdaSpace{}%
\AgdaDatatype{==}\AgdaSpace{}%
\AgdaFunction{𝟚₀}\AgdaSymbol{\}}\AgdaSpace{}%
\AgdaSymbol{→}\AgdaSpace{}%
\AgdaSymbol{(}\AgdaBound{γ}\AgdaSpace{}%
\AgdaSymbol{:}\AgdaSpace{}%
\AgdaBound{p}\AgdaSpace{}%
\AgdaDatatype{==}\AgdaSpace{}%
\AgdaBound{p}\AgdaSymbol{)}\AgdaSpace{}%
\AgdaSymbol{→}\AgdaSpace{}%
\AgdaBound{γ}\AgdaSpace{}%
\AgdaDatatype{==}\AgdaSpace{}%
\AgdaInductiveConstructor{refl}\AgdaSpace{}%
\AgdaBound{p}\<%
\\
\>[0]\AgdaFunction{all{-}2{-}loops}\AgdaSpace{}%
\AgdaSymbol{\{}\AgdaBound{p}\AgdaSymbol{\}}\AgdaSpace{}%
\AgdaBound{γ}\AgdaSpace{}%
\AgdaSymbol{=}\AgdaSpace{}%
\AgdaFunction{Ω𝟚₀{-}is{-}set}\AgdaSpace{}%
\AgdaBound{p}\AgdaSpace{}%
\AgdaBound{p}\AgdaSpace{}%
\AgdaBound{γ}\AgdaSpace{}%
\AgdaSymbol{(}\AgdaInductiveConstructor{refl}\AgdaSpace{}%
\AgdaBound{p}\AgdaSymbol{)}\<%
\end{code}

In the next section, we will use {\small\AgdaFunction{all-1-loops}}
and {\small\AgdaFunction{all-2-loops}} as crucial ingredients for
showing the correspondence between {\small\AgdaFunction{U[𝟚]}} and
\PiTwo.

Note that most of the results in this section are generic.  However
when we move beyond {\small\AgdaFunction{𝟚}}, the combinatorial
explosion of the path space is such that explicit enumeration quickly
becomes impractical, and other techniques will become necessary.




\section{Correspondence between {\normalfont\AgdaFunction{U[𝟚]}} and \PiTwo}
\label{sec:correspondence}

Formalizing, in a precise sense, the connection between reversible
functions in a programming language and paths in a univalent universe,
as intuitive as it may seem, is rather subtle. Paths in HoTT come
equipped with principles like the ``contractibility of singletons'',
``transport'', and ``path induction'' and none of these principles
seem to have any direct counterpart in the world of reversible
programming. We will however demonstrate how the semantics of an
entire (but admittedly small) reversible programming language such as
$\PiTwo$ can be captured by a specification as compact as
{\small\AgdaRecord{Σ[} \AgdaBound{X} \AgdaRecord{∶} \AgdaFunction{𝒰}
  \AgdaRecord{]} \AgdaPostulate{∥} \AgdaBound{X} \AgdaDatatype{==}
  \AgdaDatatype{𝟚} \AgdaPostulate{∥}}. Our precise correspondence will
consist of building mappings between~\PiTwo{} and
{\small{\AgdaFunction{Ũ[𝟚]}}}, for points, 1-paths, 2-paths, and
3-paths, such that each map is invertible up to the appropriate notion
of equality. This gives a notion of soundness and completeness for
each level.


\subsection{Mappings}

The mappings for points (level-0) are straightforward, as both
{\small\AgdaDatatype{Π₂}} and {\small\AgdaFunction{U[𝟚]}} are
singletons:

\begin{code}%
\>[0]\AgdaFunction{⟦\_⟧₀}\AgdaSpace{}%
\AgdaSymbol{:}\AgdaSpace{}%
\AgdaDatatype{Π₂}\AgdaSpace{}%
\AgdaSymbol{→}\AgdaSpace{}%
\AgdaFunction{U[𝟚]}\<%
\\
\>[0]\AgdaFunction{⟦}\AgdaSpace{}%
\AgdaInductiveConstructor{`𝟚}\AgdaSpace{}%
\AgdaFunction{⟧₀}\AgdaSpace{}%
\AgdaSymbol{=}\AgdaSpace{}%
\AgdaFunction{𝟚₀}\<%
\\
\\
\>[0]\AgdaFunction{⌜\_⌝₀}\AgdaSpace{}%
\AgdaSymbol{:}\AgdaSpace{}%
\AgdaFunction{U[𝟚]}\AgdaSpace{}%
\AgdaSymbol{→}\AgdaSpace{}%
\AgdaDatatype{Π₂}\<%
\\
\>[0]\AgdaFunction{⌜}\AgdaSpace{}%
\AgdaSymbol{\_}\AgdaSpace{}%
\AgdaFunction{⌝₀}\AgdaSpace{}%
\AgdaSymbol{=}\AgdaSpace{}%
\AgdaInductiveConstructor{`𝟚}\<%
\end{code}

Level-1 is the first non-trivial level. To each syntactic combinator
{\small\AgdaFunction{c}~\AgdaSymbol{:}~\AgdaBound{A}~\AgdaDatatype{⟷₁}~\AgdaBound{B}},
we associate a path from
{\small{\AgdaFunction{⟦}~\AgdaBound{A}~\AgdaFunction{⟧₀}}} to
{\small{\AgdaFunction{⟦}~\AgdaBound{B}~\AgdaFunction{⟧₀}}} and
vice-versa. The mapping from the univalent universe back to the syntax
of the reversible language is only possible because we have a complete
characterization of the paths in the universe (captured in the
construction of {\small\AgdaFunction{all-1-loops}} in the previous
section):

\begin{code}%
\>[0]\AgdaFunction{⟦\_⟧₁}\AgdaSpace{}%
\AgdaSymbol{:}\AgdaSpace{}%
\AgdaSymbol{\{}\AgdaBound{A}\AgdaSpace{}%
\AgdaBound{B}\AgdaSpace{}%
\AgdaSymbol{:}\AgdaSpace{}%
\AgdaDatatype{Π₂}\AgdaSymbol{\}}\AgdaSpace{}%
\AgdaSymbol{→}\AgdaSpace{}%
\AgdaBound{A}\AgdaSpace{}%
\AgdaDatatype{⟷₁}\AgdaSpace{}%
\AgdaBound{B}\AgdaSpace{}%
\AgdaSymbol{→}\AgdaSpace{}%
\AgdaFunction{⟦}\AgdaSpace{}%
\AgdaBound{A}\AgdaSpace{}%
\AgdaFunction{⟧₀}\AgdaSpace{}%
\AgdaDatatype{==}\AgdaSpace{}%
\AgdaFunction{⟦}\AgdaSpace{}%
\AgdaBound{B}\AgdaSpace{}%
\AgdaFunction{⟧₀}\<%
\\
\>[0]\AgdaFunction{⟦}\AgdaSpace{}%
\AgdaInductiveConstructor{`id}\AgdaSpace{}%
\AgdaFunction{⟧₁}%
\>[15]\AgdaSymbol{=}\AgdaSpace{}%
\AgdaFunction{id𝟚}\<%
\\
\>[0]\AgdaFunction{⟦}\AgdaSpace{}%
\AgdaInductiveConstructor{`not}\AgdaSpace{}%
\AgdaFunction{⟧₁}%
\>[15]\AgdaSymbol{=}\AgdaSpace{}%
\AgdaFunction{not𝟚}\<%
\\
\>[0]\AgdaFunction{⟦}\AgdaSpace{}%
\AgdaInductiveConstructor{!₁}\AgdaSpace{}%
\AgdaBound{p}\AgdaSpace{}%
\AgdaFunction{⟧₁}%
\>[15]\AgdaSymbol{=}\AgdaSpace{}%
\AgdaFunction{!}\AgdaSpace{}%
\AgdaFunction{⟦}\AgdaSpace{}%
\AgdaBound{p}\AgdaSpace{}%
\AgdaFunction{⟧₁}\<%
\\
\>[0]\AgdaFunction{⟦}\AgdaSpace{}%
\AgdaBound{p}\AgdaSpace{}%
\AgdaInductiveConstructor{⊙₁}\AgdaSpace{}%
\AgdaBound{q}\AgdaSpace{}%
\AgdaFunction{⟧₁}%
\>[15]\AgdaSymbol{=}\AgdaSpace{}%
\AgdaFunction{⟦}\AgdaSpace{}%
\AgdaBound{p}\AgdaSpace{}%
\AgdaFunction{⟧₁}\AgdaSpace{}%
\AgdaFunction{◾}\AgdaSpace{}%
\AgdaFunction{⟦}\AgdaSpace{}%
\AgdaBound{q}\AgdaSpace{}%
\AgdaFunction{⟧₁}\<%
\\
\\
\>[0]\AgdaFunction{⌜\_⌝₁}\AgdaSpace{}%
\AgdaSymbol{:}\AgdaSpace{}%
\AgdaFunction{𝟚₀}\AgdaSpace{}%
\AgdaDatatype{==}\AgdaSpace{}%
\AgdaFunction{𝟚₀}\AgdaSpace{}%
\AgdaSymbol{→}\AgdaSpace{}%
\AgdaFunction{⌜}\AgdaSpace{}%
\AgdaFunction{𝟚₀}\AgdaSpace{}%
\AgdaFunction{⌝₀}\AgdaSpace{}%
\AgdaDatatype{⟷₁}\AgdaSpace{}%
\AgdaFunction{⌜}\AgdaSpace{}%
\AgdaFunction{𝟚₀}\AgdaSpace{}%
\AgdaFunction{⌝₀}\<%
\\
\>[0]\AgdaFunction{⌜}\AgdaSpace{}%
\AgdaBound{p}\AgdaSpace{}%
\AgdaFunction{⌝₁}\AgdaSpace{}%
\AgdaKeyword{with}\AgdaSpace{}%
\AgdaPostulate{all{-}1{-}loops}\AgdaSpace{}%
\AgdaBound{p}\<%
\\
\>[0]\AgdaSymbol{...}\AgdaSpace{}%
\AgdaSymbol{|}\AgdaSpace{}%
\AgdaInductiveConstructor{inl}\AgdaSpace{}%
\AgdaBound{pid}%
\>[16]\AgdaSymbol{=}\AgdaSpace{}%
\AgdaInductiveConstructor{`id}\<%
\\
\>[0]\AgdaSymbol{...}\AgdaSpace{}%
\AgdaSymbol{|}\AgdaSpace{}%
\AgdaInductiveConstructor{inr}\AgdaSpace{}%
\AgdaBound{pnot}%
\>[16]\AgdaSymbol{=}\AgdaSpace{}%
\AgdaInductiveConstructor{`not}\<%
\end{code}

At level-2, we know by the construction of
{\small\AgdaFunction{all-2-loops}} in the previous section that all
self-paths in the univalent universe are trivial. Nevertheless the
mappings back and forth require quite a bit of (tedious) work. We show
below a few cases of the mapping from 2-combinators to 2-paths and the
full definition of the reverse mapping. In the first direction, it is
a matter of using the necessary properties of paths in the univalent
universe (e.g, each path has an inverse). These properties are proved
by path induction. The reverse direction crucially relies again on the
characterization of 1-loops and the fact that the identity equivalence
and the equivalence that swaps the two booleans are distinct:

\AgdaHide{
\begin{code}%
\>[0]\AgdaKeyword{postulate}\<%
\\
\>[0][@{}l@{\AgdaIndent{0}}]%
\>[2]\AgdaPostulate{!not𝟚=not𝟚}\AgdaSpace{}%
\AgdaSymbol{:}\AgdaSpace{}%
\AgdaFunction{!}\AgdaSpace{}%
\AgdaFunction{not𝟚}\AgdaSpace{}%
\AgdaDatatype{==}\AgdaSpace{}%
\AgdaFunction{not𝟚}\<%
\\
\>[0][@{}l@{\AgdaIndent{0}}]%
\>[2]\AgdaPostulate{id𝟚≠not𝟚}\AgdaSpace{}%
\AgdaSymbol{:}\AgdaSpace{}%
\AgdaFunction{id𝟚}\AgdaSpace{}%
\AgdaDatatype{==}\AgdaSpace{}%
\AgdaFunction{not𝟚}\AgdaSpace{}%
\AgdaSymbol{→}\AgdaSpace{}%
\AgdaDatatype{⊥}\<%
\end{code}
}

\begin{AgdaMultiCode}{2}
\begin{code}%
\>[0]\AgdaFunction{⟦\_⟧₂}\AgdaSpace{}%
\AgdaSymbol{:}\AgdaSpace{}%
\AgdaSymbol{\{}\AgdaBound{A}\AgdaSpace{}%
\AgdaBound{B}\AgdaSpace{}%
\AgdaSymbol{:}\AgdaSpace{}%
\AgdaDatatype{Π₂}\AgdaSymbol{\}}\AgdaSpace{}%
\AgdaSymbol{\{}\AgdaBound{p}\AgdaSpace{}%
\AgdaBound{q}\AgdaSpace{}%
\AgdaSymbol{:}\AgdaSpace{}%
\AgdaBound{A}\AgdaSpace{}%
\AgdaDatatype{⟷₁}\AgdaSpace{}%
\AgdaBound{B}\AgdaSymbol{\}}\AgdaSpace{}%
\AgdaSymbol{→}\AgdaSpace{}%
\AgdaSymbol{(}\AgdaBound{u}\AgdaSpace{}%
\AgdaSymbol{:}\AgdaSpace{}%
\AgdaBound{p}\AgdaSpace{}%
\AgdaDatatype{⟷₂}\AgdaSpace{}%
\AgdaBound{q}\AgdaSymbol{)}\AgdaSpace{}%
\AgdaSymbol{→}\AgdaSpace{}%
\AgdaFunction{⟦}\AgdaSpace{}%
\AgdaBound{p}\AgdaSpace{}%
\AgdaFunction{⟧₁}\AgdaSpace{}%
\AgdaDatatype{==}\AgdaSpace{}%
\AgdaFunction{⟦}\AgdaSpace{}%
\AgdaBound{q}\AgdaSpace{}%
\AgdaFunction{⟧₁}\<%
\\
\>[0]\AgdaFunction{⟦}\AgdaSpace{}%
\AgdaInductiveConstructor{`id₂}\AgdaSpace{}%
\AgdaSymbol{\{}\AgdaArgument{p}\AgdaSpace{}%
\AgdaSymbol{=}\AgdaSpace{}%
\AgdaBound{p}\AgdaSymbol{\}}\AgdaSpace{}%
\AgdaFunction{⟧₂}%
\>[20]\AgdaSymbol{=}\AgdaSpace{}%
\AgdaInductiveConstructor{refl}\AgdaSpace{}%
\AgdaFunction{⟦}\AgdaSpace{}%
\AgdaBound{p}\AgdaSpace{}%
\AgdaFunction{⟧₁}\<%
\\
\>[0]\AgdaFunction{⟦}\AgdaSpace{}%
\AgdaInductiveConstructor{!₂}\AgdaSpace{}%
\AgdaBound{u}\AgdaSpace{}%
\AgdaFunction{⟧₂}%
\>[20]\AgdaSymbol{=}\AgdaSpace{}%
\AgdaFunction{!}\AgdaSpace{}%
\AgdaFunction{⟦}\AgdaSpace{}%
\AgdaBound{u}\AgdaSpace{}%
\AgdaFunction{⟧₂}\<%
\\
\>[0]\AgdaFunction{⟦}\AgdaSpace{}%
\AgdaBound{u₁}\AgdaSpace{}%
\AgdaInductiveConstructor{⊙₂}\AgdaSpace{}%
\AgdaBound{u₂}\AgdaSpace{}%
\AgdaFunction{⟧₂}%
\>[20]\AgdaSymbol{=}\AgdaSpace{}%
\AgdaFunction{⟦}\AgdaSpace{}%
\AgdaBound{u₁}\AgdaSpace{}%
\AgdaFunction{⟧₂}\AgdaSpace{}%
\AgdaFunction{◾}\AgdaSpace{}%
\AgdaFunction{⟦}\AgdaSpace{}%
\AgdaBound{u₂}\AgdaSpace{}%
\AgdaFunction{⟧₂}\<%
\\
\>[0]\AgdaFunction{⟦}\AgdaSpace{}%
\AgdaInductiveConstructor{`idl}\AgdaSpace{}%
\AgdaBound{p}\AgdaSpace{}%
\AgdaFunction{⟧₂}%
\>[20]\AgdaSymbol{=}\AgdaSpace{}%
\AgdaFunction{◾unitl}\AgdaSpace{}%
\AgdaFunction{⟦}\AgdaSpace{}%
\AgdaBound{p}\AgdaSpace{}%
\AgdaFunction{⟧₁}\<%
\\
\>[0]\AgdaFunction{⟦}\AgdaSpace{}%
\AgdaInductiveConstructor{`idr}\AgdaSpace{}%
\AgdaBound{p}\AgdaSpace{}%
\AgdaFunction{⟧₂}%
\>[20]\AgdaSymbol{=}\AgdaSpace{}%
\AgdaFunction{◾unitr}\AgdaSpace{}%
\AgdaFunction{⟦}\AgdaSpace{}%
\AgdaBound{p}\AgdaSpace{}%
\AgdaFunction{⟧₁}\<%
\\
\>[0]\AgdaFunction{⟦}\AgdaSpace{}%
\AgdaInductiveConstructor{`!}\AgdaSpace{}%
\AgdaBound{u}\AgdaSpace{}%
\AgdaFunction{⟧₂}%
\>[20]\AgdaSymbol{=}\AgdaSpace{}%
\AgdaFunction{ap}\AgdaSpace{}%
\AgdaFunction{!\_}\AgdaSpace{}%
\AgdaFunction{⟦}\AgdaSpace{}%
\AgdaBound{u}\AgdaSpace{}%
\AgdaFunction{⟧₂}\<%
\\
\>[0]\AgdaComment{{-}{-} remaining cases are omitted}\<%
\end{code}
\AgdaHide{
\begin{code}%
\>[0]\AgdaFunction{⟦}\AgdaSpace{}%
\AgdaSymbol{\_}\AgdaSpace{}%
\AgdaFunction{⟧₂}%
\>[20]\AgdaSymbol{=}\AgdaSpace{}%
\AgdaSymbol{?}\<%
\end{code}
}
\end{AgdaMultiCode}

\begin{code}%
\>[0]\AgdaFunction{⌜\_⌝₂}\AgdaSpace{}%
\AgdaSymbol{:}\AgdaSpace{}%
\AgdaSymbol{\{}\AgdaBound{p}\AgdaSpace{}%
\AgdaBound{q}\AgdaSpace{}%
\AgdaSymbol{:}\AgdaSpace{}%
\AgdaFunction{𝟚₀}\AgdaSpace{}%
\AgdaDatatype{==}\AgdaSpace{}%
\AgdaFunction{𝟚₀}\AgdaSymbol{\}}\AgdaSpace{}%
\AgdaSymbol{→}\AgdaSpace{}%
\AgdaBound{p}\AgdaSpace{}%
\AgdaDatatype{==}\AgdaSpace{}%
\AgdaBound{q}\AgdaSpace{}%
\AgdaSymbol{→}\AgdaSpace{}%
\AgdaFunction{⌜}\AgdaSpace{}%
\AgdaBound{p}\AgdaSpace{}%
\AgdaFunction{⌝₁}\AgdaSpace{}%
\AgdaDatatype{⟷₂}\AgdaSpace{}%
\AgdaFunction{⌜}\AgdaSpace{}%
\AgdaBound{q}\AgdaSpace{}%
\AgdaFunction{⌝₁}\<%
\\
\>[0]\AgdaFunction{⌜\_⌝₂}\AgdaSpace{}%
\AgdaSymbol{\{}\AgdaBound{p}\AgdaSymbol{\}}\AgdaSpace{}%
\AgdaSymbol{\{}\AgdaBound{q}\AgdaSymbol{\}}\AgdaSpace{}%
\AgdaBound{u}\AgdaSpace{}%
\AgdaKeyword{with}\AgdaSpace{}%
\AgdaPostulate{all{-}1{-}loops}\AgdaSpace{}%
\AgdaBound{p}\AgdaSpace{}%
\AgdaSymbol{|}\AgdaSpace{}%
\AgdaPostulate{all{-}1{-}loops}\AgdaSpace{}%
\AgdaBound{q}\<%
\\
\>[0]\AgdaSymbol{...}\AgdaSpace{}%
\AgdaSymbol{|}\AgdaSpace{}%
\AgdaInductiveConstructor{inl}\AgdaSpace{}%
\AgdaBound{p=id}%
\>[17]\AgdaSymbol{|}\AgdaSpace{}%
\AgdaInductiveConstructor{inl}\AgdaSpace{}%
\AgdaBound{q=id}%
\>[30]\AgdaSymbol{=}\AgdaSpace{}%
\AgdaInductiveConstructor{`id₂}\<%
\\
\>[0]\AgdaSymbol{...}\AgdaSpace{}%
\AgdaSymbol{|}\AgdaSpace{}%
\AgdaInductiveConstructor{inl}\AgdaSpace{}%
\AgdaBound{p=id}%
\>[17]\AgdaSymbol{|}\AgdaSpace{}%
\AgdaInductiveConstructor{inr}\AgdaSpace{}%
\AgdaBound{q=not}%
\>[30]\AgdaSymbol{=}\AgdaSpace{}%
\AgdaFunction{⊥{-}elim}\AgdaSpace{}%
\AgdaSymbol{(}\AgdaPostulate{id𝟚≠not𝟚}\AgdaSpace{}%
\AgdaSymbol{((}\AgdaFunction{!}\AgdaSpace{}%
\AgdaBound{p=id}\AgdaSymbol{)}\AgdaSpace{}%
\AgdaFunction{◾}\AgdaSpace{}%
\AgdaBound{u}\AgdaSpace{}%
\AgdaFunction{◾}\AgdaSpace{}%
\AgdaBound{q=not}\AgdaSymbol{))}\<%
\\
\>[0]\AgdaSymbol{...}\AgdaSpace{}%
\AgdaSymbol{|}\AgdaSpace{}%
\AgdaInductiveConstructor{inr}\AgdaSpace{}%
\AgdaBound{p=not}%
\>[17]\AgdaSymbol{|}\AgdaSpace{}%
\AgdaInductiveConstructor{inl}\AgdaSpace{}%
\AgdaBound{q=id}%
\>[30]\AgdaSymbol{=}\AgdaSpace{}%
\AgdaFunction{⊥{-}elim}\AgdaSpace{}%
\AgdaSymbol{(}\AgdaPostulate{id𝟚≠not𝟚}\AgdaSpace{}%
\AgdaSymbol{((}\AgdaFunction{!}\AgdaSpace{}%
\AgdaBound{q=id}\AgdaSymbol{)}\AgdaSpace{}%
\AgdaFunction{◾}\AgdaSpace{}%
\AgdaFunction{!}\AgdaSpace{}%
\AgdaBound{u}\AgdaSpace{}%
\AgdaFunction{◾}\AgdaSpace{}%
\AgdaBound{p=not}\AgdaSymbol{))}\<%
\\
\>[0]\AgdaSymbol{...}\AgdaSpace{}%
\AgdaSymbol{|}\AgdaSpace{}%
\AgdaInductiveConstructor{inr}\AgdaSpace{}%
\AgdaBound{p=not}%
\>[17]\AgdaSymbol{|}\AgdaSpace{}%
\AgdaInductiveConstructor{inr}\AgdaSpace{}%
\AgdaBound{q=not}%
\>[30]\AgdaSymbol{=}\AgdaSpace{}%
\AgdaInductiveConstructor{`id₂}\<%
\end{code}

For the final level-3, mapping from the univalent universe to $\PiTwo$
is trivial as the latter has only one constructor at level-3. The
other direction requires some involved reasoning in the univalent
universe to construct the required 3-path:

\begin{code}%
\>[0]\AgdaFunction{lemma}%
\>[2717I]\AgdaSymbol{:}\AgdaSpace{}%
\AgdaSymbol{\{}\AgdaBound{p}\AgdaSpace{}%
\AgdaBound{q}\AgdaSpace{}%
\AgdaBound{r}\AgdaSpace{}%
\AgdaSymbol{:}\AgdaSpace{}%
\AgdaFunction{𝟚₀}\AgdaSpace{}%
\AgdaDatatype{==}\AgdaSpace{}%
\AgdaFunction{𝟚₀}\AgdaSymbol{\}}\AgdaSpace{}%
\AgdaSymbol{(}\AgdaBound{p=r}\AgdaSpace{}%
\AgdaSymbol{:}\AgdaSpace{}%
\AgdaBound{p}\AgdaSpace{}%
\AgdaDatatype{==}\AgdaSpace{}%
\AgdaBound{r}\AgdaSymbol{)}\AgdaSpace{}%
\AgdaSymbol{(}\AgdaBound{q=r}\AgdaSpace{}%
\AgdaSymbol{:}\AgdaSpace{}%
\AgdaBound{q}\AgdaSpace{}%
\AgdaDatatype{==}\AgdaSpace{}%
\AgdaBound{r}\AgdaSymbol{)}\AgdaSpace{}%
\AgdaSymbol{(}\AgdaBound{u}\AgdaSpace{}%
\AgdaSymbol{:}\AgdaSpace{}%
\AgdaBound{p}\AgdaSpace{}%
\AgdaDatatype{==}\AgdaSpace{}%
\AgdaBound{q}\AgdaSymbol{)}\<%
\\
\>[0][@{}l@{\AgdaIndent{0}}]\<[2717I]%
\>[6]\AgdaSymbol{→}\AgdaSpace{}%
\AgdaBound{u}\AgdaSpace{}%
\AgdaDatatype{==}\AgdaSpace{}%
\AgdaBound{p=r}\AgdaSpace{}%
\AgdaFunction{◾}\AgdaSpace{}%
\AgdaSymbol{((}\AgdaFunction{!}\AgdaSpace{}%
\AgdaBound{p=r}\AgdaSymbol{)}\AgdaSpace{}%
\AgdaFunction{◾}\AgdaSpace{}%
\AgdaBound{u}\AgdaSpace{}%
\AgdaFunction{◾}\AgdaSpace{}%
\AgdaBound{q=r}\AgdaSymbol{)}\AgdaSpace{}%
\AgdaFunction{◾}\AgdaSpace{}%
\AgdaSymbol{(}\AgdaFunction{!}\AgdaSpace{}%
\AgdaBound{q=r}\AgdaSymbol{)}\<%
\\
\\
\>[0]\AgdaFunction{⟦\_⟧₃}\AgdaSpace{}%
\AgdaSymbol{:}\AgdaSpace{}%
\AgdaSymbol{\{}\AgdaBound{A}\AgdaSpace{}%
\AgdaBound{B}\AgdaSpace{}%
\AgdaSymbol{:}\AgdaSpace{}%
\AgdaDatatype{Π₂}\AgdaSymbol{\}}\AgdaSpace{}%
\AgdaSymbol{\{}\AgdaBound{p}\AgdaSpace{}%
\AgdaBound{q}\AgdaSpace{}%
\AgdaSymbol{:}\AgdaSpace{}%
\AgdaBound{A}\AgdaSpace{}%
\AgdaDatatype{⟷₁}\AgdaSpace{}%
\AgdaBound{B}\AgdaSymbol{\}}\AgdaSpace{}%
\AgdaSymbol{\{}\AgdaBound{u}\AgdaSpace{}%
\AgdaBound{v}\AgdaSpace{}%
\AgdaSymbol{:}\AgdaSpace{}%
\AgdaBound{p}\AgdaSpace{}%
\AgdaDatatype{⟷₂}\AgdaSpace{}%
\AgdaBound{q}\AgdaSymbol{\}}\AgdaSpace{}%
\AgdaSymbol{→}\AgdaSpace{}%
\AgdaSymbol{(}\AgdaBound{α}\AgdaSpace{}%
\AgdaSymbol{:}\AgdaSpace{}%
\AgdaBound{u}\AgdaSpace{}%
\AgdaDatatype{⟷₃}\AgdaSpace{}%
\AgdaBound{v}\AgdaSymbol{)}\AgdaSpace{}%
\AgdaSymbol{→}\AgdaSpace{}%
\AgdaFunction{⟦}\AgdaSpace{}%
\AgdaBound{u}\AgdaSpace{}%
\AgdaFunction{⟧₂}\AgdaSpace{}%
\AgdaDatatype{==}\AgdaSpace{}%
\AgdaFunction{⟦}\AgdaSpace{}%
\AgdaBound{v}\AgdaSpace{}%
\AgdaFunction{⟧₂}\<%
\\
\>[0]\AgdaFunction{⟦\_⟧₃}\AgdaSpace{}%
\AgdaSymbol{\{}\AgdaInductiveConstructor{`𝟚}\AgdaSymbol{\}}\AgdaSpace{}%
\AgdaSymbol{\{}\AgdaInductiveConstructor{`𝟚}\AgdaSymbol{\}}\AgdaSpace{}%
\AgdaSymbol{\{}\AgdaBound{p}\AgdaSymbol{\}}\AgdaSpace{}%
\AgdaSymbol{\{}\AgdaBound{q}\AgdaSymbol{\}}\AgdaSpace{}%
\AgdaSymbol{\{}\AgdaBound{u}\AgdaSymbol{\}}\AgdaSpace{}%
\AgdaSymbol{\{}\AgdaBound{v}\AgdaSymbol{\}}\AgdaSpace{}%
\AgdaInductiveConstructor{`trunc}\AgdaSpace{}%
\AgdaKeyword{with}\AgdaSpace{}%
\AgdaPostulate{all{-}1{-}loops}\AgdaSpace{}%
\AgdaFunction{⟦}\AgdaSpace{}%
\AgdaBound{p}\AgdaSpace{}%
\AgdaFunction{⟧₁}\AgdaSpace{}%
\AgdaSymbol{|}\AgdaSpace{}%
\AgdaPostulate{all{-}1{-}loops}\AgdaSpace{}%
\AgdaFunction{⟦}\AgdaSpace{}%
\AgdaBound{q}\AgdaSpace{}%
\AgdaFunction{⟧₁}\<%
\\
\>[0]\AgdaSymbol{...}\AgdaSpace{}%
\AgdaSymbol{|}\AgdaSpace{}%
\AgdaInductiveConstructor{inl}\AgdaSpace{}%
\AgdaBound{p=id}%
\>[16]\AgdaSymbol{|}\AgdaSpace{}%
\AgdaInductiveConstructor{inl}\AgdaSpace{}%
\AgdaBound{q=id}%
\>[28]\AgdaSymbol{=}\<%
\\
\>[0][@{}l@{\AgdaIndent{0}}]%
\>[2]\AgdaFunction{lemma}\AgdaSpace{}%
\AgdaBound{p=id}\AgdaSpace{}%
\AgdaBound{q=id}\AgdaSpace{}%
\AgdaFunction{⟦}\AgdaSpace{}%
\AgdaBound{u}\AgdaSpace{}%
\AgdaFunction{⟧₂}\<%
\\
\>[0][@{}l@{\AgdaIndent{0}}]%
\>[2]\AgdaFunction{◾}\AgdaSpace{}%
\AgdaFunction{ap}%
\>[8]\AgdaSymbol{(λ}\AgdaSpace{}%
\AgdaBound{x}\AgdaSpace{}%
\AgdaSymbol{→}\AgdaSpace{}%
\AgdaBound{p=id}\AgdaSpace{}%
\AgdaFunction{◾}\AgdaSpace{}%
\AgdaBound{x}\AgdaSpace{}%
\AgdaFunction{◾}\AgdaSpace{}%
\AgdaFunction{!}\AgdaSpace{}%
\AgdaBound{q=id}\AgdaSymbol{)}\<%
\\
\>[8][@{}l@{\AgdaIndent{0}}]%
\>[9]\AgdaSymbol{(}\AgdaFunction{all{-}2{-}loops}\AgdaSpace{}%
\AgdaSymbol{(}\AgdaFunction{!}\AgdaSpace{}%
\AgdaBound{p=id}\AgdaSpace{}%
\AgdaFunction{◾}\AgdaSpace{}%
\AgdaFunction{⟦}\AgdaSpace{}%
\AgdaBound{u}\AgdaSpace{}%
\AgdaFunction{⟧₂}\AgdaSpace{}%
\AgdaFunction{◾}\AgdaSpace{}%
\AgdaBound{q=id}\AgdaSymbol{)}\AgdaSpace{}%
\AgdaFunction{◾}\AgdaSpace{}%
\AgdaFunction{!}\AgdaSpace{}%
\AgdaSymbol{(}\AgdaFunction{all{-}2{-}loops}\AgdaSpace{}%
\AgdaSymbol{(}\AgdaFunction{!}\AgdaSpace{}%
\AgdaBound{p=id}\AgdaSpace{}%
\AgdaFunction{◾}\AgdaSpace{}%
\AgdaFunction{⟦}\AgdaSpace{}%
\AgdaBound{v}\AgdaSpace{}%
\AgdaFunction{⟧₂}\AgdaSpace{}%
\AgdaFunction{◾}\AgdaSpace{}%
\AgdaBound{q=id}\AgdaSymbol{)))}\<%
\\
\>[0][@{}l@{\AgdaIndent{0}}]%
\>[2]\AgdaFunction{◾}\AgdaSpace{}%
\AgdaFunction{!}\AgdaSpace{}%
\AgdaSymbol{(}\AgdaFunction{lemma}\AgdaSpace{}%
\AgdaBound{p=id}\AgdaSpace{}%
\AgdaBound{q=id}\AgdaSpace{}%
\AgdaFunction{⟦}\AgdaSpace{}%
\AgdaBound{v}\AgdaSpace{}%
\AgdaFunction{⟧₂}\AgdaSymbol{)}\<%
\\
\>[0]\AgdaSymbol{...}\AgdaSpace{}%
\AgdaSymbol{|}\AgdaSpace{}%
\AgdaInductiveConstructor{inl}\AgdaSpace{}%
\AgdaBound{p=id}%
\>[16]\AgdaSymbol{|}\AgdaSpace{}%
\AgdaInductiveConstructor{inr}\AgdaSpace{}%
\AgdaBound{q=not}\AgdaSpace{}%
\AgdaSymbol{=}%
\>[31]\AgdaFunction{⊥{-}elim}\AgdaSpace{}%
\AgdaSymbol{(}\AgdaPostulate{id𝟚≠not𝟚}\AgdaSpace{}%
\AgdaSymbol{((}\AgdaFunction{!}\AgdaSpace{}%
\AgdaBound{p=id}\AgdaSymbol{)}\AgdaSpace{}%
\AgdaFunction{◾}\AgdaSpace{}%
\AgdaFunction{⟦}\AgdaSpace{}%
\AgdaBound{u}\AgdaSpace{}%
\AgdaFunction{⟧₂}\AgdaSpace{}%
\AgdaFunction{◾}\AgdaSpace{}%
\AgdaBound{q=not}\AgdaSymbol{))}\<%
\\
\>[0]\AgdaSymbol{...}\AgdaSpace{}%
\AgdaSymbol{|}\AgdaSpace{}%
\AgdaInductiveConstructor{inr}\AgdaSpace{}%
\AgdaBound{p=not}\AgdaSpace{}%
\AgdaSymbol{|}\AgdaSpace{}%
\AgdaInductiveConstructor{inl}\AgdaSpace{}%
\AgdaBound{q=id}%
\>[28]\AgdaSymbol{=}%
\>[31]\AgdaFunction{⊥{-}elim}\AgdaSpace{}%
\AgdaSymbol{(}\AgdaPostulate{id𝟚≠not𝟚}\AgdaSpace{}%
\AgdaSymbol{((}\AgdaFunction{!}\AgdaSpace{}%
\AgdaBound{q=id}\AgdaSymbol{)}\AgdaSpace{}%
\AgdaFunction{◾}\AgdaSpace{}%
\AgdaFunction{!}\AgdaSpace{}%
\AgdaFunction{⟦}\AgdaSpace{}%
\AgdaBound{u}\AgdaSpace{}%
\AgdaFunction{⟧₂}\AgdaSpace{}%
\AgdaFunction{◾}\AgdaSpace{}%
\AgdaBound{p=not}\AgdaSymbol{))}\<%
\\
\>[0]\AgdaSymbol{...}\AgdaSpace{}%
\AgdaSymbol{|}\AgdaSpace{}%
\AgdaInductiveConstructor{inr}\AgdaSpace{}%
\AgdaBound{p=not}\AgdaSpace{}%
\AgdaSymbol{|}\AgdaSpace{}%
\AgdaInductiveConstructor{inr}\AgdaSpace{}%
\AgdaBound{q=not}\AgdaSpace{}%
\AgdaSymbol{=}\<%
\\
\>[0][@{}l@{\AgdaIndent{0}}]%
\>[2]\AgdaFunction{lemma}\AgdaSpace{}%
\AgdaBound{p=not}\AgdaSpace{}%
\AgdaBound{q=not}\AgdaSpace{}%
\AgdaFunction{⟦}\AgdaSpace{}%
\AgdaBound{u}\AgdaSpace{}%
\AgdaFunction{⟧₂}\<%
\\
\>[0][@{}l@{\AgdaIndent{0}}]%
\>[2]\AgdaFunction{◾}\AgdaSpace{}%
\AgdaFunction{ap}%
\>[8]\AgdaSymbol{(λ}\AgdaSpace{}%
\AgdaBound{x}\AgdaSpace{}%
\AgdaSymbol{→}\AgdaSpace{}%
\AgdaBound{p=not}\AgdaSpace{}%
\AgdaFunction{◾}\AgdaSpace{}%
\AgdaBound{x}\AgdaSpace{}%
\AgdaFunction{◾}\AgdaSpace{}%
\AgdaFunction{!}\AgdaSpace{}%
\AgdaBound{q=not}\AgdaSymbol{)}\<%
\\
\>[8][@{}l@{\AgdaIndent{0}}]%
\>[9]\AgdaSymbol{(}\AgdaFunction{all{-}2{-}loops}\AgdaSpace{}%
\AgdaSymbol{(}\AgdaFunction{!}\AgdaSpace{}%
\AgdaBound{p=not}\AgdaSpace{}%
\AgdaFunction{◾}\AgdaSpace{}%
\AgdaFunction{⟦}\AgdaSpace{}%
\AgdaBound{u}\AgdaSpace{}%
\AgdaFunction{⟧₂}\AgdaSpace{}%
\AgdaFunction{◾}\AgdaSpace{}%
\AgdaBound{q=not}\AgdaSymbol{)}\AgdaSpace{}%
\AgdaFunction{◾}\AgdaSpace{}%
\AgdaFunction{!}\AgdaSpace{}%
\AgdaSymbol{(}\AgdaFunction{all{-}2{-}loops}\AgdaSpace{}%
\AgdaSymbol{(}\AgdaFunction{!}\AgdaSpace{}%
\AgdaBound{p=not}\AgdaSpace{}%
\AgdaFunction{◾}\AgdaSpace{}%
\AgdaFunction{⟦}\AgdaSpace{}%
\AgdaBound{v}\AgdaSpace{}%
\AgdaFunction{⟧₂}\AgdaSpace{}%
\AgdaFunction{◾}\AgdaSpace{}%
\AgdaBound{q=not}\AgdaSymbol{)))}\<%
\\
\>[0][@{}l@{\AgdaIndent{0}}]%
\>[2]\AgdaFunction{◾}\AgdaSpace{}%
\AgdaFunction{!}\AgdaSpace{}%
\AgdaSymbol{(}\AgdaFunction{lemma}\AgdaSpace{}%
\AgdaBound{p=not}\AgdaSpace{}%
\AgdaBound{q=not}\AgdaSpace{}%
\AgdaFunction{⟦}\AgdaSpace{}%
\AgdaBound{v}\AgdaSpace{}%
\AgdaFunction{⟧₂}\AgdaSymbol{)}\<%
\\
\\
\>[0]\AgdaFunction{⌜\_⌝₃}\AgdaSpace{}%
\AgdaSymbol{:}\AgdaSpace{}%
\AgdaSymbol{\{}\AgdaBound{p}\AgdaSpace{}%
\AgdaBound{q}\AgdaSpace{}%
\AgdaSymbol{:}\AgdaSpace{}%
\AgdaFunction{𝟚₀}\AgdaSpace{}%
\AgdaDatatype{==}\AgdaSpace{}%
\AgdaFunction{𝟚₀}\AgdaSymbol{\}}\AgdaSpace{}%
\AgdaSymbol{\{}\AgdaBound{u}\AgdaSpace{}%
\AgdaBound{v}\AgdaSpace{}%
\AgdaSymbol{:}\AgdaSpace{}%
\AgdaBound{p}\AgdaSpace{}%
\AgdaDatatype{==}\AgdaSpace{}%
\AgdaBound{q}\AgdaSymbol{\}}\AgdaSpace{}%
\AgdaSymbol{→}\AgdaSpace{}%
\AgdaBound{u}\AgdaSpace{}%
\AgdaDatatype{==}\AgdaSpace{}%
\AgdaBound{v}\AgdaSpace{}%
\AgdaSymbol{→}\AgdaSpace{}%
\AgdaFunction{⌜}\AgdaSpace{}%
\AgdaBound{u}\AgdaSpace{}%
\AgdaFunction{⌝₂}\AgdaSpace{}%
\AgdaDatatype{⟷₃}\AgdaSpace{}%
\AgdaFunction{⌜}\AgdaSpace{}%
\AgdaBound{v}\AgdaSpace{}%
\AgdaFunction{⌝₂}\<%
\\
\>[0]\AgdaFunction{⌜}\AgdaSpace{}%
\AgdaSymbol{\_}\AgdaSpace{}%
\AgdaFunction{⌝₃}\AgdaSpace{}%
\AgdaSymbol{=}\AgdaSpace{}%
\AgdaInductiveConstructor{`trunc}\<%
\end{code}

\AgdaHide{
\begin{code}%
\>[0]\AgdaFunction{lemma}\AgdaSpace{}%
\AgdaSymbol{=}\AgdaSpace{}%
\AgdaSymbol{?}\<%
\end{code}
}

\subsection{Coherence}

It now remains to show that all these mapping are coherent with each
other in the sense that each round trip produces a term that is
identifiable with the original term, effectively showing soundness and
completness of the univalent universe with respect to $\PiTwo$. At
level-0, this is trivial.

At level-1, \emph{soundness} means that the mappings are inverses:
\begin{itemize}
\item any 1-combinator {\small\AgdaBound{p}} mapped to a 1-path and
  back is 2-equivalent to itself, and
\item there is always a 2-path between a 1-path {\small\AgdaBound{p}}
sent to a 1-combinator and back.
\end{itemize}
This is rather more succinct in code:
\begin{code}%
\>[0]\AgdaFunction{⌜⟦\_⟧₁⌝₁}\AgdaSpace{}%
\AgdaSymbol{:}\AgdaSpace{}%
\AgdaSymbol{(}\AgdaBound{p}\AgdaSpace{}%
\AgdaSymbol{:}\AgdaSpace{}%
\AgdaInductiveConstructor{`𝟚}\AgdaSpace{}%
\AgdaDatatype{⟷₁}\AgdaSpace{}%
\AgdaInductiveConstructor{`𝟚}\AgdaSymbol{)}\AgdaSpace{}%
\AgdaSymbol{→}\AgdaSpace{}%
\AgdaBound{p}\AgdaSpace{}%
\AgdaDatatype{⟷₂}\AgdaSpace{}%
\AgdaFunction{⌜}\AgdaSpace{}%
\AgdaFunction{⟦}\AgdaSpace{}%
\AgdaBound{p}\AgdaSpace{}%
\AgdaFunction{⟧₁}\AgdaSpace{}%
\AgdaFunction{⌝₁}\<%
\\
\>[0]\AgdaFunction{⌜⟦}\AgdaSpace{}%
\AgdaBound{p}\AgdaSpace{}%
\AgdaFunction{⟧₁⌝₁}\AgdaSpace{}%
\AgdaKeyword{with}\AgdaSpace{}%
\AgdaFunction{canonical}\AgdaSpace{}%
\AgdaBound{p}\AgdaSpace{}%
\AgdaSymbol{|}\AgdaSpace{}%
\AgdaPostulate{all{-}1{-}loops}\AgdaSpace{}%
\AgdaFunction{⟦}\AgdaSpace{}%
\AgdaBound{p}\AgdaSpace{}%
\AgdaFunction{⟧₁}\<%
\\
\>[0]\AgdaSymbol{...}\AgdaSpace{}%
\AgdaSymbol{|}\AgdaSpace{}%
\AgdaInductiveConstructor{ID}%
\>[10]\AgdaInductiveConstructor{,}\AgdaSpace{}%
\AgdaBound{p⇔id}%
\>[19]\AgdaSymbol{|}\AgdaSpace{}%
\AgdaInductiveConstructor{inl}\AgdaSpace{}%
\AgdaBound{p=id}%
\>[32]\AgdaSymbol{=}\AgdaSpace{}%
\AgdaBound{p⇔id}\<%
\\
\>[0]\AgdaSymbol{...}\AgdaSpace{}%
\AgdaSymbol{|}\AgdaSpace{}%
\AgdaInductiveConstructor{ID}%
\>[10]\AgdaInductiveConstructor{,}\AgdaSpace{}%
\AgdaBound{p⇔id}%
\>[19]\AgdaSymbol{|}\AgdaSpace{}%
\AgdaInductiveConstructor{inr}\AgdaSpace{}%
\AgdaBound{p=not}%
\>[32]\AgdaSymbol{=}\AgdaSpace{}%
\AgdaFunction{⊥{-}elim}\AgdaSpace{}%
\AgdaSymbol{(}\AgdaPostulate{id𝟚≠not𝟚}\AgdaSpace{}%
\AgdaSymbol{(}\AgdaFunction{!}\AgdaSpace{}%
\AgdaSymbol{((}\AgdaFunction{!}\AgdaSpace{}%
\AgdaBound{p=not}\AgdaSymbol{)}\AgdaSpace{}%
\AgdaFunction{◾}\AgdaSpace{}%
\AgdaFunction{⟦}\AgdaSpace{}%
\AgdaBound{p⇔id}\AgdaSpace{}%
\AgdaFunction{⟧₂}\AgdaSymbol{)))}\<%
\\
\>[0]\AgdaSymbol{...}\AgdaSpace{}%
\AgdaSymbol{|}\AgdaSpace{}%
\AgdaInductiveConstructor{NOT}\AgdaSpace{}%
\AgdaInductiveConstructor{,}\AgdaSpace{}%
\AgdaBound{p⇔not}%
\>[19]\AgdaSymbol{|}\AgdaSpace{}%
\AgdaInductiveConstructor{inl}\AgdaSpace{}%
\AgdaBound{p=id}%
\>[32]\AgdaSymbol{=}\AgdaSpace{}%
\AgdaFunction{⊥{-}elim}\AgdaSpace{}%
\AgdaSymbol{(}\AgdaPostulate{id𝟚≠not𝟚}\AgdaSpace{}%
\AgdaSymbol{((}\AgdaFunction{!}\AgdaSpace{}%
\AgdaBound{p=id}\AgdaSymbol{)}\AgdaSpace{}%
\AgdaFunction{◾}\AgdaSpace{}%
\AgdaFunction{⟦}\AgdaSpace{}%
\AgdaBound{p⇔not}\AgdaSpace{}%
\AgdaFunction{⟧₂}\AgdaSymbol{))}\<%
\\
\>[0]\AgdaSymbol{...}\AgdaSpace{}%
\AgdaSymbol{|}\AgdaSpace{}%
\AgdaInductiveConstructor{NOT}\AgdaSpace{}%
\AgdaInductiveConstructor{,}\AgdaSpace{}%
\AgdaBound{p⇔not}%
\>[19]\AgdaSymbol{|}\AgdaSpace{}%
\AgdaInductiveConstructor{inr}\AgdaSpace{}%
\AgdaBound{p=not}%
\>[32]\AgdaSymbol{=}\AgdaSpace{}%
\AgdaBound{p⇔not}\<%
\\
\\
\>[0]\AgdaFunction{⟦⌜\_⌝₁⟧₁}\AgdaSpace{}%
\AgdaSymbol{:}\AgdaSpace{}%
\AgdaSymbol{(}\AgdaBound{p}\AgdaSpace{}%
\AgdaSymbol{:}\AgdaSpace{}%
\AgdaFunction{𝟚₀}\AgdaSpace{}%
\AgdaDatatype{==}\AgdaSpace{}%
\AgdaFunction{𝟚₀}\AgdaSymbol{)}\AgdaSpace{}%
\AgdaSymbol{→}\AgdaSpace{}%
\AgdaBound{p}\AgdaSpace{}%
\AgdaDatatype{==}\AgdaSpace{}%
\AgdaFunction{⟦}\AgdaSpace{}%
\AgdaFunction{⌜}\AgdaSpace{}%
\AgdaBound{p}\AgdaSpace{}%
\AgdaFunction{⌝₁}\AgdaSpace{}%
\AgdaFunction{⟧₁}\<%
\\
\>[0]\AgdaFunction{⟦⌜}\AgdaSpace{}%
\AgdaBound{p}\AgdaSpace{}%
\AgdaFunction{⌝₁⟧₁}%
\>[11]\AgdaKeyword{with}\AgdaSpace{}%
\AgdaPostulate{all{-}1{-}loops}\AgdaSpace{}%
\AgdaBound{p}\AgdaSpace{}%
\AgdaSymbol{|}\AgdaSpace{}%
\AgdaFunction{canonical}\AgdaSpace{}%
\AgdaFunction{⌜}\AgdaSpace{}%
\AgdaBound{p}\AgdaSpace{}%
\AgdaFunction{⌝₁}\<%
\\
\>[0]\AgdaSymbol{...}\AgdaSpace{}%
\AgdaSymbol{|}\AgdaSpace{}%
\AgdaInductiveConstructor{inl}\AgdaSpace{}%
\AgdaBound{p=id}%
\>[17]\AgdaSymbol{|}\AgdaSpace{}%
\AgdaInductiveConstructor{ID}%
\>[23]\AgdaInductiveConstructor{,}\AgdaSpace{}%
\AgdaBound{p⇔id}%
\>[32]\AgdaSymbol{=}\AgdaSpace{}%
\AgdaBound{p=id}\<%
\\
\>[0]\AgdaSymbol{...}\AgdaSpace{}%
\AgdaSymbol{|}\AgdaSpace{}%
\AgdaInductiveConstructor{inl}\AgdaSpace{}%
\AgdaBound{p=id}%
\>[17]\AgdaSymbol{|}\AgdaSpace{}%
\AgdaInductiveConstructor{NOT}\AgdaSpace{}%
\AgdaInductiveConstructor{,}\AgdaSpace{}%
\AgdaBound{p⇔not}%
\>[32]\AgdaSymbol{=}\AgdaSpace{}%
\AgdaFunction{⊥{-}elim}\AgdaSpace{}%
\AgdaSymbol{(}\AgdaPostulate{id𝟚≠not𝟚}\AgdaSpace{}%
\AgdaFunction{⟦}\AgdaSpace{}%
\AgdaBound{p⇔not}\AgdaSpace{}%
\AgdaFunction{⟧₂}\AgdaSymbol{)}\<%
\\
\>[0]\AgdaSymbol{...}\AgdaSpace{}%
\AgdaSymbol{|}\AgdaSpace{}%
\AgdaInductiveConstructor{inr}\AgdaSpace{}%
\AgdaBound{p=not}%
\>[17]\AgdaSymbol{|}\AgdaSpace{}%
\AgdaInductiveConstructor{ID}%
\>[23]\AgdaInductiveConstructor{,}\AgdaSpace{}%
\AgdaBound{p⇔id}%
\>[32]\AgdaSymbol{=}\AgdaSpace{}%
\AgdaFunction{⊥{-}elim}\AgdaSpace{}%
\AgdaSymbol{(}\AgdaPostulate{id𝟚≠not𝟚}\AgdaSpace{}%
\AgdaSymbol{(}\AgdaFunction{!}\AgdaSpace{}%
\AgdaFunction{⟦}\AgdaSpace{}%
\AgdaBound{p⇔id}\AgdaSpace{}%
\AgdaFunction{⟧₂}\AgdaSymbol{))}\<%
\\
\>[0]\AgdaSymbol{...}\AgdaSpace{}%
\AgdaSymbol{|}\AgdaSpace{}%
\AgdaInductiveConstructor{inr}\AgdaSpace{}%
\AgdaBound{p=not}%
\>[17]\AgdaSymbol{|}\AgdaSpace{}%
\AgdaInductiveConstructor{NOT}\AgdaSpace{}%
\AgdaInductiveConstructor{,}\AgdaSpace{}%
\AgdaBound{p⇔not}%
\>[32]\AgdaSymbol{=}\AgdaSpace{}%
\AgdaBound{p=not}\<%
\end{code}

They are also \emph{complete} in the following sense:
\begin{itemize}
\item for any two 1-combinators which map to 1-paths which are
related by a 2-path, the 1-combinators are related by a 2-combinator, and
\item for any two 1-paths which map to 1-combinators which are
related by a 2-combinator these are related by a 2-path.
\end{itemize}
Normally, completeness is a rather difficult result to prove.  But in
our case, the infrastructure from the previous section makes the proof
immediate: For the first proof, the key is \emph{reversibility} of the
level-2 combinators using {\small\AgdaInductiveConstructor{!₂}}; for
the second proof it is the reversibility of paths in the univalent
universe that is critical:

\begin{code}%
\>[0]\AgdaFunction{completeness₁}\AgdaSpace{}%
\AgdaSymbol{:}\AgdaSpace{}%
\AgdaSymbol{\{}\AgdaBound{p}\AgdaSpace{}%
\AgdaBound{q}\AgdaSpace{}%
\AgdaSymbol{:}\AgdaSpace{}%
\AgdaInductiveConstructor{`𝟚}\AgdaSpace{}%
\AgdaDatatype{⟷₁}\AgdaSpace{}%
\AgdaInductiveConstructor{`𝟚}\AgdaSymbol{\}}\AgdaSpace{}%
\AgdaSymbol{→}\AgdaSpace{}%
\AgdaFunction{⟦}\AgdaSpace{}%
\AgdaBound{p}\AgdaSpace{}%
\AgdaFunction{⟧₁}\AgdaSpace{}%
\AgdaDatatype{==}\AgdaSpace{}%
\AgdaFunction{⟦}\AgdaSpace{}%
\AgdaBound{q}\AgdaSpace{}%
\AgdaFunction{⟧₁}\AgdaSpace{}%
\AgdaSymbol{→}\AgdaSpace{}%
\AgdaBound{p}\AgdaSpace{}%
\AgdaDatatype{⟷₂}\AgdaSpace{}%
\AgdaBound{q}\<%
\\
\>[0]\AgdaFunction{completeness₁}\AgdaSpace{}%
\AgdaSymbol{\{}\AgdaBound{p}\AgdaSymbol{\}}\AgdaSpace{}%
\AgdaSymbol{\{}\AgdaBound{q}\AgdaSymbol{\}}\AgdaSpace{}%
\AgdaBound{u}\AgdaSpace{}%
\AgdaSymbol{=}\AgdaSpace{}%
\AgdaFunction{⌜⟦}\AgdaSpace{}%
\AgdaBound{p}\AgdaSpace{}%
\AgdaFunction{⟧₁⌝₁}\AgdaSpace{}%
\AgdaInductiveConstructor{⊙₂}\AgdaSpace{}%
\AgdaSymbol{(}\AgdaFunction{⌜}\AgdaSpace{}%
\AgdaBound{u}\AgdaSpace{}%
\AgdaFunction{⌝₂}\AgdaSpace{}%
\AgdaInductiveConstructor{⊙₂}\AgdaSpace{}%
\AgdaInductiveConstructor{!₂}\AgdaSpace{}%
\AgdaFunction{⌜⟦}\AgdaSpace{}%
\AgdaBound{q}\AgdaSpace{}%
\AgdaFunction{⟧₁⌝₁}\AgdaSymbol{)}\<%
\\
\\
\>[0]\AgdaFunction{completeness₁⁻¹}\AgdaSpace{}%
\AgdaSymbol{:}\AgdaSpace{}%
\AgdaSymbol{\{}\AgdaBound{p}\AgdaSpace{}%
\AgdaBound{q}\AgdaSpace{}%
\AgdaSymbol{:}\AgdaSpace{}%
\AgdaFunction{𝟚₀}\AgdaSpace{}%
\AgdaDatatype{==}\AgdaSpace{}%
\AgdaFunction{𝟚₀}\AgdaSymbol{\}}\AgdaSpace{}%
\AgdaSymbol{→}\AgdaSpace{}%
\AgdaFunction{⌜}\AgdaSpace{}%
\AgdaBound{p}\AgdaSpace{}%
\AgdaFunction{⌝₁}\AgdaSpace{}%
\AgdaDatatype{⟷₂}\AgdaSpace{}%
\AgdaFunction{⌜}\AgdaSpace{}%
\AgdaBound{q}\AgdaSpace{}%
\AgdaFunction{⌝₁}\AgdaSpace{}%
\AgdaSymbol{→}\AgdaSpace{}%
\AgdaBound{p}\AgdaSpace{}%
\AgdaDatatype{==}\AgdaSpace{}%
\AgdaBound{q}\<%
\\
\>[0]\AgdaFunction{completeness₁⁻¹}\AgdaSpace{}%
\AgdaSymbol{\{}\AgdaBound{p}\AgdaSymbol{\}}\AgdaSpace{}%
\AgdaSymbol{\{}\AgdaBound{q}\AgdaSymbol{\}}\AgdaSpace{}%
\AgdaBound{u}\AgdaSpace{}%
\AgdaSymbol{=}\AgdaSpace{}%
\AgdaFunction{⟦⌜}\AgdaSpace{}%
\AgdaBound{p}\AgdaSpace{}%
\AgdaFunction{⌝₁⟧₁}\AgdaSpace{}%
\AgdaFunction{◾}\AgdaSpace{}%
\AgdaFunction{⟦}\AgdaSpace{}%
\AgdaBound{u}\AgdaSpace{}%
\AgdaFunction{⟧₂}\AgdaSpace{}%
\AgdaFunction{◾}\AgdaSpace{}%
\AgdaSymbol{(}\AgdaFunction{!}\AgdaSpace{}%
\AgdaFunction{⟦⌜}\AgdaSpace{}%
\AgdaBound{q}\AgdaSpace{}%
\AgdaFunction{⌝₁⟧₁}\AgdaSymbol{)}\<%
\end{code}

For level-2, the statements are informally quite similar (with all
levels bumped up by one).  For 2-combinators, the result is
trivial. For the other direction starting from 2-paths in the
univalent universe soundness is tricky to even state, mostly because
the types involved in {\small\AgdaFunction{⌜
    ⟦}~\AgdaBound{u}~\AgdaFunction{⟧₂ ⌝₂}} and {\small\AgdaFunction{⟦
    ⌜}~\AgdaBound{u}~\AgdaFunction{⌝₂ ⟧₂}} are non-trivial. But
enumeration of 1-loops reduces the complexity of the problem to
``unwinding'' complex expressions for identity paths:

\begin{code}%
\>[0]\AgdaFunction{⌜⟦\_⟧₂⌝₂}%
\>[3148I]\AgdaSymbol{:}%
\>[11]\AgdaSymbol{\{}\AgdaBound{p}\AgdaSpace{}%
\AgdaBound{q}\AgdaSpace{}%
\AgdaSymbol{:}\AgdaSpace{}%
\AgdaInductiveConstructor{`𝟚}\AgdaSpace{}%
\AgdaDatatype{⟷₁}\AgdaSpace{}%
\AgdaInductiveConstructor{`𝟚}\AgdaSymbol{\}}\<%
\\
\>[3148I][@{}l@{\AgdaIndent{0}}]%
\>[11]\AgdaSymbol{(}\AgdaBound{u}\AgdaSpace{}%
\AgdaSymbol{:}\AgdaSpace{}%
\AgdaBound{p}\AgdaSpace{}%
\AgdaDatatype{⟷₂}\AgdaSpace{}%
\AgdaBound{q}\AgdaSymbol{)}\AgdaSpace{}%
\AgdaSymbol{→}\AgdaSpace{}%
\AgdaBound{u}\AgdaSpace{}%
\AgdaDatatype{⟷₃}\AgdaSpace{}%
\AgdaSymbol{(}\AgdaFunction{⌜⟦}\AgdaSpace{}%
\AgdaBound{p}\AgdaSpace{}%
\AgdaFunction{⟧₁⌝₁}\AgdaSpace{}%
\AgdaInductiveConstructor{⊙₂}\AgdaSpace{}%
\AgdaSymbol{(}\AgdaFunction{⌜}\AgdaSpace{}%
\AgdaFunction{⟦}\AgdaSpace{}%
\AgdaBound{u}\AgdaSpace{}%
\AgdaFunction{⟧₂}\AgdaSpace{}%
\AgdaFunction{⌝₂}\AgdaSpace{}%
\AgdaInductiveConstructor{⊙₂}\AgdaSpace{}%
\AgdaSymbol{(}\AgdaInductiveConstructor{!₂}\AgdaSpace{}%
\AgdaFunction{⌜⟦}\AgdaSpace{}%
\AgdaBound{q}\AgdaSpace{}%
\AgdaFunction{⟧₁⌝₁}\AgdaSymbol{)))}\<%
\\
\>[0]\AgdaFunction{⌜⟦}\AgdaSpace{}%
\AgdaBound{u}\AgdaSpace{}%
\AgdaFunction{⟧₂⌝₂}\AgdaSpace{}%
\AgdaSymbol{=}\AgdaSpace{}%
\AgdaInductiveConstructor{`trunc}\<%
\\
\\
\>[0]\AgdaFunction{⟦⌜\_⌝₂⟧₂}\AgdaSpace{}%
\AgdaSymbol{:}\AgdaSpace{}%
\AgdaSymbol{\{}\AgdaBound{p}\AgdaSpace{}%
\AgdaBound{q}\AgdaSpace{}%
\AgdaSymbol{:}\AgdaSpace{}%
\AgdaFunction{𝟚₀}\AgdaSpace{}%
\AgdaDatatype{==}\AgdaSpace{}%
\AgdaFunction{𝟚₀}\AgdaSymbol{\}}\AgdaSpace{}%
\AgdaSymbol{(}\AgdaBound{u}\AgdaSpace{}%
\AgdaSymbol{:}\AgdaSpace{}%
\AgdaBound{p}\AgdaSpace{}%
\AgdaDatatype{==}\AgdaSpace{}%
\AgdaBound{q}\AgdaSymbol{)}\AgdaSpace{}%
\AgdaSymbol{→}\AgdaSpace{}%
\AgdaBound{u}\AgdaSpace{}%
\AgdaDatatype{==}\AgdaSpace{}%
\AgdaFunction{⟦⌜}\AgdaSpace{}%
\AgdaBound{p}\AgdaSpace{}%
\AgdaFunction{⌝₁⟧₁}\AgdaSpace{}%
\AgdaFunction{◾}\AgdaSpace{}%
\AgdaFunction{⟦}\AgdaSpace{}%
\AgdaFunction{⌜}\AgdaSpace{}%
\AgdaBound{u}\AgdaSpace{}%
\AgdaFunction{⌝₂}\AgdaSpace{}%
\AgdaFunction{⟧₂}\AgdaSpace{}%
\AgdaFunction{◾}\AgdaSpace{}%
\AgdaSymbol{(}\AgdaFunction{!}\AgdaSpace{}%
\AgdaFunction{⟦⌜}\AgdaSpace{}%
\AgdaBound{q}\AgdaSpace{}%
\AgdaFunction{⌝₁⟧₁}\AgdaSymbol{)}\<%
\\
\>[0]\AgdaFunction{⟦⌜\_⌝₂⟧₂}\AgdaSpace{}%
\AgdaSymbol{\{}\AgdaBound{p}\AgdaSymbol{\}}\AgdaSpace{}%
\AgdaSymbol{\{}\AgdaBound{q}\AgdaSymbol{\}}\AgdaSpace{}%
\AgdaBound{u}\AgdaSpace{}%
\AgdaKeyword{with}\AgdaSpace{}%
\AgdaPostulate{all{-}1{-}loops}\AgdaSpace{}%
\AgdaBound{p}\AgdaSpace{}%
\AgdaSymbol{|}\AgdaSpace{}%
\AgdaPostulate{all{-}1{-}loops}\AgdaSpace{}%
\AgdaBound{q}\<%
\\
\>[0]\AgdaSymbol{...}\AgdaSpace{}%
\AgdaSymbol{|}\AgdaSpace{}%
\AgdaInductiveConstructor{inl}\AgdaSpace{}%
\AgdaBound{p=id}%
\>[17]\AgdaSymbol{|}\AgdaSpace{}%
\AgdaInductiveConstructor{inl}\AgdaSpace{}%
\AgdaBound{q=id}%
\>[30]\AgdaSymbol{=}%
\>[33]\AgdaSymbol{(}\AgdaFunction{lemma}\AgdaSpace{}%
\AgdaBound{p=id}\AgdaSpace{}%
\AgdaBound{q=id}\AgdaSpace{}%
\AgdaBound{u}\AgdaSymbol{)}\<%
\\
\>[30][@{}l@{\AgdaIndent{0}}]%
\>[33]\AgdaFunction{◾}\AgdaSpace{}%
\AgdaSymbol{(}\AgdaFunction{ap}\AgdaSpace{}%
\AgdaSymbol{(λ}\AgdaSpace{}%
\AgdaBound{x}\AgdaSpace{}%
\AgdaSymbol{→}\AgdaSpace{}%
\AgdaBound{p=id}\AgdaSpace{}%
\AgdaFunction{◾}\AgdaSpace{}%
\AgdaBound{x}\AgdaSpace{}%
\AgdaFunction{◾}\AgdaSpace{}%
\AgdaFunction{!}\AgdaSpace{}%
\AgdaBound{q=id}\AgdaSymbol{)}\AgdaSpace{}%
\AgdaSymbol{(}\AgdaFunction{all{-}2{-}loops}\AgdaSpace{}%
\AgdaSymbol{(}\AgdaFunction{!}\AgdaSpace{}%
\AgdaBound{p=id}\AgdaSpace{}%
\AgdaFunction{◾}\AgdaSpace{}%
\AgdaBound{u}\AgdaSpace{}%
\AgdaFunction{◾}\AgdaSpace{}%
\AgdaBound{q=id}\AgdaSymbol{)))}\<%
\\
\>[0]\AgdaSymbol{...}\AgdaSpace{}%
\AgdaSymbol{|}\AgdaSpace{}%
\AgdaInductiveConstructor{inl}\AgdaSpace{}%
\AgdaBound{p=id}%
\>[17]\AgdaSymbol{|}\AgdaSpace{}%
\AgdaInductiveConstructor{inr}\AgdaSpace{}%
\AgdaBound{q=not}%
\>[30]\AgdaSymbol{=}\AgdaSpace{}%
\AgdaFunction{⊥{-}elim}\AgdaSpace{}%
\AgdaSymbol{(}\AgdaPostulate{id𝟚≠not𝟚}\AgdaSpace{}%
\AgdaSymbol{((}\AgdaFunction{!}\AgdaSpace{}%
\AgdaBound{p=id}\AgdaSymbol{)}\AgdaSpace{}%
\AgdaFunction{◾}\AgdaSpace{}%
\AgdaBound{u}\AgdaSpace{}%
\AgdaFunction{◾}\AgdaSpace{}%
\AgdaBound{q=not}\AgdaSymbol{))}\<%
\\
\>[0]\AgdaSymbol{...}\AgdaSpace{}%
\AgdaSymbol{|}\AgdaSpace{}%
\AgdaInductiveConstructor{inr}\AgdaSpace{}%
\AgdaBound{p=not}%
\>[17]\AgdaSymbol{|}\AgdaSpace{}%
\AgdaInductiveConstructor{inl}\AgdaSpace{}%
\AgdaBound{q=id}%
\>[30]\AgdaSymbol{=}\AgdaSpace{}%
\AgdaFunction{⊥{-}elim}\AgdaSpace{}%
\AgdaSymbol{(}\AgdaPostulate{id𝟚≠not𝟚}\AgdaSpace{}%
\AgdaSymbol{(}\AgdaFunction{!}\AgdaSpace{}%
\AgdaSymbol{((}\AgdaFunction{!}\AgdaSpace{}%
\AgdaBound{p=not}\AgdaSymbol{)}\AgdaSpace{}%
\AgdaFunction{◾}\AgdaSpace{}%
\AgdaBound{u}\AgdaSpace{}%
\AgdaFunction{◾}\AgdaSpace{}%
\AgdaBound{q=id}\AgdaSymbol{)))}\<%
\\
\>[0]\AgdaSymbol{...}\AgdaSpace{}%
\AgdaSymbol{|}\AgdaSpace{}%
\AgdaInductiveConstructor{inr}\AgdaSpace{}%
\AgdaBound{p=not}%
\>[17]\AgdaSymbol{|}\AgdaSpace{}%
\AgdaInductiveConstructor{inr}\AgdaSpace{}%
\AgdaBound{q=not}%
\>[30]\AgdaSymbol{=}%
\>[33]\AgdaSymbol{(}\AgdaFunction{lemma}\AgdaSpace{}%
\AgdaBound{p=not}\AgdaSpace{}%
\AgdaBound{q=not}\AgdaSpace{}%
\AgdaBound{u}\AgdaSymbol{)}\<%
\\
\>[30][@{}l@{\AgdaIndent{0}}]%
\>[33]\AgdaFunction{◾}\AgdaSpace{}%
\AgdaSymbol{(}\AgdaFunction{ap}\AgdaSpace{}%
\AgdaSymbol{(λ}\AgdaSpace{}%
\AgdaBound{x}\AgdaSpace{}%
\AgdaSymbol{→}\AgdaSpace{}%
\AgdaBound{p=not}\AgdaSpace{}%
\AgdaFunction{◾}\AgdaSpace{}%
\AgdaBound{x}\AgdaSpace{}%
\AgdaFunction{◾}\AgdaSpace{}%
\AgdaFunction{!}\AgdaSpace{}%
\AgdaBound{q=not}\AgdaSymbol{)}\AgdaSpace{}%
\AgdaSymbol{(}\AgdaFunction{all{-}2{-}loops}\AgdaSpace{}%
\AgdaSymbol{(}\AgdaFunction{!}\AgdaSpace{}%
\AgdaBound{p=not}\AgdaSpace{}%
\AgdaFunction{◾}\AgdaSpace{}%
\AgdaBound{u}\AgdaSpace{}%
\AgdaFunction{◾}\AgdaSpace{}%
\AgdaBound{q=not}\AgdaSymbol{)))}\<%
\end{code}

Level-2 completeness offers no new difficulties:

\begin{code}%
\>[0]\AgdaFunction{completeness₂}\AgdaSpace{}%
\AgdaSymbol{:}\AgdaSpace{}%
\AgdaSymbol{\{}\AgdaBound{p}\AgdaSpace{}%
\AgdaBound{q}\AgdaSpace{}%
\AgdaSymbol{:}\AgdaSpace{}%
\AgdaInductiveConstructor{`𝟚}\AgdaSpace{}%
\AgdaDatatype{⟷₁}\AgdaSpace{}%
\AgdaInductiveConstructor{`𝟚}\AgdaSymbol{\}}\AgdaSpace{}%
\AgdaSymbol{\{}\AgdaBound{u}\AgdaSpace{}%
\AgdaBound{v}\AgdaSpace{}%
\AgdaSymbol{:}\AgdaSpace{}%
\AgdaBound{p}\AgdaSpace{}%
\AgdaDatatype{⟷₂}\AgdaSpace{}%
\AgdaBound{q}\AgdaSymbol{\}}\AgdaSpace{}%
\AgdaSymbol{→}\AgdaSpace{}%
\AgdaFunction{⟦}\AgdaSpace{}%
\AgdaBound{u}\AgdaSpace{}%
\AgdaFunction{⟧₂}\AgdaSpace{}%
\AgdaDatatype{==}\AgdaSpace{}%
\AgdaFunction{⟦}\AgdaSpace{}%
\AgdaBound{v}\AgdaSpace{}%
\AgdaFunction{⟧₂}\AgdaSpace{}%
\AgdaSymbol{→}\AgdaSpace{}%
\AgdaBound{u}\AgdaSpace{}%
\AgdaDatatype{⟷₃}\AgdaSpace{}%
\AgdaBound{v}\<%
\\
\>[0]\AgdaFunction{completeness₂}\AgdaSpace{}%
\AgdaBound{u}\AgdaSpace{}%
\AgdaSymbol{=}\AgdaSpace{}%
\AgdaInductiveConstructor{`trunc}\<%
\\
\\
\>[0]\AgdaFunction{completeness₂⁻¹}\AgdaSpace{}%
\AgdaSymbol{:}\AgdaSpace{}%
\AgdaSymbol{\{}\AgdaBound{p}\AgdaSpace{}%
\AgdaBound{q}\AgdaSpace{}%
\AgdaSymbol{:}\AgdaSpace{}%
\AgdaFunction{𝟚₀}\AgdaSpace{}%
\AgdaDatatype{==}\AgdaSpace{}%
\AgdaFunction{𝟚₀}\AgdaSymbol{\}}\AgdaSpace{}%
\AgdaSymbol{\{}\AgdaBound{u}\AgdaSpace{}%
\AgdaBound{v}\AgdaSpace{}%
\AgdaSymbol{:}\AgdaSpace{}%
\AgdaBound{p}\AgdaSpace{}%
\AgdaDatatype{==}\AgdaSpace{}%
\AgdaBound{q}\AgdaSymbol{\}}\AgdaSpace{}%
\AgdaSymbol{→}\AgdaSpace{}%
\AgdaFunction{⌜}\AgdaSpace{}%
\AgdaBound{u}\AgdaSpace{}%
\AgdaFunction{⌝₂}\AgdaSpace{}%
\AgdaDatatype{⟷₃}\AgdaSpace{}%
\AgdaFunction{⌜}\AgdaSpace{}%
\AgdaBound{v}\AgdaSpace{}%
\AgdaFunction{⌝₂}\AgdaSpace{}%
\AgdaSymbol{→}\AgdaSpace{}%
\AgdaBound{u}\AgdaSpace{}%
\AgdaDatatype{==}\AgdaSpace{}%
\AgdaBound{v}\<%
\\
\>[0]\AgdaFunction{completeness₂⁻¹}\AgdaSpace{}%
\AgdaSymbol{\{}\AgdaBound{p}\AgdaSymbol{\}}\AgdaSpace{}%
\AgdaSymbol{\{}\AgdaBound{q}\AgdaSymbol{\}}\AgdaSpace{}%
\AgdaSymbol{\{}\AgdaBound{u}\AgdaSymbol{\}}\AgdaSpace{}%
\AgdaSymbol{\{}\AgdaBound{v}\AgdaSymbol{\}}%
\>[3351I]\AgdaBound{α}%
\>[3352I]\AgdaSymbol{=}\AgdaSpace{}%
\AgdaFunction{⟦⌜}\AgdaSpace{}%
\AgdaBound{u}\AgdaSpace{}%
\AgdaFunction{⌝₂⟧₂}\<%
\\
\>[3351I][@{}l@{\AgdaIndent{0}}]\<[3352I]%
\>[34]\AgdaFunction{◾}\AgdaSpace{}%
\AgdaFunction{ap}\AgdaSpace{}%
\AgdaSymbol{(λ}\AgdaSpace{}%
\AgdaBound{x}\AgdaSpace{}%
\AgdaSymbol{→}\AgdaSpace{}%
\AgdaFunction{⟦⌜}\AgdaSpace{}%
\AgdaBound{p}\AgdaSpace{}%
\AgdaFunction{⌝₁⟧₁}\AgdaSpace{}%
\AgdaFunction{◾}\AgdaSpace{}%
\AgdaBound{x}\AgdaSpace{}%
\AgdaFunction{◾}\AgdaSpace{}%
\AgdaFunction{!}\AgdaSpace{}%
\AgdaFunction{⟦⌜}\AgdaSpace{}%
\AgdaBound{q}\AgdaSpace{}%
\AgdaFunction{⌝₁⟧₁}\AgdaSymbol{)}\AgdaSpace{}%
\AgdaFunction{⟦}\AgdaSpace{}%
\AgdaBound{α}\AgdaSpace{}%
\AgdaFunction{⟧₃}\<%
\\
\>[3351I][@{}l@{\AgdaIndent{0}}]%
\>[34]\AgdaFunction{◾}\AgdaSpace{}%
\AgdaSymbol{(}\AgdaFunction{!}\AgdaSpace{}%
\AgdaFunction{⟦⌜}\AgdaSpace{}%
\AgdaBound{v}\AgdaSpace{}%
\AgdaFunction{⌝₂⟧₂}\AgdaSymbol{)}\<%
\end{code}

\section{Discussion and Related Work}
\label{sec:discussion}

\paragraph*{Reversible Languages.}
\noindent The practice of programming languages is replete with \emph{ad hoc}
instances of reversible computations: database transactions, mechanisms for data
provenance, checkpoints, stack and exception traces, logs, backups, rollback
recoveries, version control systems, reverse engineering, software transactional
memories, continuations, backtracking search, and multiple-level ``undo''
features in commercial applications. In the early nineties,
Baker~\cite{Baker:1992:LLL,Baker:1992:NFT} argued for a systematic, first-class,
treatment of reversibility. But intensive research in full-fledged reversible
models of computations and reversible programming languages was only sparked by
the discovery of deep connections between physics and
computation~\cite{Landauer:1961,PhysRevA.32.3266,Toffoli:1980,bennett1985fundamental,Frank:1999:REC:930275},
and by the potential for efficient quantum
computation~\cite{springerlink:10.1007/BF02650179}.

The early developments of reversible programming languages started with a
conventional programming language, e.g., an extended $\lambda$-calculus, and either
\begin{enumerate}
\item extended the language with a history
mechanism~\cite{vanTonder:2004,Kluge:1999:SEMCD,lorenz,danos2004reversible}, or
\item imposed constraints on the control flow constructs to make them
reversible~\cite{Yokoyama:2007:RPL:1244381.1244404}.
\end{enumerate}
More modern approaches recognize that reversible programming languages require
a fresh approach and should be designed from first principles without the
detour via conventional irreversible
languages~\cite{Yokoyama:2008:PRP,Mu:2004:ILRC,abramsky2005structural,DiPierro:2006:RCL:1166042.1166047}.

\paragraph*{The $\Pi$ Family of Languages}
\noindent In previous work, Carette, Bowman, James, and
Sabry~\cite{rc2011,James:2012:IE:2103656.2103667,Carette2016}
introduced the~$\Pi$ family of reversible languages based on type
isomorphisms and commutative semiring identities. The fragment without
recursive types is universal for reversible boolean
circuits~\cite{James:2012:IE:2103656.2103667} and the extension with
recursive types and trace
operators~\cite{Hasegawa:1997:RCS:645893.671607} is a Turing-complete
reversible language~\cite{James:2012:IE:2103656.2103667,rc2011}. While
at first sight, $\Pi$ might appear \emph{ad hoc}, it really arises
naturally from an ``extended'' view of the Curry-Howard
correspondence~\cite{Carette2016}: rather than looking at mere
\emph{inhabitation} as the main source of analogy between logic and
computation, \emph{type equivalences} becomes the source of analogy.
This allows one to see an analogy between algebra and reversible
computation.  Furthermore, this works at multiple levels: that of
$1$-algebra (types form a semiring under isomorphism), but also
$2$-algebra (types and equivalences form a weak Rig Groupoid).  In
other words, by taking ``weak Rig Groupoid'' as the starting
semantics, one naturally gets $\Pi$ as the syntax for the language of
proofs of isomorphisms -- in the same way that many terms of the
$\lambda$-calculus arise from Cartesian Closed Categories.

One can also flip this around, and use the $\lambda$-calculus as the
internal language for Cartesian Closed Categories.  However, as
Shulman explains well in his draft book on approaching Categorical
Logic via Type Theory~\cite{shulman}, this works for many other kinds
of categories.  As we are interested in \emph{reversibility}, it is
most natural to look at Groupoids.  Thus $\PiTwo$ represents the
simplest non-trivial case of a (reversible) programming language
distilled from such ideas.

What is more surprising is how this also turns out to be a sound
and complete language for describing the univalent universe
{\small\AgdaFunction{U[𝟚]}}.

\paragraph*{The infinite real projective space $\mathbf{RP}^∞$}
\noindent Buchholtz and Rijke~\cite{buchholtz2017real} use the ``type
of two element sets,'' {\small\AgdaRecord{Σ[} \AgdaBound{X}
  \AgdaRecord{∶} \AgdaFunction{𝒰} \AgdaRecord{]} \AgdaPostulate{∥}
  \AgdaBound{X} \AgdaDatatype{==} \AgdaDatatype{𝕊⁰}
  \AgdaPostulate{∥}}, where~{\small\AgdaDatatype{𝕊⁰}} is the 0-sphere,
or the 0-iterated suspension of {\small\AgdaDatatype{𝟚}}, that is,
{\small\AgdaDatatype{𝟚}} itself. They construct the infinite real
projective space $\mathbf{RP}^∞$ by using universal covering spaces,
and show that it is homotopy equivalent to the Eilenberg-Maclane space
$K(\mathbb{Z}/2\mathbb{Z},1)$ which classifies all the 0-sphere
bundles. Our reversible programming language is exactly the syntactic
presentation of this classifying space. If we choose
{\small\AgdaDatatype{𝕊¹}} instead of {\small\AgdaDatatype{𝕊⁰}}, we get
the infinite complex projective space $\mathbf{CP}^∞$, but it remains
to investigate what kind of reversible programming language this would
lead to.

If we consider the $\Pi$ language over all finite types, we conjecture
that we should get a representation of
$\coprod_{n\in\mathbb{N}}K(S_n,1)$ where $S_n$ is the symmetric
group. The idea is that the $n^\mathrm{th}$ homotopy group of an
Eilenberg-Maclane space $K(G,n)$ is isomorphic to $G$ (and every other
homotopy group is trivial). Thus, all necessary information about
paths and equivalences between finite types is captured in this model.






\ack We would like to thank Robert Rose for developing the model based
on univalent fibrations, for extensive contributions to the code, and for
numerous discussions.

\bibliographystyle{acm}
{
  \footnotesize
  \bibliography{cites}
}

\end{document}